\documentclass[a4paper,11pt]{article}
\pdfoutput=1 

\usepackage{jheppub} 

\usepackage{cleveref}%
\usepackage[scr=rsfs]{mathalfa}%
\usepackage{mathtools}%
\usepackage{physics}%
\usepackage{subcaption}%
\usepackage{subfiles}%
\usepackage{tensor}%
\usepackage{tikz-cd}%
\usepackage{xspace}%

\graphicspath{{figures/}}%
\newcommand{\cf}{\textit{c.f.\@\xspace}}%
\newcommand{\Cf}{\textit{C.f.\@\xspace}}%
\newcommand{\eg}{\textit{e.g.\@\xspace}}%
\newcommand{\Eg}{\textit{E.g.\@\xspace}}%
\newcommand{\ie}{\textit{i.e.\@\xspace}}%
\title{\boldmath Thermodynamic stability from Lorentzian path integrals and
  codimension-two singularities}

\author{Hong Zhe (Vincent) Chen}

\affiliation{Department of Physics, University of California,\\
  Santa Barbara, CA 93106, USA}

\emailAdd{hzchen@ucsb.edu}

\abstract{It has previously been shown how the gravitational thermal partition
  function can be obtained from a Lorentzian path integral. Unlike the Euclidean
  case, the integration contour over Lorentzian metrics is not immediately ruled
  out by the conformal factor problem. One can then ask whether this contour can
  be deformed to pick up nontrivial contributions from various saddle points. In
  Einstein-Maxwell theory, we argue that the relevance of each black hole saddle
  to the thermal partition function depends on its thermodynamic stability
  against variations in energy, angular momentum, and charge. The argument
  involves consideration of constrained saddles where area and quantities
  associated with angular momentum and charge are fixed on a codimension-two
  surface. Consequently, this surface possesses not only a conical singularity,
  but two other types of singularities. The latter are characterized by shifts
  along the surface and along the Maxwell gauge group acquired as one winds
  around near the surface in a metric-orthogonal and connection-horizontal
  manner. We first study this enlarged class of codimension-two singularities in
  generality and propose an action for singular configurations. We then
  incorporate these configurations into the path integral calculation of the
  partition function, focusing on three-dimensional spacetimes to simplify the
  treatment of angular momentum.}

\begin{document}
\maketitle
\flushbottom

\section{Introduction}
\label{sec:intro}

Euclidean path integrals provide useful representations of various quantities in
quantum field theories. With the inclusion of gravity, however, the Euclidean
action becomes unbounded from below, due to the conformal mode of the metric.
This conformal factor problem precludes the integration contour over Euclidean
metrics as a possible choice for the gravitational path integral. One treatment
of this problem is to simply rotate the contour of integration for the conformal
mode \cite{Gibbons:1978ac}, but this prescription seems rather ad hoc. An
arguably more natural starting point is Lorentz signature
\cite{Schleich:1987fm,Mazur:1989by,Hartle:2020glw}. In contrast to the Euclidean
case, the integration contour over Lorentzian metrics is not immediately ruled
out by the conformal factor problem. It then becomes a reasonable question to
ask whether this integration contour can be deformed appropriately to pick up
nontrivial contributions from various saddle points. A goal of this paper is to
better understand the connection between this Lorentzian starting point and some
established intuition about Euclidean gravitational saddles.

In particular, it has long been recognized that Euclidean gravitational
solutions serve as important saddle points for path integrals in the
semiclassical approximation. Consider, for example, the gravitational analogue
of the grand canonical partition function
\begin{align}
  Z(\beta,\Omega,\Phi)
  &= \tr\left(e^{-\beta\,H_{\xi,\Phi}}\right)
    \;,
    \label{eq:euctrace}
\end{align}
where
\begin{align}
    H_{\xi,\Phi}
  &= H - \Omega\, J - \Phi\, Q
    \;,
\end{align}
\(H\) is the Hamiltonian generating evolution in a static
(Lorentzian\footnote{For the purposes of maintaining consistent notation, we
  will take the vector \(\zeta\) to be real in Lorentz signature and imaginary
  in Euclidean signature. The vector \(\varphi\) is real in both Euclidean and
  Lorentz signature. \Cref{eq:euctrace} has the usual interpretation as a grand
  canonical partition function when \(\Omega\) and \(\Phi\) are real; to obtain
  real Euclidean boundary conditions for the path integral, \(\Omega\) and
  \(\Phi\) should instead be imaginary. See
  \cref{sec:pathintegral,sec:lorpathint} for more details.}) time direction
\(\zeta\), \(J\) is the angular momentum generating rotation in a spatial
direction \(\varphi\), and \(Q\) is electric charge. More precisely, if the
gravitational theory is holographically dual to a theory on its boundary, then
\cref{eq:euctrace} would be the partition function of this boundary theory with
operators \(H\), \(J\), and \(Q\). Notwithstanding the conformal factor problem,
the gravitational partition function has been historically defined by a path
integral over a set of bulk geometries with a given boundary manifold. In
particular, the boundary has Killing vectors \(-i\,\zeta\) and \(\varphi\), and
contains a circle of length \(\beta\) generated by
\(-i\,\xi=-i\,(\zeta+\Omega\,\varphi)\). Other bulk fields are also integrated
over and the boundary conditions for a Maxwell field in particular are
parametrized by the electric potential \(\Phi\). As Gibbons and Hawking
\cite{Gibbons:1976ue} point out, such a path integral can then be
semiclassically approximated by using Euclidean black holes as saddle points.
Moreover, taking black hole thermodynamics seriously, one might expect only
thermodynamically stable black holes to be relevant. However, it is difficult to
verify this intuition from the Euclidean path integral without first specifying
a viable choice of integration contour.

More recently, ref.~\cite{Marolf:2022ybi} has made significant strides towards
showing how these results can arise starting from a purely Lorentzian path
integral. A key step in performing this path integral involves initially fixing
the area of a codimension-two surface \(\gamma\), giving rise to constrained
saddles with conical singularities on \(\gamma\). In this paper, we will study a
larger class of codimension-two singularities and find that they appear in
constrained saddles where area and quantities associated to angular momentum and
charge are fixed on \(\gamma\). To appreciate why this extended analysis is
interesting and what it teaches us about the thermodynamic stability of relevant
saddles, it is worthwhile to review some of the key points of
ref.~\cite{Marolf:2022ybi} below in \cref{sec:marolfreview}. Readers already
familiar the analysis of ref.~\cite{Marolf:2022ybi} may safely skip to
\cref{sec:newstuff}, where we sketch the new ideas to be explored in this paper.

\subsection{Lorentzian path integral to thermal partition function: a review}
\label{sec:marolfreview}

While the goal is to recover the thermal partition function
\(Z(\beta,\Omega,\Phi)\) of \cref{eq:euctrace}, the starting point of
ref.~\cite{Marolf:2022ybi} is a purely Lorentzian path integral
\(Z_{\mathrm{L}}(T,\Omega,\Phi)\). For simplicity and concreteness, let us focus
on Einstein-Maxwell theory with a negative cosmological constant. The path
integral \(Z_{\mathrm{L}}(T,\Omega,\Phi)\) is then an integral over real
Lorentzian geometries and Maxwell connections subject to certain boundary
conditions parametrized by \((T,\Omega,\Phi)\). For example, the spacetime
boundary is required to have a Lorentzian time circle \(S^1_{\mathrm{time}}\) of
parameter length \(T\) generated by the Killing vector
\(\xi=\zeta+\Omega\,\varphi\). The Maxwell field is also subject to boundary
conditions that are again parametrized by an electric potential \(\Phi\). Of
course, the integrand of the path integral is \(e^{i\,I}\) where \(I\) is the
Lorentzian action. At first glance, this integrand appears purely oscillatory,
but, in contrast to the Euclidean conformal factor problem, such integrals can
converge in an appropriate distributional sense as we will describe shortly.

\begin{figure}
  \centering
    \includegraphics{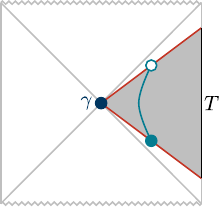}
    \caption{A conically singular constrained saddle constructed from a
      stationary black hole. The spacetime of the constrained saddle is the
      shaded region between the two red surfaces related by a boost around the
      bifurcation surface \(\gamma\). The red surfaces are identified. The
      bifurcation surface \(\gamma\) is now conically singular, characterized by
      a hyperbolic opening angle. The teal curve shows a closed curve which is
      contractible to a point on \(\gamma\).}
    \label{fig:lorconicalbh}
\end{figure}

A key point emphasized by ref.~\cite{Marolf:2022ybi}\footnote{See also similar
  comments made by ref.~\cite{Dittrich:2024awu} in the context of a Lorentzian
  simplicial path integral.} is that singular configurations must be included
in the path integral \(Z_{\mathrm{L}}(T,\Omega,\Phi)\). In particular, to
include configurations analogous to Euclidean black holes, the boundary cycle
\(S^1_{\mathrm{time}}\) must be allowed to contract in the bulk to a point on
some codimension-two surface \(\gamma\). The causal structure breaks down on
\(\gamma\) and this surface is conically singular, characterized by the opening
angle around \(\gamma\). \Cref{fig:lorconicalbh} illustrates one such
configuration constructed as a quotient of the exterior of a Lorentzian
stationary black hole with angular velocity \(\Omega\) and electric potential
\(\Phi\). This configuration is not a saddle for the full integral
\(Z_{\mathrm{L}}(T,\Omega,\Phi)\) because of the conical singularity on
\(\gamma\). Rather, it is a \emph{constrained} saddle which extremizes the
action \(I\) for a fixed value of the area of \(\gamma\) or, for later
convenience, let us say fixed
\begin{align}
  \mathcal{S}=\frac{\mathrm{Area}(\gamma)}{4G_{\mathrm{N}}}
  \;,
  \label{eq:bhentropy}
\end{align}
where \(G_{\mathrm{N}}\) is Newton's constant. Leaving the integral over
\(\mathcal{S}\) for last, we can decompose
\begin{align}
  Z_{\mathrm{L}}(T,\Omega,\Phi)
  &= \int \dd{\mathcal{S}}
    Z_{\mathrm{L}}(T,\Omega,\Phi;\mathcal{S})
    \label{eq:lorz}
\end{align}
in terms of path integrals \(Z_{\mathrm{L}}(T,\Omega,\Phi;\mathcal{S})\) over
subcontours of fixed \(\mathcal{S}\). The constrained saddle in
\cref{fig:lorconicalbh} is then a saddle for \(Z_{\mathrm{L}}(T,\Omega,\Phi;\mathcal{S})\).
In fact, using Morse theory, ref.~\cite{Marolf:2022ybi} argues that the contour
of integration for \(Z_{\mathrm{L}}(T,\Omega,\Phi;\mathcal{S})\) can always be deformed to
pick up nontrivial contributions from constrained saddles constructed in this
way.\footnote{See \cref{sec:bhstability} for more details.}

To leading order in the semiclassical approximation, the contribution
\(Z_{\mathrm{LBH}}(T,\Omega,\Phi;\mathcal{S})\) of such a constrained saddle is
determined by the value of the action \(I\),
\begin{align}
  Z_{\mathrm{LBH}}(T,\Omega,\Phi;\mathcal{S})
  &\sim e^{i I}
    = e^{\mathcal{S} - i\,T\,E_{\xi,\Phi}(\mathcal{S})}
    \;,
    \label{eq:ztsconstrsaddle}
\end{align}
where
\begin{align}
  E_{\xi,\Phi}(\mathcal{S})
  &=
    E(\Omega,\Phi;\mathcal{S})
    - \Omega\, J(\Omega,\Phi;\mathcal{S})
    - \Phi\, Q(\Omega,\Phi;\mathcal{S})
    \;,
\end{align}
with \(E(\Omega,\Phi;\mathcal{S})\), \(J(\Omega,\Phi;\mathcal{S})\), and
\(Q(\Omega,\Phi;\mathcal{S})\) being the energy, angular momentum, and charge of
the original black hole in \cref{fig:lorconicalbh} with angular velocity
\(\Omega\), electric potential \(\Phi\), and Bekenstein-Hawking entropy
\(\mathcal{S}\). Note, in particular, that the action has acquired an imaginary
part \(\Im(I_{\mathrm{LBH}})=-\mathcal{S}\), as can be seen from applying a
Lorentzian version of the Gauss-Bonnet theorem to the conical singularity
\cite{Colin-Ellerin:2020mva,Marolf:2022ybi}. Substituting
\cref{eq:ztsconstrsaddle} into \cref{eq:lorz},
\begin{align}
  Z_{\mathrm{LBH}}(T,\Omega,\Phi)
  &\sim
    \int \dd{\mathcal{S}}
    e^{\mathcal{S} - i\,T\,E_{\xi,\Phi}(\mathcal{S})
    }
    \;,
    \label{eq:ztconstrsaddle}
\end{align}
one finds an integrand which grows exponentially with the integration variable
\(\mathcal{S}\). As written, the above integral does not converge, at least to a
function of \(T\).

However, unlike the conformal factor problem in Euclidean
signature, this is both expected and manageable. It is expected because we are
computing the gravitational analogue of the trace
\begin{align}
  Z_{\mathrm{L}}(T,\Omega,\Phi)
  &= \tr(e^{-i\,T\,H_{\xi,\Phi}})\;.
    \label{eq:lortrace}
\end{align}
If each energy window contains \(\sim{e^{\mathcal{S}}}\)-many states where
\(\mathcal{S}\) grows polynomially with energy, then this trace, written as a
sum over energy windows, behaves similarly to \cref{eq:ztconstrsaddle}.

\Cref{eq:lortrace} also suggests that \(Z_{\mathrm{L}}(T,\Omega,\Phi)\) has a good
interpretation as a distribution in \(T\), rather than as a function. For example,
if we smear \cref{eq:lortrace} against a function \(f_\beta(T)\), with the
property that
\begin{align}
  \int_{-\infty}^\infty \dd{T} f_\beta(T)\, e^{-i\, T\, \omega}
  &= e^{-\beta\,\omega}
\end{align}
for all \(\omega\) above the ground state eigenvalue of \(H_\xi\), then we
expect to recover the usual thermal partition function \labelcref{eq:euctrace}:
\begin{align}
  \int_{-\infty}^\infty \dd{T} f_\beta(T)\, Z_{\mathrm{L}}(T,\Omega,\Phi)
  &= Z(\beta,\Omega,\Phi)
    \;.
    \label{eq:inttransintro}
\end{align}
Despite notational appearances, let us emphasize that the convention will
always be to perform the \(T\) integral \emph{before} the \(\mathcal{S}\)
integral in \(Z_{\mathrm{LBH}}(T,\Omega,\Phi)\) in order to realize its distributional
meaning. Applying this integral transform to \labelcref{eq:ztconstrsaddle}, one
might then expect to find the corresponding contribution
\(Z_{\mathrm{BH}}(\beta,\Omega,\Phi)\) to the thermal partition function \(Z(\beta,\Omega,\Phi)\) to be
given by
\begin{align}
  Z_{\mathrm{BH}}(\beta,\Omega,\Phi)
  &\stackrel{?}{\sim} \int \dd{\mathcal{S}}
    e^{\mathcal{S} - \beta\,E_{\xi,\Phi}(\mathcal{S})}
    \;.
    \label{eq:zbetaconstrsaddle}
\end{align}
Indeed, the above expression is what one would expect from the thermal trace
\labelcref{eq:euctrace} if we again identify \(e^{\mathcal{S}}\) as an
approximate density of states. (However, the validity of
\cref{eq:zbetaconstrsaddle} will later be called into question.)

At last, we may approximate the final integral over \(\mathcal{S}\) by
evaluating the integrand at its local maxima, \ie{} at the local minima
of the free energy
\begin{align}
  F(\beta,\Omega,\Phi;\mathcal{S})
  &= E_{\xi,\Phi}(\mathcal{S})
    - \frac{1}{\beta}\, \mathcal{S}
    \\
    &= E(\Omega,\Phi;\mathcal{S})
    - \Omega\, J(\Omega,\Phi;\mathcal{S})
    - \Phi\, Q(\Omega,\Phi;\mathcal{S})
    - \frac{1}{\beta}\, \mathcal{S}
\end{align}
with respect to \(\mathcal{S}\). Extrema occur whenever the temperature of the original
black hole in \cref{fig:lorconicalbh},
\begin{align}
  \frac{1}{\beta_{\mathrm{BH}}(\Omega,\Phi;\mathcal{S})}
  &= \dv{E_{\xi,\Phi}(\mathcal{S})}{\mathcal{S}}
    \;,
\end{align}
coincides with the ensemble temperature \(1/\beta\). In this sense, saddles for
the path integral \(Z(\beta,\Omega,\Phi)\) correspond to black holes with
temperature \(1/\beta\), angular velocity \(\Omega\), and electric potential
\(\Phi\). Indeed, the extremal values of the exponent in
\cref{eq:zbetaconstrsaddle} are simply minus the Euclidean action evaluated on
these black holes.

Actually, \cref{eq:zbetaconstrsaddle} suggests that not all black holes are
relevant saddles for the final partition function \(Z(\beta)\). In particular,
the relevant saddles must not merely be extrema, but in fact are local
\emph{maxima} of the integrand in \cref{eq:zbetaconstrsaddle} or, equivalently,
local \emph{minima} of the free energy \(F\) with respect to
\(\mathcal{S}\).\footnote{Let us refrain here from precisely defining the
  meaning of ``relevant''. It should be evident that the local maxima of
  integrand in \cref{eq:zbetaconstrsaddle} unambiguously contribute with nonzero
  weights to the integral on the RHS. Whether the integral receives
  contributions from extrema that are not local maxima of the integrand is a
  more subtle question. The answer depends on the precise definition of Lefshetz
  thimbles (\ie{}, steepest descent contours flowing from extrema in the
  complexified space of the integration variables). We will postpone this
  discussion for \cref{sec:toyexample2}. \label{foot:unstablesaddles}} Changing
variables from \(\mathcal{S}\) to \(E_{\xi,\Phi}\), this can be understood
physically as additionally requiring the positivity of specific heat \(C\), given by
\begin{align}
  \frac{1}{C}
  &= \dv{E_{\xi,\Phi}} \frac{1}{\beta_{\mathrm{BH}}(\Omega,\Phi;E_{\xi,\Phi})}
    \;.
    \label{eq:specheat}
\end{align}
The conclusion of ref.~\cite{Marolf:2022ybi} is therefore that relevant saddles
for the thermal partition function \(Z(\beta)\) include only those black holes
that are thermodynamically stable in the sense of having positive specific heat.

\subsection{Questions to be addressed and new ideas}
\label{sec:newstuff}

Although the results reviewed above seem satisfying, they leave some things to
be desired. In particular, following thermodynamic intuition, shouldn't we have
instead expected to find a more complete stability condition? For example, as
ref.~\cite{Marolf:2022ybi} remarks in its discussion, it would have been more
natural to expect the free energy to be locally minimized with respect to
\emph{independent} variations of a complete set of thermodynamic variables,
\eg{} energy, angular momentum, and charge. Relatedly, the special treatment of
\(\mathcal{S}\) seems rather undemocratic --- what happens if we also fix
other quantities in the path integral until after the integral transform on the
Lorentzian time \(T\)? These are the main questions which will motivate our
extended analysis in this paper.

In particular, we will consider the consequences of fixing three independent
quantities \((\mathcal{S},\mathcal{J},\mathcal{Q})\) on a generically singular
codimension-two surface \(\gamma\) in a subcontour path integral
\(Z(T,\Omega,\Phi;\mathcal{S},\mathcal{J},\mathcal{Q})\). Here, \(\mathcal{S}\)
is again proportional to the area of \(\gamma\) as given in
\labelcref{eq:bhentropy}; \(\mathcal{J}\) is the integral of a local momentum
density \((p_{\mathrm{BY}})_i\) on \(\gamma\) arising from a Brown-York-like
construction; and
\begin{align}
  \mathcal{Q}
  &= \frac{1}{g_{\mathrm{M}}^2} \int_\gamma *F
\end{align}
is an electric charge as given by Gauss's law from the electric flux
\(\int_\gamma *F\) across \(\gamma\). If evaluated on the bifurcation surface
\(\gamma\) of a stationary black hole solution, these quantities would coincide
with the Bekenstein-Hawking entropy, angular momentum, and electric charge of
the black hole (with the latter two ordinarily defined at the spacetime
boundary). However, in generic spacetimes without preferred vector fields on
\(\gamma\), it is not immediately obvious how one should define \(\mathcal{J}\)
from a momentum \emph{dual-vector} density \((p_{\mathrm{BY}})_i\) on
\(\gamma\). We will therefore focus our analysis of angular momentum to \(D=3\)
spacetime dimensions, where the unit vector \(\chi^i\) on \(\gamma\) can be used
to define \(\mathcal{J}\propto\int_{\gamma}\chi^i\,(p_{\mathrm{BY}})_i\).
(Notwithstanding a few issues\footnote{See \cref{sec:angmom} for a discussion of
  these issues in higher dimensions.} such as this, most portions of this paper
are written to allow immediate generalizations to higher dimensions.)

\begin{figure}
  \centering
  \includegraphics{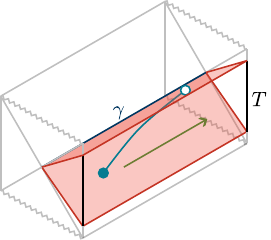}
  \caption{A conically and helically singular constrained saddle. An extra
    spatial symmetry direction is shown compared to \cref{fig:lorconicalbh};
    \eg{} this can be the (periodically identified) angular direction for a
    three-dimensional black hole. Whereas the red surfaces of constant time
    \(\hat{t}\) were presumably identified in \cref{fig:lorconicalbh} in a way
    that trivializes near \(\gamma\), they are shown here as identified after a
    relative shift, indicated by the green arrow. In particular, the teal closed
    curve does not become orthogonal to surfaces of constant \(\hat{t}\) even as
    the identified endpoints are pushed towards \(\gamma\). The bifurcation
    surface \(\gamma\) now also has a helical singularity, characterized by the
    shift indicated in green.}
    \label{fig:lorhelicalbhsimp}
\end{figure}

We will find that appropriate constrained saddles for
\(Z(T,\Omega,\Phi;\mathcal{S},\mathcal{J},\mathcal{Q})\) can be constructed from
the procedure illustrated in \cref{fig:lorhelicalbhsimp}. In particular, we
start with a stationary black hole with the prescribed values of
\((\mathcal{S},\mathcal{J},\mathcal{Q})\) on its bifurcation surface \(\gamma\)
--- such a black hole might have an angular velocity \(\Omega_0\) and electric
potential \(\Phi_0\) differing from \(\Omega\) and \(\Phi\). The boundary
Killing vectors \(\zeta\) and \(\varphi\) extend into the interior of this
configuration, and it is useful to introduce a time coordinate \(\hat{t}\)
coinciding with the Killing parameter of \(\zeta\) and preserved by \(\varphi\).
To match the boundary conditions prescribed by \((T,\Omega)\), we again quotient
one exterior of the black hole with parameter period \(T\) along the flow
generated by \(\xi=\zeta+\Omega\,\varphi\). To match the electric potential
\(\Phi\), we also shift the Maxwell field \(A\) by
\((\Phi-\Phi_0)\dd{\hat{t}}\).\footnote{We take it to be part of the definition
  of the boundary conditions (and the coordinate \(\hat{t}\) on the boundary)
  that shifting \(A\) by \((\Phi-\Phi_0)\dd{\hat{t}}\) in a neighbourhood of the
  boundary takes a configuration satisfying boundary conditions with electric
  potential \(\Phi_0\) to another configuration with electric potential \(\Phi\).}

In this constrained saddle, \(\gamma\) is now a singular surface. For example,
as before, there is a conical singularity characterized by the opening angle
around \(\gamma\). Because generically \(\Omega\ne\Omega_0\) and
\(\Phi\ne\Phi_0\), two other types of singularities now appear on \(\gamma\).

Let us first consider what ref.~\cite{Marolf:2022ybi} dubbed, and what we will
continue to call, the \emph{helical} singularity on \(\gamma\). Because
\(\Omega\ne\Omega_0\), the Killing vector \(\xi=\zeta+\Omega\,\varphi\) does not
vanish on the bifurcation surface \(\gamma\) of the original black hole
geometry. Consequently, the surfaces of constant time \(\hat{t}\) illustrated in
red in \cref{fig:lorhelicalbhsimp} are identified with a relative shift, in the
\(\varphi\) direction, which remains nontrivial even as we approach \(\gamma\).
The closed orbits generated by \(\xi\) near \(\gamma\), \eg{} the teal curve in
\cref{fig:lorhelicalbhsimp}, appears to move in the \(\varphi\) direction along
\(\gamma\), but only because the picture is drawn to be approximately faithful
to metric angles. On the other hand, in a picture faithfully showing identified
points,\footnote{For example, see \cref{fig:coordinates}. A closed loop
  parametrized by \(\tau\) is obtained at fixed \(\rho\) and \(y^i\).} the teal
curve would appear as a closed loop while another curve orthogonal to slices of
constant time \(\hat{t}\) would generically wind helically around \(\gamma\).
(We will find it useful in this paper to adopt this latter picture, where it is
the closed teal curve of \cref{fig:lorhelicalbhsimp} that hovers near a fixed
``point'' on \(\gamma\).\footnote{More precisely, we view this teal closed curve
  as being contactable to a point on \(\gamma\) in a regulated version of the
  geometry where a small neighbourhood of \(\gamma\) is used to smooth out the
  singularity, as described in \cref{sec:regsing}. Other smooth in-fillings of
  the neighbourhood can lead to physically distinct singularities, labelled by
  other values of the helical shifts, as described in \cref{sec:singdisambig},
  or even having no helical singularity, as sketched in
  \cref{fig:conicalcrtbh}.}) Some previous work which considered helical
singularities (though, not under this name) in special contexts include
refs.~\cite{Tod:1994iop,Chua:2023srl}; the Euclidean analysis of the latter
bears some resemblance to the naive Euclidean path integral calculation
presented in \cref{sec:partfunc}.

Due to the shift by \((\Phi-\Phi_0)\dd{\hat{t}}\), there is also on \(\gamma\)
what we will call a \emph{holonomic} singularity in the Maxwell field \(A\).
\Cref{fig:lorhelicalbhsimp} doubles as an illustration of this singularity if we
reinterpret what we previously viewed as the spatial angular direction now as
the fibre direction of the Maxwell principal fibre bundle. In particular, the
two red surfaces over constant \(\hat{t}\) in the principal fibre bundle are
identified with not only a helical shift in the \(\varphi\) direction as
discussed in the previous paragraph, but also a shift in the gauge fibre
direction. If we turn off the helical singularity (\eg{}, if
\(\Omega=\Omega_0\)) to isolate the effect of the holonomic singularity, then a
Wilson loop around a closed orbit of \(\xi\) infinitesimally near \(\gamma\)
will be proportional to the strength of the holonomic singularity --- hence the
name.\footnote{But more generically, this Wilson loop will be nonzero due to a
  combined effect of the helical and holonomic singularities.}
Ref.~\cite{Maxfield:2020ale} has previously studied holonomic singularities in
\(D=2\) dimensions.

The enlarged class of codimension-two singularities --- conical, helical, and
holonomic --- considered in this paper, to my knowledge, has not been treated as
thoroughly in literature relative to purely conical singularities.\footnote{For
  example, refs.~\cite{Tod:1994iop,Chua:2023srl} considered helical
  singularities in highly symmetric configurations. Ref.~\cite{Maxfield:2020ale}
  considered holonomic singularities in \(D=2\) dimensions resulting from a
  dimensional reduction of smooth \(D=3\) configurations, so any action
  contribution localized on the singularity is purposely excluded. We would
  instead like to study conical, helical, and holonomic singularities in generic
  configurations that might appear in the path integral and write down an action
  including contributions from the singularity. This in particular requires
  resolving some subtleties that become apparent only when the singularity is
  regulated, as described in \cref{sec:regsing}. Further complications can arise
  in Lorentzian signature, as described in \cref{sec:lightcones}.} A significant
portion of this paper will be dedicated to studying properties of these
singularities in general Euclidean and Lorentzian configurations. In the
presence of these new types of singularities, we must in particular revisit the
definition of the Einstein-Maxwell action. Our strategy, for deducing what the
action should be, will be to consider regulated configurations where the
singularity on \(\gamma\) has been smoothed out over a neighbourhood of
\(\gamma\). A perhaps surprising comment is that certain configurations which
seem diffeomorphic or gauge equivalent to each other away from \(\gamma\) can
actually be physically distinguished by very different in-fillings of the
regulated neighbourhood. Considering localized curvature contributions in the
regulated neighbourhood, we will propose a definition for the action of singular
configurations --- see \cref{eq:action,eq:loraction}. Admittedly, some open
questions of interpretation remain for certain infinite terms that arise in this
derivation but are omitted from our proposed action. Regardless, we find that
our action leads to a reasonable variational principle. Moreover, at fixed
\((\mathcal{S},\mathcal{J},\mathcal{Q})\), we are able to construct constrained
saddles as illustrated in \cref{fig:lorhelicalbhsimp} with conical, helical, and
holonomic singularities.

Continuing now our discussion of partition functions, the path integral
\(Z(T,\Omega,\Phi;\mathcal{S},\mathcal{J},\mathcal{Q})\) over a subcontour of
fixed \((\mathcal{S},\mathcal{J},\mathcal{Q})\) receives a nontrivial
contribution
\(Z_{\mathrm{BH}}(T,\Omega,\Phi;\mathcal{S},\mathcal{J},\mathcal{Q})\) from each
such constrained saddle constructed from a black hole. This follows from Morse
theory arguments just as in ref.~\cite{Marolf:2022ybi} --- see
\cref{sec:bhstability}. At leading order in the semiclassical approximation,
each contribution
\begin{align}
  Z_{\mathrm{LBH}}(T,\Omega,\Phi;\mathcal{S},\mathcal{J},\mathcal{Q})
  &\sim e^{i I}
    = e^{\mathcal{S} - i\,T\,E_{\xi,\Phi}(\mathcal{S},\mathcal{J},\mathcal{Q})}
    \;,
    \label{eq:ztsjqconstrsaddle}
\end{align}
is given by the value of the action, where now
\begin{align}
  E_{\xi,\Phi}(\mathcal{S},\mathcal{J},\mathcal{Q})
  &=
    E(\mathcal{S},\mathcal{J},\mathcal{Q})
    - \Omega\, \mathcal{J}
    - \Phi\, \mathcal{Q}
    \;,
\end{align}
\(E(\mathcal{S},\mathcal{J},\mathcal{Q})\) is the energy of the original black
hole in \cref{fig:lorhelicalbhsimp} with Bekenstein-Hawking entropy
\(\mathcal{S}\), (angular) momentum \(\mathcal{J}\), and electric charge
\(\mathcal{Q}\).

The analysis then proceeds similarly to \cref{sec:marolfreview}, but we now
treat \((\mathcal{S},\mathcal{J},\mathcal{Q})\) on the same footing. In
particular, by integrating \cref{eq:ztsjqconstrsaddle}, we obtain the
contribution
\begin{align}
  Z_{\mathrm{LBH}}(T,\Omega,\Phi)
  &\sim
    \int \dd{\mathcal{S}} \dd{\mathcal{J}} \dd{\mathcal{Q}}
    e^{\mathcal{S} - i\,T\,E_{\xi,\Phi}(\mathcal{S},\mathcal{J},\mathcal{Q})}
    \label{eq:ztconstrsaddle2}
\end{align}
to the path integral \(Z_{\mathrm{L}}(T,\Omega,\Phi)\) over the full contour
over Lorentzian configurations. We can further apply the integral transform
\labelcref{eq:inttransintro} to obtain the corresponding contribution
\(Z_{\mathrm{BH}}(\beta,\Omega,\Phi)\) to the thermal partition function
\(Z(\beta,\Omega,\Phi)\). Similar to our previous treatment of \(\mathcal{S}\)
in \cref{sec:marolfreview}, our prescription in this paper will be to always
leave the \((\mathcal{S},\mathcal{Q},\mathcal{J})\) integrals until \emph{after}
the \(T\) integral. We thus find:
\begin{align}
  Z_{\mathrm{BH}}(\beta,\Omega,\Phi)
  &\sim \int \dd{\mathcal{S}}\dd{\mathcal{J}}\dd{\mathcal{Q}}
    e^{\mathcal{S} - \beta\,E_{\xi,\Phi}(\mathcal{S},\mathcal{J},\mathcal{Q})}
    \;.
    \label{eq:zbetaconstrsaddle2}
\end{align}
Again, these final integrals can be evaluated with saddle points corresponding
to black holes with temperature \(1/\beta\), angular velocity \(\Omega\), and
electric potential \(\Phi\).

Note however that the relevant\footnote{See
  \cref{foot:unstablesaddles,sec:toyexample2}.} saddles are now local maxima of
the integrand in \cref{eq:zbetaconstrsaddle2}, \ie{} local minima of the free
energy
\begin{align}
  F(\beta,\Omega,\Phi;\mathcal{S},\mathcal{J},\mathcal{Q})
  &= E_{\xi,\Phi}(\mathcal{S},\mathcal{J},\mathcal{Q})
    - \frac{1}{\beta}\, \mathcal{S}
  \\
  &= E(\mathcal{S},\mathcal{J},\mathcal{Q})
    - \Omega\, \mathcal{J}
    - \Phi\, \mathcal{Q}
    - \frac{1}{\beta}\, \mathcal{S}
\end{align}
with respect to the three independent variables
\((\mathcal{S},\mathcal{J},\mathcal{Q})\). Alternatively, one may change
variables to the more standard set of thermodynamic variables
\((E,\mathcal{J},\mathcal{Q})\) and view
\(\mathcal{S}(E,\mathcal{J},\mathcal{Q})\) as a function giving the
Bekenstein-Hawking entropy of a black hole with energy \(E\), angular momentum
\(\mathcal{J}\), and charge \(\mathcal{Q}\). At any rate, we recover a stronger
thermodynamic stability condition relative to the analysis of
ref.~\cite{Marolf:2022ybi} summarized in \cref{sec:marolfreview}.

How did this happen? Relative to the previous analysis, we seem to have merely
reorganized the ordering of the \(T\) integral relative to the \(\mathcal{J}\)
and \(\mathcal{Q}\) integrals.\footnote{One might also note that we have now
  allowed a larger class of singularities in the path integral. However, even if
  the preceding analysis summarized in \cref{sec:marolfreview} included such
  configurations in the path integral, without fixing \(\mathcal{J}\) and
  \(\mathcal{Q}\), constrained saddles at fixed \(\mathcal{S}\) would still only
  have purely conical singularities.} In the discussion of
\cref{sec:discussion}, we will emphasize how the order of integration can be
particularly important in Lorentzian path integrals evaluated using approximate
saddle-point methods. Specifically,
if performed \emph{after} the integral over possibly ``unstable'' variables such
as \(\mathcal{J}\) and \(\mathcal{Q}\), then the integral transform from \(T\)
to \(\beta\) can behave rather pathologically. For example, we will see from
some toy examples how one-loop corrections to \cref{eq:ztconstrsaddle} from any
such unstable variable can, under the integral transform, map to a nonsensical
answer differing drastically from the naively expectation
\labelcref{eq:zbetaconstrsaddle}. To avoid these pathologies, one should
therefore perform the integral transform in \(T\) \emph{before} the integrals
over possibly unstable variables such as
\((\mathcal{S},\mathcal{J},\mathcal{Q})\). An interesting question which we will
leave for future work is whether saddles that are stable with respect to
variations in \((\mathcal{S},\mathcal{J},\mathcal{Q})\) might be unstable with
respect to other variables (which one should then also integrate last in the
path integral) --- see \cref{sec:unstablevars}.

\subsection{Overview of this paper}

Let us now give an overview of the remainder of this paper. While conical,
helical, and holonomic singularities will serve as important ingredients for our
gravitational path integral, the latter two types of singularities might seem
somewhat exotic and, to my knowledge, has not been analyzed in existing
literature to a sufficient degree for our purposes. A significant portion of
this paper is therefore dedicated to studying our enlarged class of
codimension-two singularities. We will therefore start studying these
singularities in the more modest Euclidean context in
\cref{sec:sing,sec:action,sec:partfunc} before graduating to Lorentzian
signature in \cref{sec:lorentzian,sec:discussion}.

\paragraph{\Cref{sec:sing}} is dedicated to specifying these singularities and studying
their properties. In particular, conical and helical geometries are defined in
\cref{sec:conhelsings} while holonomic singularities in the Maxwell
configuration are defined in \cref{sec:holsing}. Some questions about whether
singularity strengths and related parameters are allowed to vary around and
along \(\gamma\) are discussed in \cref{sec:stationarity}.

To better understand these singularities, in \cref{sec:regsing}, we consider a
procedure for regulating them in an \(\varepsilon\)-neighbourhood
\(\mathscr{N}_\varepsilon\) of the singular surface \(\gamma\). Firstly, this
demonstrates that configurations which are naively diffeomorphic or gauge
equivalent to each other away from \(\gamma\) can have physically distinct
internal structure within \(\mathscr{N}_\varepsilon\) when regulated. Secondly,
in \cref{sec:curvature}, this allows us to extract localized curvature
contributions, of both the geometry and Maxwell connection, from the regulated
\(\mathscr{N}_\varepsilon\).

\paragraph{\Cref{sec:action}} is dedicated to studying the action for the codimension-two
singularities introduced in \cref{sec:sing}. Motivated by the curvature
contributions found in \cref{sec:curvature}, we first define our proposal for
the action of singular configurations in \cref{sec:actionprop}. This action
includes terms accounting for the conical singularity, previously appearing also
in ref.~\cite{Marolf:2022ybi}, and additional terms associated to the helical
and holonomic singularities. Subtleties related to the cutoff surface
\(\partial\mathscr{N}_\varepsilon\), on which some of these terms reside, are
discussed in \cref{sec:shapeindep}.

Having specified the action, we then consider its variation in
\cref{sec:actionvar}, giving rise to equations of motions away from \(\gamma\)
and boundary terms near \(\gamma\). A Brown-York stress tensor on
\(\partial\mathscr{N}_\varepsilon\) provides a convenient expression for the
gravitational boundary terms here that arise in this variation. Its
``time-space'' component, in particular, provides a useful notion of a momentum
density \((p_{\mathrm{BY}})_i\) along \(\gamma\). By setting the variation of
the action to zero, possibly subject to fixed constraints on \(\gamma\), we
deduce in \cref{sec:saddle} conditions obeyed by saddles and constrained
saddles. In particular, we will see in \cref{sec:constrsaddle} that fixing the
area, integrated momentum density \((p_{\mathrm{BY}})_i\), and electric flux on
a codimension-two surface \(\gamma\) leads to constrained saddles which
generically have conical, helical, and holonomic singularities on \(\gamma\).

\paragraph{\Cref{sec:partfunc}} then considers a naively \emph{Euclidean} path
integral calculation of the thermal partition function \(Z(\beta,\Omega,\Phi)\).
Of course, the integral over Euclidean metrics is doomed at the outset by the
conformal factor problem; at the same time, this naive representation of
\(Z(\beta,\Omega,\Phi)\) is also perhaps familiar to most readers. Remaining
agnostic to the actual contour of integration, we will
therefore use the Euclidean framework to practice constructing
(constrained) saddles.

Specifically, in \cref{sec:bhsaddles}, we will consider evaluating
\(Z(\beta,\Omega,\Phi)\) by first fixing, and then later integrating over, quantities
\((\mathcal{S},\mathcal{J},\mathcal{Q})\) proportional to the area, integrated
momentum density, and electric flux on a generically singular codimension-two
surface \(\gamma\). We then look for constrained saddles which are saddles for
the intermediate integral
\(Z(\beta,\Omega,\Phi;\mathcal{S},\mathcal{J},\mathcal{Q})\) over a subcontour
of fixed \((\mathcal{S},\mathcal{J},\mathcal{Q})\). In particular, starting from
a smooth Euclidean black hole, we construct conically, helically, and
holonomically singular constrained saddles using a procedure analogous to
\cref{fig:lorhelicalbhsimp}. The resulting contribution to
\(Z(\beta,\Omega,\Phi)\) is \cref{eq:zbetaconstrsaddle2}, but derived with
complete disregard for the conformal factor problem. Of course, this is
unsatisfactory.

\paragraph{\Cref{sec:lorentzian}} therefore extends our analysis to Lorentz
signature. The first step, in \cref{sec:lorsing}, is to translate what we have
learned about conical, helical, and holonomic singularities from Euclidean to
Lorentz signature. As described in \cref{sec:nolightcones}, this is fairly
straightforward when the singular surface \(\gamma\) has no lightcones in the
Lorenzian configuration --- indeed, this is the case for the configuration
illustrated in \cref{fig:lorhelicalbhsimp}. For completeness, however, we also
include in \cref{sec:lightcones} a fairly detailed discussion of cases where the
spacetime contains lightcones for \(\gamma\), leading to complications, for
example, in evaluating the action.

The second step, in \cref{sec:lorpartfunc}, is to re-evaluate the thermal
partition function \(Z(\beta,\Omega,\Phi)\) using a Lorentzian path integral
\(Z(T,\Omega,\Phi)\). This analysis, already summarized in \cref{sec:newstuff},
leads to the contribution \labelcref{eq:ztconstrsaddle2} from constrained
saddles built from Lorentzian black holes as illustrated in
\cref{fig:lorhelicalbhsimp}. Upon performing the integral transform from \(T\)
to \(\beta\), \cref{eq:zbetaconstrsaddle2} is reproduced now from a purely
Lorentzian path integral without the conformal factor problem.

\paragraph{\Cref{sec:discussion}} concludes this paper with a discussion. After
a brief summary in \cref{sec:summary}, we further flesh out in
\cref{sec:bhstability} some of the ideas sketched above in \cref{sec:newstuff}
concerning the thermodynamic stability of saddles relevant to the grand
canonical partition function. For example, as an intermediate step, some basic
Morse theory is reviewed to argue that the constrained saddles built from
Lorentzian black holes do in fact contribute with nonzero weight to the path
integrals \(Z(T,\Omega,\Phi;\mathcal{S},\mathcal{J},\mathcal{Q})\) over
subcontours of fixed \((\mathcal{S},\mathcal{J},\mathcal{Q})\). Moreover, we
again highlight that, by leaving the \((\mathcal{S},\mathcal{J},\mathcal{Q})\)
integrals for last, we are able to recover a more complete stability condition
which indicates the relevance of a given black hole saddle to the final
partition function \(Z(\beta,\Omega,\Phi)\).

In \cref{sec:toyexamples}, we study toy examples firstly to explain why this
order of integration is particularly important for potentially unstable
variables like \((\mathcal{S},\mathcal{J},\mathcal{Q})\) when using approximate
saddle-point methods. Secondly, we also address the question of whether unstable
saddles contribute to the path integral. The answer can depend subtly on the
definition of Lefschetz thimbles (\ie{} steepest descent contours) for stable
saddles, such that, without precisely specifying the integrals along the latter,
contributions from unstable saddles can sometimes become meaningless.

\Cref{sec:bhsums} discusses an interesting feature that arises from the
equivalence of boundary conditions related by full rotations in space and around
a compact Maxwell gauge group. As described in
\cref{sec:partfunc,sec:lorpartfunc}, from this equivalence, our construction
naturally generates a sum over constrained saddles, which enforces the
quantization of angular momentum and charge. This then leads to a discrete
family of smooth black hole saddles for the final integrals over
\((\mathcal{S},\mathcal{J},\mathcal{Q})\). For AdS\(_3\), we identify this
family as a subset of the \(\mathrm{SL}(2,\mathbb{Z})\) black holes
\cite{Maloney:2007ud}. We also sketch how some other \(\mathrm{SL}(2,\mathbb{Z})\)
black holes might fit into our formalism by incorporating constrained saddles
generated from exotic, \eg{} CRT-twisted \cite{Harlow:2023hjb}, black holes.

Finally, we end in \cref{sec:openproblems} with a discussion of open questions.

\section{Codimension-two singularities}
\label{sec:sing}

Let us start by specifying and studying properties of conical, helical, and
holonomic singularities in Euclidean signature. (Lorentz signature will be
treated later in \cref{sec:lorentzian}.) In \cref{sec:singconfigs}, we will
first introduce these codimension-two singularities in generality, as might be
found in a generic configuration included in the path integral. Then, in
\cref{sec:regsing}, we will introduce a procedure for regulating the
singularities in a small neighbourhood \(\mathscr{N}_\varepsilon\) of the
singular surface \(\gamma\). This will firstly give us a better understanding of
whether singular configurations which are diffeomorphic or gauge equivalent to
each other away from \(\gamma\) are actually physically distinct when regulated
in \(\mathscr{N}_\varepsilon\). Secondly, in \cref{sec:curvature}, we will be
able to extract curvature contributions from the regulated
\(\mathscr{N}_\varepsilon\), in preparation for defining an action for singular
configurations in \cref{sec:action}.

\subsection{Specifying codimension-two singularities}
\label{sec:singconfigs}

We will now describe the codimension-two singularities \(\gamma\) by specifying
how the metric and Maxwell field are allowed to behave near \(\gamma\).

\subsubsection{Conical and helical singularities}
\label{sec:conhelsings}

We start with codimension-two singularities of the metric. In addition to the
more familiar purely conical singularities, we will consider also singularities
which impart a ``helical'' shift upon going around the codimension-two surface.
Refs.~\cite{Tod:1994iop,Chua:2023srl}, for example, have previously considered
highly symmetric cases of these singularities constructed by cutting and gluing
flat spacetime and black holes respectively, with the latter similar to the
constrained saddles in \cref{sec:bhsaddles}. Let us, however, give a general
description of conical and helical singularities.

\begin{figure}
  \centering
  \includegraphics{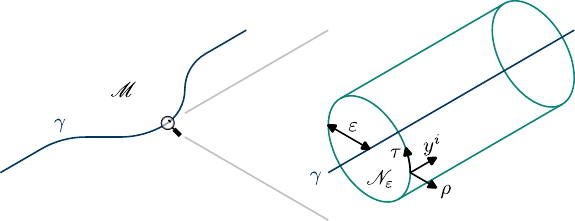}
  \caption{A codimension-two singularity \(\gamma\) and coordinates in its
    neighbourhood. Metric angles are \emph{not} accurately depicted in this
    figure.}
  \label{fig:coordinates}
\end{figure}

On a \(D\)-dimensional spacetime \(\mathscr{M}\), we consider Euclidean metric
configurations which are smooth, with the possible exception of a
codimension-two surface \(\gamma\). In a neighbourhood of \(\gamma\), we require
the metric to take the form\footnote{This way of decomposing the metric is
  inspired by ref.~\cite{Chua:2023srl}. Whereas ref.~\cite{Chua:2023srl} applies
  the ADM decomposition to a time slice, our ADM decomposition
  \labelcref{eq:metricg} is of a surface at constant radial separation from
  \(\gamma\). Our \cref{fig:coordinates} can be compared with
  ref.~\cite{Chua:2023srl}'s fig.~3.}
\begin{align}
  G_{AB} \dd{X^A} \dd{X^B}
  &= \dd{\rho}^2 + g_{ab} \dd{x}^a \dd{x}^b
    \label{eq:metricG}
  \\
  g_{ab} \dd{x}^a \dd{x}^b
  &= N^2 \dd{\tau}^2
    + h_{ij}\,
    (\dd{y}^i + N^i \dd{\tau})
    (\dd{y}^j + N^j \dd{\tau})
    \;.
    \label{eq:metricg}
\end{align}
As illustrated in \cref{fig:coordinates}, \(X^A=(\rho,x^a)\) are
\(D\)-dimensional spacetime coordinates, with \(\rho\) a radial geodesic
coordinate\footnote{Our coordinates are Gaussian normal coordinates associated
  to surfaces of constant proper distance \(\rho\) from \(\gamma\) and the
  orthogonal geodesic vector field \(u^A = (\partial_\rho)^A\).} starting at \(\eval{\rho}_\gamma=0\);
\(x^a=(\tau,y^i)\) are coordinates on constant \(\rho\) surfaces, with
\(\tau\sim \tau + 2\pi\) an angular coordinate around
\(\gamma\); and \(y^i\) are coordinates in the remaining directions, parallel to
\(\gamma\).

Smooth metric configurations have, up to partial gauge-fixing,
\begin{align}
  N
  &= \rho + \order{\rho^3} \;,
  &
    N^i
  &=\order{\rho^2}
    \;.
    \label{eq:shiftlapsesmooth}
\end{align}
Instead, we will also allow generically singular configurations\footnote{To more
  precisely map out the admissible forms of the small \(\rho\) expansions of
  metric components, one might try to generalize the analysis in the appendices
  of ref.~\cite{Dong:2019piw} which considered only conical
  singularities. \label{foot:tauindep}}
\begin{align}
  N &= \kappa\, \rho
      + (\text{subleading in }\rho)
      \;,
  &
    N^i &= v^i + \order{\rho^2}
          \;.
          \label{eq:shiftlapsesing}
\end{align}
Here, \(\kappa\) describes the rate of change in metric angle around \(\gamma\)
with respect to \(\tau\); if constant, \(2\pi\kappa\) is the proper opening
angle around \(\gamma\). Meanwhile, \(v^i\) parametrizes the ``helical'' nature
of the singularity, as we describe shortly. We will further require
\(\lim_{\rho\to0}h_{ij}\) to have a finite limit. In \cref{sec:stationarity}, we
will explain our assumptions about whether various quantities, such as
\(\kappa\), \(v^i\), and \(\lim_{\rho\to0}h_{ij}\), are allowed to depend on
\(\tau\) and/or \(y^i\).

To understand the helical nature of the singularity, note that a small closed
circle around \(\gamma\), parametrized by \(\tau\sim \tau+2\pi\) is obtained at
fixed \(\rho\) and \(y^i\). However, even as \(\rho\to 0\), such a path is not
orthogonal to \(\gamma\) in the metric sense, as a consequence of the nontrivial
shift \(N^i \to v^i\). Relatedly, such a closed circle which shrinks towards
\(\gamma\) can nonetheless retain a nonzero metric length even though the circle
``contracts to a point'' on \(\gamma\).\footnote{Topology and the notion of
  contractibility near \(\gamma\) will become more precise when considering a
  regulated version of the geometry, as described in \cref{sec:regsing}. In the
  regulated geometry, contractible cycles do shrink to zero proper size.} A
path which is metric-orthogonal to \(\gamma\) is instead one on which the
pullback of \(\dd{y^i} + N^i \dd{\tau}\) vanishes. However, this latter path is
no longer closed but rather winds helically around \(\gamma\). In particular,
upon moving by \(\delta\tau\) in \(\tau\) along such a path infinitesimally
close to \(\gamma\), one is displaced by \(-v^i\delta\tau\) along \(\gamma\).

Before proceeding, let us establish some additional notation. Let us call the
unit normals of the \((D-1)+1\) and \((D-2)+1\) decompositions in
\cref{eq:metricG,eq:metricg}
\begin{align}
  u &= \dd\rho
      \;,
  &
    n
  &= N\dd{\tau}
    \;,
    \label{eq:normals}
\end{align}
and write
\begin{align}
  G_{AB}
  &= u_A u_B + g_{AB} = u_A u_B + n_A n_B + h_{AB}
    \;,
  &
    g_{ab}
  &= n_a n_b + h_{ab}
    \;,
\end{align}
thus defining the uplifted tensors \(g_{AB}\), \(h_{AB}\), and \(h_{ab}\). These
are projected down to \(g_{ab}\) and \(h_{ij}\) by
\begin{align}
  \tensor{g}{^A_a}
  &= \frac{\partial X^A}{\partial x^a}
    \;,
  &
    \tensor{h}{^a_i}
  &= \frac{\partial x^a}{\partial y^i}
    \;.
    \label{eq:projections}
\end{align}
We will lower/raise indices \((A,B,\ldots)\),
\((a,b,\ldots)\), and \((i,j,\ldots)\) with \(G_{AB}\)/\(G^{AB}\),
\(g_{ab}\)/\(g^{ab}\), and \(h_{ij}\)/\(h^{ij}\) respectively. As an example,
\begin{align}
  n^A
  &= G^{AB} \, n_B
    = \frac{(\partial_\tau)^A - N^A}{N}
    = \tensor{g}{^A_a}\, n^a
    \;.
    \label{eq:timenormal}
\end{align}
We will use
\(\nabla_A\), \(D_a\), and \(d_i\) to respectively denote the covariant
derivatives associated to \(G_{AB}\), \(g_{ab}\), and \(h_{ij}\).

We shall denote the spacetime volume form\footnote{We do not immediately see a
  reason why our formalism requires spacetimes to be orientable, that is possess
  a \emph{globally} defined nonvanishing top form. In general, \(\epsilon\)
  should be interpreted as simply an instruction to integrate with a positive
  weight; \ie{} whatever orientation one might \emph{locally} assign to a piece
  of spacetime, \(\epsilon\) is a similarly locally defined form that is positive relative
  to that choice.} by \(\epsilon\), and write its interior product with a
vector \(\xi^A\) as
\begin{align}
  \xi^A \epsilon_A
  &= \iota_\xi\epsilon
  \;.
\end{align}
Most codimension-one surfaces we will consider are boundaries and thus have
induced orientations. Our convention will be to take the volume form on
\(\gamma\) and more generally codimension-two surfaces of constant
\((\rho,\tau)\) to be given by
\begin{align}
  \tensor[^{(D-2)}]{\epsilon}{}
  &= n^A u^B \epsilon_{AB}
    = \iota_u(\iota_n \epsilon)
    \;,
    \label{eq:codimtwovol}
\end{align}
so
\begin{align}
  \epsilon
  &= N\,\dd{\tau} \wedge \dd{\rho} \wedge \tensor[^{(D-2)}]{\epsilon}{}
    \;.
\end{align}

\subsubsection{Holonomic singularity}
\label{sec:holsing}

We will also consider a Maxwell field \(A\), which we require to be smooth away
from \(\gamma\). Just as we have introduced conical and helical singularities in
radially static coordinates \labelcref{eq:metricG} where \(G_{\rho\, a}=0\), let
us now consider a radial gauge for the Maxwell field in which
\begin{align}
  u^A A_A
  &=0 \;.
    \label{eq:statgauge}
\end{align}
At \(\gamma\), smoothness would ordinarily require
\begin{align}
  N\, n^a\, A_a
  &= \order{\rho^2}
    \;,
    \label{eq:holsmooth}
\end{align}
in analogy to having a vanishing shift \labelcref{eq:shiftlapsesmooth}. However, just
as one can turn on a helical singularity \labelcref{eq:shiftlapsesing}, we will
more generally allow an electric potential for which
\begin{align}
  N\, n^a\, A_a
  &= \mu + \order{\rho^2}
    \;.
    \label{eq:holsing}
\end{align}

In the same way that \(v^i\) describes geometric motion along \(\gamma\),
\(\mu\) describes motion along the fibres of the Maxwell principal bundle as one
moves around \(\gamma\) in a metric-orthogonal manner. This is related to a
generically nontrivial holonomy around \(\gamma\), as quantified by an
\(\varepsilon\)-size Wilson loop \(\int_{\mathscr{C}_\varepsilon} A\) encircling
\(\gamma\). However, taking \(\mathscr{C}_\varepsilon\) to be a closed circle at
fixed \(\rho=\varepsilon\) and \(y^i\), note that the value of this Wilson loop
will depend on both \(\mu\) and the components of \(A\) along any helical shift
\(v^i\) present on \(\gamma\):
\begin{align}
  \int_{\mathscr{C}_\varepsilon} A
  &= \int_{\mathscr{C}_\varepsilon} \dd{\tau} (\partial_\tau)^a A_a
    \sim \int_{\mathscr{C}_\varepsilon} \dd{\tau} \left(
    \mu + v^i A_i
    \right)
    \;.
    \label{eq:hol}
\end{align}
We will require the components \(A_i\) along \(\gamma\) to have finite
\(\rho\to0\) limits. At any rate, for brevity, we will refer to singularities
with \(\mu\ne 0\) as holonomic singularities.

Such configurations have previously appeared, for eample, in \(D=2\) dimensions
in ref.~\cite{Maxfield:2020ale}, where they were called ``KK instantons''
because they arose from the dimensional reduction of smooth \(D=3\)
configurations. In that context, however, the singularities were not truly
singularities in the fundamental \(D=3\) description and are artifacts of the
dimensional reduction; these are analogous to conical singularities which appear
when taking orbifolds of replica-symmetric saddles computing gravitational
R\'enyi entropies \cite{Lewkowycz:2013nqa}. In both situations, action
contributions localized to these ``artificial'' singularities should be
excluded. Another perspective is that such singularities each occur with fixed
strength. To impose the right boundary conditions on the singularity to fix this
strength, terms must be added to the action, which happen to cancel the
localized contributions from the singularity
\cite{Dong:2019piw,Maxfield:2020ale}.

In contrast, the conical, helical, and holonomic singularities which we will
consider in this paper are somewhat less artificial. As we will describe in
\cref{sec:regsing}, we will view each singular configuration as a limit of some
smooth regulated configuration differing from the original singular
configuration in some neighbourhood \(\mathscr{N}_\varepsilon\) of the singular
surface \(\gamma\). Just as the smooth configuration can appear in the path
integral, weighted by an action including contributions from
\(\mathscr{N}_\varepsilon\), we will similarly allow singular configuration in
the path integral, weighted by an action including contributions localized near
\(\gamma\). In the path integral, the strengths of the conical, helical, and
holonomic singularities are not fixed; in fact, as sketched in \cref{sec:intro}
and carried out in \cref{sec:partfunc,sec:lorpartfunc}, we will evaluate the
path integral by initially fixing quantities that are conjugate in some sense to
these singularity strengths.

\subsubsection{Approximate stationarity}
\label{sec:stationarity}

Before moving on, let us address how we will allow various field components to
vary around \(\gamma\). For the most part, in this paper, we will assume a very
weak notion of approximate stationarity near \(\gamma\) --- namely that certain
components of the metric and Maxwell field in the \(\rho\to 0\) expansion are
\(\tau\)-independent, as elaborated below. However, we will revisit some of these
assumptions when considering Lorentz signature in \cref{sec:lorsing}.

Firstly, it seems reasonable to require the induced fields \(h_{ij}\) and
\(A_i\) on surfaces of constant \((\rho,\tau)\) to have \(\tau\)-independent
limits as \(\rho\to 0\). These limits can then be understood as the induced
field configurations on \(\gamma\). A \(\tau\)-dependent induced field
configuration seems highly unnatural and we will find no reason to relax the
assumed \(\tau\)-independence of \(\lim_{\rho\to 0} h_{ij},A_i\), even when we
consider Lorentzian signature in \cref{sec:lorentzian}.

In \cref{sec:actionvar,sec:saddle}, it will become apparent that the volume
element on surfaces of constant \((\rho,\tau)\) is, in a certain sense,
conjugate to the variable \(\kappa\). Thus, it seems natural also to require
\(\kappa(y)\) to be \(\tau\)-independent.\footnote{At first sight, this might
  appear to be a statement purely about gauge-fixing a redundancy in the choice
  of parameter \(\tau\). Given a \(\kappa(x)\) which is a function of
  \(x^a=(\tau,y^i)\), one can remove the \(\tau\)-dependence of \(\kappa\) by
  reparametrizing \(\tau\) in a \(y^i\)-dependent manner. This
  reparametrization, however, does not generally map surfaces of constant
  \(\tau\) to each other. As we will see in \cref{sec:singdisambig}, a special
  choice of foliation by surfaces of constant \(\tau\) is picked out by the
  physical internal structure of a helical singularity, which becomes apparent
  when we regulate the singularity. Thus, under the regulation prescription we
  will adopt, \(y^i\)-dependent reparametrizations of \(\tau\) are not viewed as
  redundancies, but rather move us between physically distinct configurations.}
By analogy, we are inclined to further restrict helical shifts \(v^i(y)\) to
also be \(\tau\)-independent. In doing so, we might also require the
\(\order{\rho^2}\) term of \(N^i\) to be similarly \(\tau\)-independent, as it
will later be revealed to be conjugate to the helical shift \(v^i\). The same
can be said for the holonomic singularity strength \(\mu(y)\) and its conjugate,
the pullback of \(*F\) to surfaces of constant \((\rho,\tau)\) in the
\(\rho\to0\) limit.

However, it is not immediately obvious that configurations which do not satisfy
all the above assumptions of \(\tau\)-independence should be ruled out as a
matter of principle. Aside from some comments on why we might want to relax some
of these assumptions when considering Lorentzian signature in
\cref{sec:lorentzian,sec:discussion}, we will largely leave this question for
future work.

A partial list of issues that may arise in case the above stated assumptions of
\(\tau\)-independence fail is as follows. If the induced fields \(h_{ij}\) and
\(A_i\) are \(\tau\)-dependent near \(\gamma\), then they must be interpolated
when regulating the singularity as described in \cref{sec:regsing} and
potentially give additional curvature contributions in \cref{sec:curvature}. If
\(\kappa\), \(v^i\), \(\mu\) or their conjugate variables are
\(\tau\)-dependent, then we must refrain from trivially integrating out the
\(\tau\) direction in certain equations, \eg{}
\cref{eq:maxwelldeltafunc,eq:varhelterm,eq:varholterm,eq:fixmom,eq:fixcharge,eq:mom,eq:charge,eq:lormom,eq:lorcharge}
(but this does not seem to qualitatively alter our discussion). Moreover, in
\cref{sec:shapeindep}, we will find that the action of a helically singular
configuration is sensitive to the shape of a cutoff surface placed around
\(\gamma\), except when \(\lim_{\rho\to 0}h_{ij}\), \(\kappa\), and \(v^i\) are
\(\tau\)-independent or when \(\lim_{\rho\to 0} h_{ij}\) is \(\tau\)-independent
and \(\kappa\) is \(y^i\)-independent.

\subsection{Regulating singularities using smooth configurations}
\label{sec:regsing}

It is worthwhile to consider smooth regulated configurations which approximate
the singularities introduced in \cref{sec:singconfigs}. This will motivate the
localized curvature and action contributions we will later assign to the
singularities. Moreover, the regulated geometries will resolve some apparent
redundancies in describing the structure of the singularities.

To regulate a configuration singular on the codimension-two surface \(\gamma\),
let us consider a small neighbourhood \(\mathscr{N}_\varepsilon\) of \(\gamma\)
in the spacetime \(\mathscr{M}\) --- see \cref{fig:coordinates}. The radial
extent of \(\mathscr{N}_\varepsilon\) from \(\gamma\) is understood to be
parametrized by a small regulator \(\varepsilon\); to be concrete for now, we
will take \(\partial\mathscr{N}_\varepsilon\) in the singular configuration to
be a surface of constant \(\rho\) of order \(\varepsilon\).\footnote{The action
  we will derive for singular configurations will be expressed in terms of this
  neighbourhood \(\mathscr{N}_\varepsilon\) of the singularity \(\gamma\). As
  described in \cref{sec:shapeindep}, however, this action will turn out to be insensitive
  to the shape of \(\partial\mathscr{N}_\varepsilon\) as one moves around the
  \(\tau\) direction.} The regulated
configuration we construct will coincide with the original singular
configuration outside the neighbourhood \(\mathscr{N}_\varepsilon\), \ie{} in
\(\mathscr{M}\setminus\mathscr{N}_\varepsilon\). However, we require the
regulated configuration to be modified in \(\mathscr{N}_\varepsilon\) such that
it is smooth everywhere, in particular at the centre \(\gamma\) of
\(\mathscr{N}_\varepsilon\) where the original configuration was singular.
Practically, this amounts to an interpolation inside \(\mathscr{N}_\varepsilon\)
from the smooth behaviour \labelcref{eq:shiftlapsesmooth,eq:holsmooth} on
\(\gamma\) to the behaviour \cref{eq:shiftlapsesing,eq:holsing} matching the
original singular configuration on
\(\partial\mathscr{N}_\varepsilon\).\footnote{Note from the interpolation in
  \(N\) that, setting the proper radius of \(\mathscr{N}_\varepsilon\) in the
  original geometry to be \(\varepsilon\), the proper radius in the regulated
  geometry will generically differ (but should be of the same order). Thus, if
  we insist on the form \labelcref{eq:metricG} of the metric in the regulated
  geometry and require \(\rho\) to match continuously to the unregulated
  geometry at \(\partial\mathscr{N}_\varepsilon\), then \(\gamma\) will not
  generically lie at \(\rho=0\) in the regulated geometry. \label{foot:regrho}}
We will take the interpolation scale of each function of \(\rho\) to be
\(\order{\varepsilon}\). Moreover, we require the induced metric on \(\gamma\)
in the regulated geometry to be unmodified from the \(\rho\to 0\) limit of the
induced metric \(h_{ij}\) on surfaces of constant \((\rho,\tau)\) in the
original singular geometry. While trivial at first glance, this last statement
can be regarded as a gauge condition, as we describe below.

\subsubsection{Inequivalent resolutions of naively equivalent singularities}
\label{sec:singdisambig}

We are now in a position to discuss the question of whether singularities
described by different values of \((\kappa,v^i,\mu)\) are truly physically
distinct. In light of the above described procedure for resolving singularities
using regulated configurations, we will argue that this is indeed the case.

Let us consider, in the spacetime region away from \(\gamma\), the effect of a
\(y\)-dependent reparametrization of \(\tau\). Taking the variation
\(\delta_f\equiv \mathcal{L}_{f\partial_\tau}\), generated by a vector
\(f(y)\partial_\tau\), we find the components of the metric \labelcref{eq:metricg} vary as
\begin{align}
  \delta_f N
  &= f\, \partial_\tau N
    -N\, N^i \partial_i f
    \;,
    \label{eq:lapsereparam}
  \\
  \delta_f N^i
  &= f\, \partial_\tau N^i
    + N^2 \partial^i f
    - N^i N^j \partial_j f
    \;,
    \label{eq:shiftreparam}
  \\
  \delta_f h_{ij}
  &= f\, \partial_\tau h_{ij}
    + N_i\, \partial_j f
    + N_j\, \partial_i f
    \;.
    \label{eq:metrreparam}
\end{align}
In helically singular configurations, these transformations remain nontrivial as
one approaches \(\gamma\). Thus, unregulated configurations to which we ascribe
different values of \((\kappa,v^i, \lim_{\rho\to 0}h_{ij})\) may yet be related
to each other by diffeomorphisms away from \(\gamma\). (In contrast, in the
absence \(v^i=0\) of a helical singularity,
\cref{eq:shiftlapsesmooth,eq:shiftlapsesing}, with definite values of \(\kappa\)
and \(v^i\), as well as \(\lim_{\rho\to 0} h_{ij}\), are preserved by the
above.)

However, we will view this as a redundancy only in the description of the
unregulated configuration away from \(\gamma\) or the regulated configuration
outside \(\mathscr{N}_\varepsilon\). Let us consider the question of whether two
singular configurations related by a reparametrization generated by
\cref{eq:lapsereparam,eq:shiftreparam,eq:metrreparam} are really distinct. Away
from \(\gamma\), there appears to be no physical distinction between the two
configurations. However, applying our regulation procedure to these two
configurations leads to distinct in-fillings of the regulated neighbourhood
\(\mathscr{N}_\varepsilon\). For example, the regulated configurations have
different induced metrics on \(\gamma\) related by \cref{eq:metrreparam}. Our
regulation procedure therefore imbues the singularity with some internal
structure that distinguishes the two configurations.

Practically, we will continue to describe singularities through field
components, \eg{} \((\kappa,v^i, \lim_{\rho\to 0}h_{ij})\), associated to a
choice of \(\tau\) foliation. Implicitly, we will always have in mind that the
singularity is regulated according to our procedure for \emph{that given
  \(\tau\) foliation}. For example, the induced metric on \(\gamma\) --- made
precise in the regulated geometry --- is really the stated value of
\(\lim_{\rho\to0}h_{ij}\) and not some other metric related by
\cref{eq:metrreparam}. With this in mind, singularities assigned different
values of \((\kappa,v^i, \lim_{\rho\to 0}h_{ij})\) related by
\cref{eq:lapsereparam,eq:shiftreparam,eq:metrreparam} are physically distinct.
In particular, in \cref{sec:curvature}, we will extract curvature contributions
from the regulated neighbourhood \(\mathscr{N}_\varepsilon\) of \(\gamma\) and,
in \cref{sec:actionprop}, we will use these results to deduce an action for
singular configurations. Singularities with different values of \((\kappa,v^i,
\lim_{\rho\to 0}h_{ij})\) related by
\cref{eq:lapsereparam,eq:shiftreparam,eq:metrreparam} will have generically
different action.

Analogous remarks can be made about certain discrete shifts of \(v^i\) that
naively appear to lead to redundancies. Suppose there exists a locally Killing
vector field \(w^i\), tangential to constant \((\rho,\tau)\) surfaces near
\(\gamma\), which generates closed orbits with parameter period \(2\pi\) --- in
particular, \(e^{2\pi\, w} \sim 1\) acts identically on tensor fields. Then, any
(unregulated) configuration described by helical shift \(v^i\) on \(\gamma\) is
related, in the spacetime region away from \(\gamma\), by discrete
diffeomorphisms to configurations with helical shifts \(v^i + m w^i\) for
\(m\in\mathbb{Z}\).\footnote{For an illustration, see our construction of
  helically singular constrained saddles for the gravitational partition
  function in \cref{fig:helicalbh}. In \cref{fig:helicalbh3}, the two helical
  shifts indicated by solid and dashed green arrows describe the same
  identification of the red surfaces. However, under our regulation procedure,
  these different helical shifts lead to different contractible cycles,
  indicated by the solid and dashed teal curves respectively, in the regulated
  versions of the geometry.} (Moreover, any spin structure is preserved by the
diffeomorphisms relating \(v^i + 2m w^i\) for \(m\in\mathbb{Z}\).) However, by
our regulation procedure, we will regard these shifts as describing helical
singularities which are resolved by physically distinct regulated geometries
near \(\gamma\). In particular, a helical singularity with shift parameter
\(v^i+ m w^i\) is regulated by a smooth geometry with a shift vector \(N^i\)
interpolating between zero on \(\gamma\) and \(N^i \sim v^i + m w^i\) on
\(\partial\mathscr{N}_\varepsilon\).

We can make identical remarks about the Maxwell sector. The Maxwell gauge group
\(G_{\mathrm{M}}\) can be either compact \(\cong U(1)\) or non-compact
\(\cong\mathbb{R}\). In the former case, when \(G_{\mathrm{M}}=U(1)\), one may
have naively thought that a holonomic singularity \(\mu\) is equivalent to one
with \(\mu+n\) for \(n\in\mathbb{Z}\). However, just as in the above helical
discussion, our convention for regularization resolves these singularities using
physically distinct smooth configurations near \(\gamma\).

\subsection{Curvatures for singular configurations}
\label{sec:curvature}

We now calculate various curvatures in the singular configurations introduced in
\cref{sec:singconfigs}.

\subsubsection{Geometric curvatures}
\label{sec:curvgeo}

We first consider curvatures of the spacetime geometry. We start with the
extrinsic curvatures of \((D-1)\)- and \((D-2)\)-dimensional slices foliating
the geometries \labelcref{eq:metricG,eq:metricg}. These will be related by the
Gauss-Codazzi equation to intrinsic curvature. Using the regulated geometry
introduced in \cref{sec:regsing}, we will identify pertinent contributions to
intrinsic curvature as contact terms associated to the singularity on \(\gamma\).

We denote the extrinsic curvature of the \((D-1)\)-dimensional constant \(\rho\)
surfaces by
\begin{align}
  (K_u)_{AB}
  &= \frac{1}{2} \mathcal{L}_u g_{AB}
  = \frac{1}{2} \mathcal{L}_u G_{AB}
    = \nabla_A u_B \;,
    \label{eq:extrcurv}
\end{align}
where \(\nabla_A\) denotes the covariant derivative of the metric \(G_{AB}\).
Of course, \(u^A\, (K_u)_{AB} = u^A\, (K_u)_{BA}=0\) so no information
is lost in the projection
\begin{align}
  (K_u)_{ab}
  &= \frac{1}{2} \partial_\rho g_{ab}
    \;.
    \label{eq:extrcurvproj}
\end{align}
Of interest to us are two scalars constructed from this extrinsic curvature:
\begin{align}
  \tensor{(K_u)}{^a_a}
  &= \frac{\partial_\rho N}{N}
    + \frac{1}{2} h^{ij} \partial_\rho h_{ij}
    \label{eq:extrcurvtr}
  \\
  (K_u)^{ab} (K_u)_{ab}
  &= \left( \frac{\partial_\rho N}{N} \right)^2
    + \frac{
    h_{ij}\,
    \partial_\rho N^i\,
    \partial_\rho N^j
    }{2N^2}
    + \frac{1}{4}
    h^{ik}\, h^{j\ell}\,
    \partial_\rho h_{ij}\,
    \partial_\rho h_{k\ell}\,
    \;.
    \label{eq:extrcurv2}
\end{align}

Each constant \(\rho\) surface is in turn foliated by constant \(\tau\) surfaces
with extrinsic curvature
\begin{align}
  (k_n)_{ab}
  &= \frac{1}{2} \mathcal{L}_n h_{ab}
    \;.
\end{align}
The \((D-2)+1\) decomposition \labelcref{eq:metricg} leads to the standard
formula\footnote{To avoid confusion with the lapse, we will write
  the shift vector as \(\vec{N}\) when suppressing its index.}
\begin{align}
  (k_n)_{ij}
  &= \frac{1}{2N} \left(
    \partial_\tau h_{ij}
    - \mathcal{L}_{\vec{N}} h_{ij}
    \right)
    \;.
    \label{eq:extrcurvadm}
\end{align}

Let us turn now to the intrinsic curvature, given by the Gauss-Codazzi
equation\footnote{Recall that we are currently in Euclidean signature; the
  corresponding Lorentzian equation has extra signs on the terms quadratic in
  extrinsic curvature and the normal vector, when the normal is time-like. Note
  also that \cref{eq:codazzi1} ordinarily has a term \(2 \nabla_A(u^B \nabla_B u^A)\) on the RHS, but this vanishes because our \(u^A\) is a geodesic vector field.}
\begin{align}
  R
  &= \tensor[^{(D-1)}]{R}{}
    - (K_u)^{ab}(K_u)_{ab}
    + \tensor{(K_u)}{^a_a} \tensor{(K_u)}{^b_b}
    - 2 \nabla_A (u^A \nabla_B u^B)
    \;.
    \label{eq:codazzi1}
\end{align}
A second iteration of Gauss-Codazzi gives
\begin{align}
  \tensor[^{(D-1)}]{R}{}
  &= \tensor[^{(D-2)}]{R}{}
    - (k_n)^{ij}(k_n)_{ij}
    + \tensor{(k_n)}{^i_i} \tensor{(k_n)}{^j_j}
    + 2 D_a(n^b D_b n^a - n^a D_b n^b)
    \;.
    \label{eq:codazzi2}
\end{align}
The Ricci curvatures \(R\), \(\tensor[^{(D-1)}]{R}{}\), and
\(\tensor[^{(D-2)}]{R}{}\) are associated to the full \(D\)-dimensional
spacetime, constant \(\rho\) surfaces, and constant \((\rho,\tau)\) surfaces
respectively. The covariant derivative of \(g_{ab}\) is denoted \(D_a\).

Let us now deduce contact terms in \(R\) associated to the conical and helical
singularities on \(\gamma\). To proceed, we consider the regulated geometry
introduced in \cref{sec:regsing}. On this smooth regulated geometry, which
we will indicate by subscript \(\mathrm{reg}(\varepsilon)\), let us consider the
integral \(\int_{\mathscr{N}_\varepsilon}(\epsilon \,
R)_{\mathrm{reg}(\varepsilon)}\) of curvature over the small neighbourhood
\(\mathscr{N}_\varepsilon\) of \(\gamma\). To write the integrand in terms of
metric components, we use \cref{eq:codazzi1,eq:codazzi2} for Ricci curvature and
\cref{eq:extrcurvtr,eq:extrcurv2,eq:extrcurvadm} for extrinsic curvature. Now
recall that we have required metric components to behave like
\cref{eq:shiftlapsesing} in the unregulated configuration, but our regulation
procedure interpolates the metric in \(\mathscr{N}_\varepsilon\) to be smooth on
\(\gamma\), as expressed in \cref{eq:shiftlapsesmooth}. In light of all this,
the terms contributing to
\(\int_{\mathscr{N}_\varepsilon}(\epsilon\,R)_{\mathrm{reg}(\varepsilon)}\) that
can survive the \(\varepsilon\to 0\) limit are\footnote{We orient boundaries so
  that Stokes' theorem takes the standard form \(\int_{\mathscr{N}} \dd \omega =
  \int_{\partial\mathscr{N}} \omega\). A minus sign \(-\partial\mathscr{N}\)
  indicates reversal of orientation.}
\begin{align}
  \begin{split}
    \int_{\mathscr{N}_\varepsilon}
    (\epsilon \, R)_{\mathrm{reg}(\varepsilon)}
    &\sim
      - 2\int_{\partial\mathscr{N}_\varepsilon - \partial\mathscr{N}_0}
      (u^A \epsilon_A \nabla_B u^B)_{\mathrm{reg}(\varepsilon)}
    - \int_{\mathscr{N}_\varepsilon} \left( \epsilon
    \frac{
    h_{ij}\,
    \partial_\rho N^i\,
    \partial_\rho N^j
    }{2N^2}
      \right)_{\mathrm{reg}(\varepsilon)}
  \\
  &\phantom{{}\sim{}}
    + \int_{\mathscr{N}_\varepsilon}
    \left[  \epsilon
    \frac{
    (h^{ij}h^{k\ell}- h^{ik}h^{j\ell})\mathcal{L}_{\vec{N}} h_{ij} \mathcal{L}_{\vec{N}} h_{k\ell}
    }{4N^2}  \right]_{\mathrm{reg}(\varepsilon)}
    \;.
  \end{split}
  \label{eq:regcurvint}
\end{align}
We will now discuss the RHS term by term, for example, pointing out where they
come from and explaining how they can survive the \(\varepsilon\to 0\) limit even
though the neighbourhood \(\mathscr{N}_\varepsilon\) shrinks to zero size.

The first term of \cref{eq:regcurvint} arises from the total derivative in
\cref{eq:codazzi1}. In particular,
\begin{align}
  2\int_{\partial \mathscr{N}_0}
  (u^A \epsilon_A\,  \nabla_B u^B)_{\mathrm{reg}(\varepsilon)}
  &\sim 4\pi\, \mathrm{Area}(\gamma)
    \label{eq:concontact1}
\end{align}
results from the divergence \(\nabla_B u^B\) being singular at \(\gamma\) even
on smooth geometries. Here, \(\mathscr{N}_0\) is an infinitesimal neighbourhood
of \(\gamma\) in the regulated geometry.\footnote{\Cref{eq:concontact1} is not
  written as an equality at finite \(\varepsilon\) because \(\gamma\) (and thus
  \(\partial \mathscr{N}_0\)) may not necessarily lie exactly at \(\rho=0\) in the
  regulated geometry as commented in \cref{foot:regrho}.} From
\cref{eq:extrcurv}, we see that we can also write the integral on
\(\partial\mathscr{N}_\varepsilon\) as
\begin{align}
  - 2\int_{\partial\mathscr{N}_\varepsilon}
  u^A \epsilon_A \, \nabla_B u^B
  &= - 2\int_{\partial\mathscr{N}_\varepsilon}
    u^A \epsilon_A \, \tensor{(K_u)}{^b_b}
    \;.
    \label{eq:concontact2}
\end{align}

The second term of \cref{eq:regcurvint} comes from the second term in
\cref{eq:extrcurv2} and is more subtle to interpret. Due to the regulating
interpolation, the derivatives \(\partial_\rho N^i_{\mathrm{reg}(\varepsilon)}\)
become large and cause this term to diverge in the \(\varepsilon \to 0\) limit.
Note, however, that the divergence is quadratic in the helical singularity
parameter \(v^i\) setting the size of the shift \(N^i\).

Let us recall that a similar situation arises for conical singularities in
higher curvature theories, say with an \(R^2\) term in the Lagrangian. The
resulting action diverges on conical singularities \(\gamma\) of finite
strength. However, taking the action around \(\gamma\) and linearizing with
respect to the singularity strength\footnote{As ref.~\cite{Dong:2013qoa}
  emphasizes, linearization of the action requires greater care than naively
  linearizing the Lagrangian. Specifically, ref.~\cite{Dong:2013qoa} considers
  linearizing the difference in action between a conically singular geometry and
  its smooth regulated counterpart,
  \begin{align}
    \partial_\kappa\left(
    \int_{\mathscr{M}} L_{\mathrm{reg}(\varepsilon)}
    - \int_{\mathscr{M}\setminus\mathscr{N}_0} L
    \right)_{\kappa=1}
    &=
      \partial_\kappa\left(
      \int_{\mathscr{N}_\varepsilon} L_{\mathrm{reg}(\varepsilon)}
      - \int_{\mathscr{N}_\varepsilon\setminus\mathscr{N}_0} L
      \right)_{\kappa=1}
  \end{align}
  with respect to the opening angle \(2\pi\kappa\) varied from its smooth value
  \(2\pi\). Due to the smoothing of the configuration, \(\partial_\kappa L_{\mathrm{reg}(\varepsilon)}\) remains uniformly bounded over
  \(\mathscr{N}_\varepsilon\) as \(\kappa \to 1\), so
  \begin{align}
    \partial_\kappa\left(
    \int_{\mathscr{N}_\varepsilon} L_{\mathrm{reg}(\varepsilon)}
    \right)_{\kappa=1}
    &= \int_{\mathscr{N}_\varepsilon}
      \left(
      \partial_\kappa
      L_{\mathrm{reg}(\varepsilon)}\right)_{\kappa=1}
      \;.
  \end{align}
  However, the same is not true for the Lagrangian \(L\) of the original
  singular configuration. In particular, ref.~\cite{Dong:2013qoa} showed that
  higher curvature terms in \(L\) can be equal to \((\kappa-1)^2\) times
  integrands which become increasingly singular near \(\gamma\) until they are
  nonintegrable in the \(\kappa\to 1\) limit. The integration can therefore
  produce an inverse power of \(\kappa-1\), promoting the \(\order{(\kappa-1)^2}\)
  terms of \(L\) to \(\order{\kappa-1}\) terms of the action.

  An analogous mechanism does not appear to be at play in the current
  calculation, but perhaps a more careful analysis is warranted.}, one can
extract a linear coefficient which is finite and can be identified as the
geometric entropy of the surface \(\gamma\) \cite{Dong:2013qoa}. Conical
singularities of finite strength do appear in the calculation of R\'enyi
entropies using orbifold geometries \cite{Lewkowycz:2013nqa} and in states of
fixed geometric entropy. In the former case, the singularity strength is fixed
and a good variational principle is obtained from the bulk action evaluated over
\(\mathscr{M}\setminus\mathscr{N}_\varepsilon\), excluding a neighbourhood
\(\mathscr{N}_\varepsilon\) of the singular surface \(\gamma\), possibly
supplemented by counterterms near \(\gamma\) \cite{Dong:2019piw}. Indeed, the
conical singularity is an artifact of the orbifold and should not contribute to
the action in the R\'enyi calculation \cite{Lewkowycz:2013nqa}. In contrast, to
obtain a good variational principle with fixed geometric entropy, one must
perform a Legendre transform which effectively reinstates, in the action, a
contribution from the conical singularity equal to the singularity strength
times the geometric entropy \cite{Dong:2019piw}.

Later in this paper, we will study path integrals where quantities conjugate to
the conical, helical, and holonomic singularity strengths are fixed on
\(\gamma\), analogous to the situation of fixed geometric entropy described
above. One might therefore guess that the appropriate action should only include
contributions from the singularity that are \emph{linear} in the singularity
strengths. This guess will be partially justified in
\cref{sec:actionvar,sec:saddle} by examining the variational principle that
results from such an action. For now, let us simply treat the second term of
\cref{eq:regcurvint} by keeping only the contribution linear in the strength
\(v^i\) of the helical singularity (of the original unregulated geometry). The
result is:
\begin{align}
  -\int_{\mathscr{N}_\varepsilon} \left( \epsilon\,
  \frac{
  h_{ij}\,
  \partial_\rho N^i\,
  \partial_\rho N^j
  }{2N^2}
  \right)_{\mathrm{reg}(\varepsilon)}
  &\sim
    -\int_{\partial \mathscr{N}_\varepsilon} u^A \epsilon_A
    \frac{
    N_i
    \partial_\rho N^i
    }{N^2}
  & (\text{part linear in \(v^i\)})
    \label{eq:hellinear}
  \\
  &= -2\int_{\partial \mathscr{N}_\varepsilon} u^A \epsilon_A
    \,
    (\dd\tau)_a \, N_i\, (K_u)^{a i}
    \;,
    \label{eq:helcontact}
\end{align}
which remains finite in the \(\varepsilon\to 0\) limit, as can be seen from
\cref{eq:shiftlapsesing}. The linear dependence on \(v^i\) enters through the
undifferentiated shift \(N^i\).

Finally, the second line of \cref{eq:regcurvint} comes from the second term of
\cref{eq:extrcurvadm}. In \(D=3\) spacetime dimensions, the codimension-two
metric \(h_{ij}\) has only one component and, consequently, the numerator causes
the second line of \cref{eq:regcurvint} to vanish identically. This case of
\(D=3\) will be the primary focus in later sections of this paper.

However, let us also comment briefly on the case of \(D>3\), where this term
does not vanish identically. For \(D>3\), the integrand would in fact
generically diverge like \(1/\rho\) at \(\gamma\), were it not for the
regulation \(\mathrm{reg}(\varepsilon)\) which interpolates the integrand to
zero on \(\gamma\). If we view this interpolation as a fuzzy cutoff for the
integral at some \(\rho = \varepsilon'\) between \(0\) and \(\varepsilon\),
then the term on the second line of \cref{eq:regcurvint} merely shifts the
cutoff surface for its counterpart integrated over
\(\mathscr{M}\setminus\mathscr{N}_\varepsilon\):
\begin{align}
  \begin{split}
    \MoveEqLeft
    \int_{\mathscr{M} = \mathscr{N}_\varepsilon + (\mathscr{M}\setminus\mathscr{N}_\varepsilon)}
  \left[ \epsilon
  \frac{
  (h^{ij}h^{k\ell}- h^{ik}h^{j\ell})
  \mathcal{L}_{\vec{N}} h_{ij} \mathcal{L}_{\vec{N}} h_{k\ell}
  }{4N^2} \right]_{\mathrm{reg}(\varepsilon)}
  \\
  &\sim
    \int_{\mathscr{M}\setminus\mathscr{N}_{\varepsilon'}}
    \epsilon
    \,
    \frac{
    (h^{ij}h^{k\ell}- h^{ik}h^{j\ell})\mathcal{L}_{\vec{N}} h_{ij} \mathcal{L}_{\vec{N}} h_{k\ell}
    }{4N^2}
    \;.
    \label{eq:regdepterm}
  \end{split}
\end{align}
The fact that the second line of \cref{eq:regcurvint} remains \emph{finite} in
the \(\varepsilon\to 0\) limit simply reflects the fact that the above is
\emph{logarithmically divergent} in the \(\varepsilon'\to 0\) limit. How should
we handle this potential divergence in
\(\int_{\mathscr{M}\setminus\mathscr{N}_\varepsilon}\epsilon\,R\) as
\(\varepsilon\to 0\)? Should we perhaps only allow those helical singularities
for which
\begin{align}
  (h^{ij}h^{k\ell}- h^{ik}h^{j\ell})
  \mathcal{L}_v h_{ij} \mathcal{L}_v h_{k\ell}
  &= 0
    \label{eq:divergentbulk}
\end{align}
on \(\gamma\)? We will leave these questions for future work and retreat to
cases where \cref{eq:divergentbulk} is satisfied, as in \(D=3\).

In summary, we argue that the natural analogue of \(\int_{\mathscr{M}} \epsilon
R\) in the presence of the conically and helically singular surface \(\gamma\)
is \(\int_{\mathscr{M}\setminus\mathscr{N}_\varepsilon} \epsilon R\) plus the
contributions \labelcref{eq:concontact1,eq:concontact2,eq:helcontact} accounting
for contact terms from the singularity. Aside from the above derivation, we will
see in \cref{sec:action,sec:partfunc} how the incorporation of such contact
terms in the gravitational action leads to a reasonable variational principle
and thermodynamically intuitive interpretations of the gravitational partition
function.

\subsubsection{Maxwell field strength}
\label{sec:curvmax}

It is relatively straightforward to also deduce a contact term in the Maxwell
field strength \(F= \dd{A}\) resulting from \cref{eq:hol}:
\begin{align}
  \int_{\mathscr{D}_\varepsilon}
  F
  &= \int_{\partial\mathscr{D}_\varepsilon} A
    \sim 2\pi (\mu + v^i A_i)
    \;.
    \label{eq:maxwelldeltafunc}
\end{align}
Here, \(\mathscr{D}_\varepsilon\) is an \(\varepsilon\)-disk punctured by
\(\gamma\) (with orientation chosen so that
\(\int_{\partial\mathscr{D}_\varepsilon} \dd\tau = 2\pi\)). Clearly, \(F\)
possesses a \(\delta\)-function contact term at \(\gamma\) in its orthogonal plane.

We would also like to understand how to integrate the squared field strength
\(\int_{\mathscr{M}} F \wedge * F\) on such singular configurations. The naive
integral diverges because the integrand contains a squared \(\delta\)-function,
but this pathology can be treated analogously to \cref{eq:hellinear}, where we
linearized with respect to the strength of the singularity. Let us again
consider an \(\varepsilon\)-regulated configuration as described in
\cref{sec:regsing}. Writing out
\begin{align}
  F \wedge * F
  &= \frac{1}{2}\, \epsilon\, F_{AB}\, F^{AB}
    = \epsilon \left[
    (n^a\, \partial_\rho A_a)^2
    + n^a\, n^b\, h^{ij}\, F_{ai}\, F_{bj}
    + h^{ij}\, \partial_\rho A_i\, \partial_\rho A_j
    + \frac{1}{2} F_{ij}\, F^{ij}
    \right]
    \label{eq:maxwellfsquared}
\end{align}
in radial gauge \cref{eq:statgauge}, we again recognize the dangerous first term
as a squared \(\rho\)-derivative of the interpolated field
\(A_{\mathrm{reg}(\varepsilon)}\) in the regulated configuration. As in the
helical case, linearization with respect to the (unregulated) singularity
strength, here \(\mu\) in \cref{eq:holsing}, produces the desired result:
\begin{align}
  \int_{\mathscr{N}_\varepsilon}
  (F \wedge *F)_{\mathrm{reg}(\varepsilon)}
  &\sim 2\int_{\partial\mathscr{N}_\varepsilon}
    A \wedge *F
    \;.
  & (\text{part linear in \(\mu\)})
    \label{eq:holcontact}
\end{align}

Actually, in writing \cref{eq:holcontact}, we have made an assumption analogous
to \cref{eq:divergentbulk}, namely that the second term on the RHS of
\cref{eq:maxwellfsquared} is sufficiently tame near \(\gamma\) that
\(\int_{\mathscr{M}\setminus\mathscr{N}_{\varepsilon}} \epsilon\, n^a\, n^b\,
h^{ij}\, F_{ai}\, F_{bj}\) converges as \(\varepsilon\to 0\). This can be
understood as the physical statement that the electric field \(n^a\,F_{a i}\) on
constant time \(\tau\) surfaces does not diverge like \(\rho^{-1}\) or worse as
\(\rho\to 0\). (Moreover, note that the normal electric field \(n^A u^B F_{AB} =
\order{\rho^0}\), which is the relevant component of the field strength on the
RHS of \cref{eq:holcontact}, is finite by virtue of
\cref{eq:shiftlapsesing,eq:timenormal,eq:holsing}.)

\section{Action for singular configurations}
\label{sec:action}

In \cref{sec:sing}, we have specified a class of codimension-two singularities
which we want to include in the gravitational path integral. The weight of each
configuration in the path integral is of course determined by an action. Our goal
in this section will firstly be to define an action for singular configurations
and secondly, by varying the action, to determine sufficient conditions that
pick out saddles and constrained saddles for the path integral.

\subsection{Proposal for the action}
\label{sec:actionprop}

Making use of the curvature contact terms suggested in \cref{sec:curvature} for
the types of singularities described in \cref{sec:singconfigs} on the
codimension-two surface \(\gamma\), we now propose an action for
Einstein-Hilbert-Maxwell theory including such singularities. Again we denote
the full \(D\)-dimensional Euclidean spacetime by \(\mathscr{M}\) and an
\(\varepsilon\)-neighbourhood of \(\gamma\) by \(\mathscr{N}_\varepsilon\). Then
the action we propose is
\begin{align}
  I
  &= \int_{\mathscr{M}\setminus\mathscr{N}_\varepsilon} (L_{\mathrm{EH}} + L_{\mathrm{M}})
    + \int_{\partial\mathscr{M}} L_{\partial \mathscr{M}}
    - \frac{\mathrm{Area}(\gamma)}{4 G_{\mathrm{N}}}
    + \int_{-\partial\mathscr{N}_\varepsilon}( L_{\mathrm{GH}} + L_{\mathrm{hel}} + L_{\mathrm{hol}} )
    \;,
    \label{eq:action}
\end{align}
where we will now define each term on the RHS.

The first integral contains the standard bulk Einstein-Hilbert and Maxwell
Lagrangian densities,
\begin{align}
  L_{\mathrm{EH}}
  &= - \frac{1}{16\pi G_{\mathrm{N}}}
    (R - 2\Lambda) \epsilon
    \;,
  &
    L_{\mathrm{M}}
  &= \frac{1}{2 g_{\mathrm{M}}^2} F\wedge * F
    \;.
\end{align}
We allow also for a spacetime boundary where a boundary Lagrangian
\(L_{\partial\mathscr{M}}\) resides. Restricted to a set of boundary conditions
to be specified later, we assume that the bulk
\(\mathscr{M}\setminus\mathscr{N}_\varepsilon\) and boundary
\(\partial\mathscr{M}\) actions give a good variational principle away from
\(\gamma\). In particular, surface terms resulting from the variation cancel on
\(\partial\mathscr{M}\). Of primary interest to us are the remaining terms of
\cref{eq:action} which correspond directly to the contact terms we deduced in
\cref{sec:curvature}.

Together with the area term, the integral of the Gibbons-Hawking Lagrangian density
\begin{align}
  L_{\mathrm{GH}}
  &= - \frac{1}{8\pi G_{\mathrm{N}}}
    u^A \epsilon_A\, \tensor{(K_u)}{^b_b}
    \label{eq:laggh}
\end{align}
over \(-\partial\mathscr{N}_\varepsilon\) gives the action of the conical
singularity at \(\gamma\), as described by
\cref{eq:regcurvint,eq:concontact1,eq:concontact2}. These terms have already
made an appearance in this context, for example in
ref.~\cite{Marolf:2022ybi}\footnote{See \cref{sec:lorentzian} for more direct
  comparison to \cite{Marolf:2022ybi} in Lorentzian signature.}.

The next term in \cref{eq:action} has Lagrangian density
\begin{align}
  L_{\mathrm{hel}}
  &= - \frac{1}{8\pi G_\mathrm{N}}
    u^A \epsilon_A\, (\dd{\tau})^a \, N^i\, (K_u)_{a i}
  = - \dd{\tau} \wedge (p_{\mathrm{BY}})_i\, N^i
    \;,
    \label{eq:laghel}
\end{align}
where we will later interpret the dual vector density on surfaces of constant
\((\rho,\tau)\),
\begin{align}
  (p_{\mathrm{BY}})_i
  = -\frac{1}{8\pi G_{\mathrm{N}}}\,
  \tensor[^{(D-2)}]{\epsilon}{}\,
  \tensor{(K_u)}{^{ab}}\,
  n_a \tensor{h}{_{b i}}
  \;,
  \label{eq:pby1}
\end{align}
as a momentum density. (Recall, that \(\tensor[^{(D-2)}]{\epsilon}{}\) is the
volume form on surfaces of constant \((\rho,\tau)\), as introduced in
\cref{eq:codimtwovol}.) We have shown in \cref{eq:regcurvint,eq:helcontact} how
this contact term in the action arises from a direct evaluation of the
Einstein-Hilbert action on a regulated helical singularity.
Ref.~\cite{Chua:2023srl} suggested an identical term, evaluated on a time slice,
for the purposes of obtaining a good variational principle when fixing
\((p_{\mathrm{BY}})_i\). We will make use of some of these results in
\cref{sec:actionvar}, where the connection to angular momentum will also become
clear.

One might worry about the diffeomorphism invariance of the action, given the
explicit appearance of \(\tau\) and \(N^i\) in \cref{eq:laghel}. Firstly, recall
from \cref{sec:singdisambig} that a choice of constant-\(\tau\)-foliation near a
helical singularity \(\gamma\) can be viewed as part of the specification of the
physical internal structure of the singularity. That is, different choices of
constant-\(\tau\) surfaces lead to physically distinct configurations through
our procedure for regulating the singularity. Secondly, on
\(\partial\mathscr{N}_\varepsilon\), note that diffeomorphisms along each
surface of constant \(\tau\) can be dressed to \(\gamma\) by following radial
geodesics orthogonal to \(\gamma\); in particular, over a given point on
\(\gamma\), a circle \(\mathscr{C}_\varepsilon\subset \mathscr{N}_\varepsilon\)
is unambiguously picked out by these radial geodesics. The shift vector \(N^i\)
is the projection of the tangent \(\partial_\tau\) of this circle onto surfaces
of constant \(\tau\). The only possible remaining redundancies are
reparametrizations of the circle \(\mathscr{C}_\varepsilon\), which correspond
to purely \(\tau\)-dependent reparametrizations \(\tau\mapsto f(\tau)\). We have
already implicitly fixed this redundancy (up to a constant) by requiring the
lapse \(N(y)\) to be constant in \(\tau\). But, even if we had not, the
combination \(\dd{\tau}_a N^i\) would be invariant under such reparametrizations
regardless. On a cutoff surface \(\partial\mathscr{N}_\varepsilon\) at constant
proper separation from \(\gamma\), the Lagrangian density \cref{eq:laghel} is
thus unambiguously defined. (We will explore the possibility of other choices of
cutoff surfaces in \cref{sec:shapeindep}.)

Finally, the last term in \cref{eq:action} with Lagrangian density
\begin{align}
  L_{\mathrm{hol}}
  &= - \frac{1}{g_{\mathrm{M}}^2} A \wedge * F
    \label{eq:laghol}
\end{align}
corresponds to the contact term \cref{eq:holcontact} resulting from the
generically nontrivial holonomic singularity \labelcref{eq:holsing} around
\(\gamma\).

\subsubsection{(In)dependence on the cutoff shape}
\label{sec:shapeindep}

As initially defined in \cref{sec:regsing}, the neighbourhood
\(\mathscr{N}_\varepsilon\) around the singularity \(\gamma\) has been chosen
thus far to be a solid tube of constant proper radius from \(\gamma\) as
measured by \(\rho\). What about other choices \(\tilde{\mathscr{N}}_\varepsilon\), where the cutoff
surface \(\partial\tilde{\mathscr{N}}_\varepsilon\) is not a surface of constant
\(\rho\)? In particular, suppose instead that
\(\partial\tilde{\mathscr{N}}_\varepsilon\) is a surface of constant \(\tilde{\rho}\) of
order \(\varepsilon\), where the new coordinate \(\tilde{\rho}\) is related to
\(\rho\) by
\begin{align}
  \rho
  &= \tilde{\rho}\, e^{f(x)}
    \;,
    \label{eq:wigglyrho}
\end{align}
for some function \(f(x)\) of \(x^a=(\tau,y^i)\). We would like to understand
whether the action \labelcref{eq:action} is sensitive to the choice of \(f\). We
will find below that, provided that \(f\) is a function of only \(\tau\) (\ie{}
is constant in \(y^i\)) in some neighbourhood of each connected component of
\(\gamma\), the action is insensitive to the choice of \(f\) in the
\(\varepsilon\to 0\) limit. This can become particularly useful in the
Lorentzian context where it might be more natural to consider a cutoff surface
\(\partial\tilde{\mathscr{N}}_\varepsilon\) at non-constant proper separation
from \(\gamma\).

As we will explicitly see below, the surface integral
\(\int_{-\partial\tilde{\mathscr{N}}_\varepsilon}\) in the action \labelcref{eq:action}
converges in the \(\varepsilon\to 0\) limit, for arbitrary \(f\) and the class
of singularities allowed on \(\gamma\) as described in \cref{sec:singconfigs}.
If we restrict to configurations with finite action per finite area on
\(\gamma\), then the bulk integral
\(\int_{\mathscr{M}\setminus\tilde{\mathscr{N}}_\varepsilon}\) should therefore also
converge in the \(\varepsilon\to 0\) limit. Moreover, from our study of
curvatures in \cref{sec:curvature}, it becomes apparent that the \(\varepsilon\to
0\) limit of these bulk terms, should it exist, is insensitive to the choice of
\(f\). We need therefore only consider the \(f\)-dependence of the surface
terms on
\(\partial\tilde{\mathscr{N}}_\varepsilon\) which is now a surface of constant
\(\tilde{\rho}\).

To be explicit, let \(\tilde{u}^A\), \(\tilde{g}_{ab}\), \(\tilde{g}^{ab}\), and
\((\tilde{K}_{\tilde{u}})_{ab}\) respectively be the unit normal, induced metric, its
inverse, and extrinsic curvature on surfaces of constant \(\tilde{\rho}\). Using
the same \(\tau\) coordinate as before, we obtain a \((D-2)+1\) decomposition of
\(\tilde{g}_{ab}\) analogous to \cref{eq:metricg}, in terms of a lapse
\(\tilde{N}\), shift \(\tilde{N}^i\), and \((D-2)\)-dimensional metric
\(\tilde{h}_{ij}\).

Consider first the gravitational surface terms. The Gibbons-Hawking and helical
surface Lagrangian densities on \(-\partial\tilde{\mathscr{N}}_\varepsilon\) are
taken to be
\begin{align}
  \tilde{L}_{\mathrm{GH}}
 &= -\frac{1}{8\pi G_{\mathrm{N}}}\,
   \tilde{u}^A \epsilon_A\,
   \tilde{g}^{ab}\, (\tilde{K}_{\tilde{u}})_{ab}
   \;,
   \\
  \tilde{L}_{\mathrm{hel}}
  &= -\frac{1}{8\pi G_{\mathrm{N}}}\,
    \tilde{u}^A \epsilon_A\,
    (\dd\tau)_a\, \tilde{g}^{ab}\,
    \tilde{N}^i\, (\tilde{K}_{\tilde{u}})_{bi}
    \;.
\end{align}
We want to compare these densities with the \(f=0\) case intended in \cref{eq:laggh,eq:laghel} on
a surface \(-\partial\mathscr{N}_\varepsilon\) of constant proper separation
\(\rho\) from \(\gamma\). To do so, it is helpful to pull back all densities to
a common space \(S^1\times \gamma\) with coordinates \(x^a=(\tau,y^i)\):
\begin{equation}
  \begin{tikzcd}
    S^1\times\gamma
    \arrow[r,"\phi_\varepsilon"]
    \arrow{d}{\tilde{\phi}_\varepsilon}
    & \partial\mathscr{N}_\varepsilon
    \\
    \partial\tilde{\mathscr{N}}_\varepsilon
    \;.
  \end{tikzcd}
\end{equation}
We will denote the respective pullbacks by \(\phi_\varepsilon^*\) and
\(\tilde{\phi}_\varepsilon^*\). For brevity, we will write
\(\phi_{\varepsilon\to 0}^*=\lim_{\varepsilon\to 0}\phi_\varepsilon^*\) and
\(\tilde{\phi}_{\varepsilon\to 0}^*=\lim_{\varepsilon\to
  0}\tilde{\phi}_\varepsilon^*\).

With the original choice of cutoff
surface \(\partial\mathscr{N}_\varepsilon\) at constant proper separation from
\(\gamma\), we find
\begin{align}
  \phi_{\varepsilon\to 0}^*\,
  L_{\mathrm{GH}}
  &= \frac{1}{8\pi G_{\mathrm{N}}}\,
    \kappa\,\dd{\tau}\wedge
    \tensor[^{(D-2)}]{\epsilon}{}
    \;,\label{eq:lagghexplicit}
  \\
  \phi_{\varepsilon\to 0}^*\,
  L_{\mathrm{hel}}
  &= - v^i\, \dd{\tau} \wedge
    (p_{\mathrm{BY}})_i
    \;.\label{eq:laghelexplicit}
\end{align}
(Here, \(\tensor[^{(D-2)}]{\epsilon}{}\) as introduced in \cref{eq:codimtwovol}
is the volume form on surfaces of constant \((\rho,\tau)\), implicitly evaluated
above at \(\rho\to 0\), thus giving the volume form on \(\gamma\). Similarly,
\((p_{\mathrm{BY}})_i\) is implicitly the \(\rho\to 0\) limit of the momentum
density \cref{eq:pby1}. These are then interpreted in the obvious way as forms
on \(S^1\times\gamma\) in the above RHSs.) However, with a more general cutoff
surface \(\partial\tilde{\mathscr{N}}_\varepsilon\),
\begin{align}
  \tilde{\phi}^*_{\varepsilon\to 0}
  \tilde{L}_{\mathrm{GH}}
  &=
    \phi^*_{\varepsilon\to 0} \left(
    L_{\mathrm{GH}}
    -\frac{u^A \epsilon_A\, \rho}{8\pi G_{\mathrm{N}}}\,
    \left\{
    \frac{1}{2}\,
    n^a\, \partial_a f\,
    n^b\, \partial_b \log\left[
    1 + (\rho\, n^c\, \partial_c f)^2
    \right]
    - D_a D^a f
    \right\}
    \right)
    \label{eq:wigglygh}
  \\
  \begin{split}
    \tilde{\phi}^*_{\varepsilon\to 0}
  \tilde{L}_{\mathrm{hel}}
    &= \begin{multlined}[t]
      \phi^*_{\varepsilon\to 0} \Bigg(
    L_{\mathrm{hel}}
    -\frac{u^A \epsilon_A\, \rho}{16\pi G_{\mathrm{N}}}\,
    \Bigg\{
    n^a\, \partial_a f\,
    \frac{N^i}{N}\,
    \partial_i \log\left[
    1 + (\rho\,n^c\, \partial_c f)^2
    \right]
    \\
    -N^i\, \partial_i \frac{n^a\, \partial_a f}{N}
  - \frac{N^a\, h_{ab}}{N} \mathcal{L}_n (h^{bc}\,\partial_c f)
    \Bigg\}
    \Bigg)
    \;,
  \end{multlined}
  \end{split}
  \label{eq:wigglyhel}
\end{align}
where \(\mathcal{L}_\bullet\) denotes a Lie derivative.

In the absence of a helical singularity, in the \(\varepsilon\to 0\) limit, the Gibbons-Hawking action
\(\int_{-\partial\tilde{\mathscr{N}}_\varepsilon} \tilde{L}_{\mathrm{GH}}\) becomes
insensitive to \(f(x)\) and the helical action
\(\int_{-\partial\tilde{\mathscr{N}}_\varepsilon} \tilde{L}_{\mathrm{hel}}\)
vanishes. To see the insensitivity of
\(\int_{-\partial\tilde{\mathscr{N}}_\varepsilon} \tilde{L}_{\mathrm{GH}}\),
consider the terms in braces in \cref{eq:wigglygh}, of which the second is already manifestly a total
derivative. Meanwhile, using the identity
\begin{align}
  \dv{x} [F(x)-\arctan F(x)]
  &= \frac{1}{2} F(x) \log\left[1+F^2(x)\right]
    \;,
\end{align}
applied with \(F=\rho\, n^c\partial_c f\), the other term can be recast, to
non-vanishing order, as a total \(\tau\)-derivative that integrates to zero over
the \(S^1\) circle.\footnote{As below, we have used the fact that
  \(\sqrt{\det g}/N=\sqrt{\det h}\) becomes \(\tau\)-independent as
  \(\rho\to 0\).} This is necessarily true only because
\(N\,n^a\sim(\partial_\tau)^a\) when the helical shift \(v^i\) vanishes.

In the presence of a helical singularity, we must combine the first terms in the
braces on the RHSs of \cref{eq:wigglygh,eq:wigglyhel} to obtain a total
\(\tau\)-derivative. Together, the sum of \cref{eq:wigglygh,eq:wigglyhel} can be
expressed as
\begin{align}
  \begin{split}
    \tilde{\phi}^*_{\varepsilon\to 0}\left(
    \tilde{L}_{\mathrm{GH}}
    + \tilde{L}_{\mathrm{hel}}
    \right)
    &= \begin{multlined}[t]
      \phi^*_{\varepsilon\to 0} \Bigg(
      L_{\mathrm{GH}}
      + L_{\mathrm{hel}}
      -\frac{u^A \epsilon_A}{8\pi G_{\mathrm{N}}}\,
      \Bigg\{
      -\frac{1}{N}\, \partial_\tau
      \arctan\left(
        \rho\, n^c\, \partial_c f
      \right)
      \\
      + \rho\, \partial_a f \,
      D_b \frac{n^a N^b}{N}
      \Bigg\}
      \Bigg)
      \;,
  \end{multlined}
  \end{split}
  \label{eq:ghhelshapeindep}
\end{align}
where we have used the fact that \(h_{ij}\) and thus \(\sqrt{\det
  g}/N=\sqrt{\det h}\) are \(\tau\)-independent as \(\rho\to 0\). Also because
of this fact, the first term in the braces gives a non-contributing total
\(\tau\)-derivative. However, for generic \(f(x)\), the remaining extraneous
term seems to contribute an unwanted mismatch between
\(\int_{-\partial\tilde{\mathscr{N}}_\varepsilon}(\tilde{L}_{\mathrm{GH}}+\tilde{L}_{\mathrm{hel}})\)
and
\(\int_{-\partial\mathscr{N}_\varepsilon}(L_{\mathrm{GH}}+L_{\mathrm{hel}})\).

This mismatch disappears when \(f\), in some neighbourhood of each connected
component of \(\gamma\), is a function only of \(\tau\) and not the directions
\(y^i\) along \(\gamma\). To see this, we express the extraneous term as
\begin{align}
  -\frac{u^A \epsilon_A\, \rho}{8\pi G_{\mathrm{N}}}
  \,
  \partial_a f \,
  D_b \frac{n^a N^b}{N}
  &=
    -\frac{\rho}{8\pi G_{\mathrm{N}}} \frac{u^A \epsilon_A}{N}
    \,
    \partial_a f \,
    \left(
    N^b\, D_b\, n^a
    + n^a\, d_i\, N^i
    \right)
    \;,
    \label{eq:extraterm}
\end{align}
where \(d_i\) is the covariant derivative associated to \(h_{ij}\). On the RHS,
the first term vanishes because \(n^a\) is a unit normal so \(N^b\,D_b\,n^a\) is
tangential to surfaces of constant \(\tau\). The second term, involving
\(n^a\,\partial_a f(\tau) = \partial_\tau f(\tau)/N\) for each connected
component of \(\gamma\), again contributes a total \(\tau\)-derivative, provided
\(d_i N^i/N\) is also \(\tau\)-independent at leading order in the \(\rho\to 0\)
limit. Alternatively, this term is a total \(y^i\)-derivative to finite order if
\(N\) is invariant under the flow generated by \(N^i\) at leading order as
\(\rho\to 0\) (\ie{} \(N^i\partial_i N\) vanishes faster than \(\rho\)). Thus,
we conclude that, given the \(\tau\)-independence of \(\lim_{\rho\to0}h_{ij}\),
\(\kappa\), and \(v^i\) or given the \(\tau\)-independence of
\(\lim_{\rho\to0}h_{ij}\) and the shift invariance \(v^i\partial_i\kappa=0\),
the surface terms
\(\int_{-\partial\tilde{\mathscr{N}}_\varepsilon}(\tilde{L}_{\mathrm{GH}}+\tilde{L}_{\mathrm{hel}})\)
of the action are insensitive to the time profile \(f(\tau)\) of each connected
component of the cutoff surface \(\partial\tilde{\mathscr{N}}_\varepsilon\).

Finally, let us consider the Maxwell surface term
\(\int_{-\partial\mathscr{\tilde{N}}_\varepsilon} L_{\mathrm{hol}}\), where the
Lagrangian density \(L_{\mathrm{hol}}\) is again given by \cref{eq:laghol}. We
simply have
\begin{align}
  \tilde{\phi}^*_{\varepsilon\to 0}
  L_{\mathrm{hol}}
  &=
    \phi^*_{\varepsilon\to 0}
    L_{\mathrm{hol}}
    = - \frac{1}{g_{\mathrm{M}}^2} \mu \dd{\tau} \wedge
    * F
    \;,
    \label{eq:helshapeindep}
\end{align}
provided, as previously mentioned below \cref{eq:holcontact}, the electric field
\(n^a F_{ai}\) does not diverge like \(\rho^{-1}\) or worse as \(\rho\to 0\).
(On the RHS, \(*F\) is implicitly understood in the \(\rho\to 0\) limit
and viewed as a form on \(S^1\times\gamma\) in the obvious way.) Thus,
\(\int_{-\partial\mathscr{\tilde{N}}_\varepsilon} L_{\mathrm{hol}}\) is
insensitive, in the \(\varepsilon\to 0\) limit, to the profile \(f(x)\) of the
cutoff surface \(\partial\mathscr{N}_\varepsilon\).

In summary, assuming \(\lim_{\rho\to 0}h_{ij}\) is \(\tau\)-independent
as required for \(\gamma\) to have a well-defined induced metric, we altogether have
\begin{multline}
  \tilde{\phi}^*_{\varepsilon\to 0}\left(
    \tilde{L}_{\mathrm{GH}}
    + \tilde{L}_{\mathrm{hel}}
    + L_{\mathrm{hol}}
  \right)
  \\
  = \dd{\tau}\wedge\left[
  \frac{
    \kappa
  }{8\pi G_{\mathrm{N}}}\,
  \tensor[^{(D-2)}]{\epsilon}{}
  - v^i\, (p_{\mathrm{BY}})_i
  - \frac{\mu}{g_{\mathrm{M}}^2}
  * F
  -\frac{
    f'(\tau)\, d_i\, v^i
  }{8\pi G_{\mathrm{N}}\,\kappa}
  \,
  \,
  \tensor[^{(D-2)}]{\epsilon}{}
\right]
\label{eq:wigglyall}
\end{multline}
for any profile \(f(\tau)\) of the cutoff surface
\(\partial\tilde{\mathscr{N}}_\varepsilon\) with arbitrary dependence only on time
\(\tau\). We conclude that, if \(\kappa\) and \(v^i\) are also
\(\tau\)-independent or if \(\kappa\) is invariant under the helical shift
\(v^i\), then the last term above is a total derivative and any time profile
\(f(\tau)\) for each connected component of the cutoff surface
\(\partial\tilde{\mathscr{N}}_\varepsilon\) can be used to evaluate the action
\labelcref{eq:action}, without altering the value of the action. This fact will
be useful when we eventually turn to Lorentzian signature in
\cref{sec:lorentzian}. For the sake of simplicity, however, for now we will
return to working with cutoff surfaces \(\partial\mathscr{N}_\varepsilon\) at
constant proper separation \(\rho\) from the singularity \(\gamma\).

\subsection{Variation of the action}
\label{sec:actionvar}

We will now evaluate the variation of the action \cref{eq:action}. In the
process, conjugacy between various quantities on
\(\partial\mathscr{N}_\varepsilon\) will become apparent. A surface stress
tensor, as well as angular momentum and electric charge densities here, will be
introduced.

Varying the bulk Lagrangians,
\begin{align}
  \delta (L_{\mathrm{EH}} + L_{\mathrm{M}})
  &= (E_{\mathrm{EHM}})^{AB} \delta G_{AB} + (E_{\mathrm{M}})^A \delta A_A
    + \dd(\theta_{\mathrm{EH}} + \theta_{\mathrm{M}})
    \label{eq:varlag}
\end{align}
produces the Einstein-Hilbert-Maxwell equations of motion (as densities)
\(E_{\mathrm{EHM}}\) and \(E_{\mathrm{M}}\), together with surface terms. As
previously mentioned, surface terms at \(\partial\mathscr{M}\) cancel in
variations subject to boundary conditions,
\begin{align}
  \int_{\partial\mathscr{M}} (\theta_{\mathrm{EH}} + \theta_{\mathrm{M}} + \delta L_{\partial\mathscr{M}})
  &= 0
    \;.
    \label{eq:varbdy}
\end{align}

Meanwhile, on \(\partial\mathscr{N}_\varepsilon\), we have the standard result
\begin{align}
  \pi_{\partial\mathscr{N}_\varepsilon} \theta_{\mathrm{EH}}
  + \delta L_{\mathrm{GH}}
  &= -\frac{1}{2} u^A\epsilon_A \, (T_{\mathrm{BY}})^{ab} \delta g_{ab} \;,
  \label{eq:psympeh}
\end{align}
where the surface, \ie{} Brown-York, stress tensor is given by\footnote{Note
  that we are currently in Euclidean signature --- \cf{} \cref{eq:lorstressby}.
  The signs in \cref{eq:psympeh,eq:stressby} are because \(u^A\) is directed
  radially outward, so the orientation on \(-\partial\mathscr{N}_\varepsilon\)
  as part of of \(\partial(\mathscr{M}\setminus\mathscr{N}_\varepsilon)\) is
  \(-u^A\epsilon_A\).\label{foot:stresssign}}
\begin{align}
  (T_{\mathrm{BY}})^{ab}
  &= - \frac{1}{8\pi G_{\mathrm{N}}} \left[
    (K_u)^{ab} - g^{ab} \tensor{(K_u)}{^c_c}
    \right]
    \;.
    \label{eq:stressby}
\end{align}
From a radial evolution perspective, \cref{eq:psympeh} is the standard
Einstein-Hilbert presymplectic potential density, so the coefficient of \(\delta
g_{ab}\) there is understood as the canonical momentum of the induced metric
\(g_{ab}\) on \(\partial\mathscr{N}_\varepsilon\). Note that presymplectic
potential densities related by field space exact one-forms, \eg{} \(\delta L_{\mathrm{GH}}\),
lead to the same presymplectic density \(\omega = \delta \theta\) (where \(\delta\)
is the field space exterior derivative).

We will find it helpful to apply the \((D-2)+1\) decomposition
\labelcref{eq:metricg} to \cref{eq:psympeh} \cite{Chua:2023srl}:
\begin{align}
  \pi_{\partial\mathscr{N}_\varepsilon} \theta_{\mathrm{EH}}
  + \delta L_{\mathrm{GH}}
  &= \dd{\tau}
    \wedge \left[
    e_{\mathrm{BY}}\, \delta N
    + (p_{\mathrm{BY}})_i\, \delta N^i
    \right]
    - \frac{1}{2} u^A \epsilon_A\, T^{ij} \delta h_{ij}
    \label{eq:psympadm}
\end{align}
where the surface energy and (angular) momentum densities are given
by\footnote{Our extrinsic curvature \(K_u\) is the negative of the extrinsic
  curvature \(K\) in ref.~\cite{Chua:2023srl}, because the subscript reminds
  us that \(K_u\) is defined using the vector \(u\) which, near \(\gamma\),
  points \emph{into} the spacetime region
  \(\mathscr{M}\setminus\mathscr{N}_\varepsilon\). Our \(e_{\mathrm{BY}}\) and
  \(p_{\mathrm{BY}}\) are thus equal to \(-\tensor[^{(D-2)}]{\epsilon}{}=u^A n^B \epsilon_{AB}\) times the
  \(q\) and \(j\) defined in ref.~\cite{Chua:2023srl}'s eq.~(2.9) and (2.10).
  When comparing our \cref{eq:varaction} to ref.~\cite{Chua:2023srl}'s
  eq.~(4.3), one should be wary of the orientation \(-u^A\epsilon_A\) of
  \(-\partial\mathscr{N}_\varepsilon\).}
\begin{align}
  e_{\mathrm{BY}}
  &= \tensor[^{(D-2)}]{\epsilon}{} \,
    (T_{\mathrm{BY}})^{ab} n_a n_b
    =
    \frac{1}{8\pi G_{\mathrm{N}}}\,
    \tensor[^{(D-2)}]{\epsilon}{}\,
    \tensor{(K_u)}{^i_i}
    \label{eq:eby}
\end{align}
and, as in \cref{eq:pby1},
\begin{align}
  (p_{\mathrm{BY}})^i
  &= \tensor[^{(D-2)}]{\epsilon}{} \,
    (T_{\mathrm{BY}})^{ab} n_a \tensor{h}{_b^i}
    =
    - \frac{1}{8\pi G_{\mathrm{N}}}\,
    \tensor[^{(D-2)}]{\epsilon}{}\,
    \tensor{(K_u)}{^{ab}}
    n_a \tensor{h}{_b^i}
    \;,
    \label{eq:pby2}
\end{align}
where \(\tensor[^{(D-2)}]{\epsilon}{}\) is the volume form on surfaces of
constant \((\rho,\tau)\), as defined in \cref{eq:codimtwovol}. The variation of
the above can be easily combined with \cref{eq:psympadm}:
\begin{align}
  \pi_{\partial\mathscr{N}_\varepsilon} \theta_{\mathrm{EH}}
  + \delta L_{\mathrm{GH}} + \delta L_{\mathrm{hel}}
  &= \dd{\tau}
    \wedge \left[
    e_{\mathrm{BY}}\, \delta N
    - \delta (p_{\mathrm{BY}})_i\, N^i
    \right]
    - \frac{1}{2} u^A \epsilon_A\, (T_{\mathrm{BY}})^{ij} \delta h_{ij}
    \;.
    \label{eq:varsurfeh}
\end{align}

For the Maxwell field, the presymplectic potential density is
\begin{align}
  \theta_{\mathrm{M}}
  &= \frac{1}{g_{\mathrm{M}}^2} \delta A \wedge * F
    \;,
\end{align}
or equivalently
\begin{align}
  \theta_{\mathrm{M}} + \delta L_{\mathrm{hol}}
  &= - \frac{1}{g_{\mathrm{M}}^2} A \wedge \delta(* F) \;,
    \label{eq:varsurfm}
\end{align}
which both reproduce the usual conjugacy between the vector potential and the
electric field on codimension-one surfaces.

Collecting together \cref{eq:varlag,eq:varbdy,eq:varsurfeh,eq:varsurfm}, the
variation of the action \labelcref{eq:action} is given by
\begin{align}
  \begin{split}
  \delta I
  &= \int_{\mathscr{M}\setminus\mathscr{N}_\varepsilon}
    \left[
    (E_{\mathrm{EHM}})^{AB} \delta G_{AB} + (E_{\mathrm{M}})^A \delta A_A
    \right]
    - \frac{\delta\mathrm{Area}(\gamma)}{4 G_{\mathrm{N}}}
    \\
  &\phantom{{}={}}
    + \int_{-\partial\mathscr{N}_\varepsilon} \left\{
    \dd{\tau}
    \wedge \left[
    e_{\mathrm{BY}}\, \delta N
    - \delta(p_{\mathrm{BY}})_i\, N^i
    \right]
    - \frac{1}{2} u^A \epsilon_A\, (T_{\mathrm{BY}})^{ij} \delta h_{ij}
    - \frac{1}{g_{\mathrm{M}}^2} A \wedge \delta(* F)
    \right\}
    \;.
  \end{split}
  \label{eq:varaction}
\end{align}
We have a good variational principle if the variation of the action reduces only
to equations of motion while all other terms vanish when appropriate
``boundary'' conditions are imposed, in this context, on
\(\partial\mathscr{N}_\varepsilon\) in the \(\varepsilon\to 0\) limit or
\(\gamma\).\footnote{Recall we are assuming that we already have boundary
  conditions on \(\partial\mathscr{M}\) which make the variation surface terms
  there vanish, as written in \labelcref{eq:varbdy}.} In particular, as we will
note in \cref{sec:surftermsunconstr}, taking the \(\varepsilon\to 0\) limit, the
\(\int_{-\partial\mathscr{N}_\varepsilon}\dd{\tau}\wedge e_{\mathrm{BY}}\,\delta
N\) term above vanishes and the \(\int_{-\partial\mathscr{N}_\varepsilon}u^A
\epsilon_A\,(T_{\mathrm{BY}})^{ij} \delta h_{ij}\) term becomes proportional to
\(\int_\gamma \delta\left( \tensor[^{(D-2)}]{\epsilon}{} \right)\,\kappa\) where
again \(\tensor[^{(D-2)}]{\epsilon}{}\) is the volume form on \(\gamma\).
Therefore, we can say we have a good variational principle at fixed
\(\tensor[^{(D-2)}]{\epsilon}{}\), \((p_{\mathrm{BY}})_i\), and \(*F\) along
\(\gamma\). Alternatively, in \cref{sec:constrsaddle}, we will see that the
surface terms in \cref{eq:varaction} vanish also if we fix the area, integrated
momentum density, and electric flux on \(\gamma\), and the singularity strengths
satisfy some constancy conditions
\labelcref{eq:constrsaddle1,eq:constrsaddle2,eq:constrsaddle3}.\footnote{In this
  paper, we are not necessarily restricting our path integrals to be over
  configurations with constant singularity strengths. Rather, if we call the
  constancy conditions
  \labelcref{eq:constrsaddle1,eq:constrsaddle2,eq:constrsaddle3} ``equations of
  motion'', we can also claim to have a good variational principle at fixed
  area, integrated momentum density, and electric flux on \(\gamma\). (As
  described in \cref{sec:partfunc}, path integrals which fix these integrated
  quantities will come into play in our analysis of the gravitational partition
  function.)}

\subsection{Conditions for saddles and constrained saddles}
\label{sec:saddle}

We would now like to deduce conditions that must be satisfied by saddles of the
action \labelcref{eq:action} by requiring that its variation
\labelcref{eq:varaction} vanishes. Obviously, saddles must solve the bulk
gravitational and Maxwell equations of motion away from
\(\mathscr{N}_\varepsilon\),
\begin{align}
  E_{\mathrm{EHM}}
  &= 0\;,
  &
    E_{\mathrm{M}}
  &= 0 \;.
    \label{eq:bulkonshell}
\end{align}
We would also like to understand the saddle-point conditions implied by the
other terms of \cref{eq:varaction} near the possibly singular surface
\(\gamma\). We will be further interested in constrained saddles where certain
quantities are fixed near \(\gamma\), thus restricting the set of possible field
variations there.

\subsubsection{Equations of motion near \(\gamma\)}

Before proceeding, let us observe that some nontrivial conditions on \(\gamma\)
and its embedding into (constrained) saddles can already be obtained from the
bulk equations of motion \labelcref{eq:bulkonshell}. For example, let us
consider the components of the gravitational equations of motion corresponding
to ``momentum constraints'' on constant \(\rho\) surfaces. Let us take the
expression \labelcref{eq:stressby} for the Brown-York stress tensor to define
\((T_{\mathrm{BY}})^{ab}\) on any surface of small constant \(\rho\) in terms of
the induced metric \(g_{ab}\) and extrinsic curvature \((K_u)_{ab}\) of the surface.
Then, the momentum constraints are equivalent to the conservation of
\((T_{\mathrm{BY}})^{ab}\),
\begin{align}
  D_a\, (T_{\mathrm{BY}})^{ab} &= 0
      \;,
\end{align}
where again \(D_a\) is the covariant derivative associated to the metric \(g_{ab}\) on
constant \(\rho\) surfaces. Let us focus on the component
\begin{align}
  n_a\, D_b\, (T_{\mathrm{BY}})^{ab}
  &= \frac{1}{8 \pi G_{\mathrm{N}}}
    \left[
    h^{ij} \partial_\rho (k_n)_{ij}
    + \frac{1}{2} \partial_\rho h^{ij} (k_n)_{ij}
    + \frac{1}{2} d_i \frac{\partial_\rho N^i}{N}
    - \frac{\partial_\rho N^i \partial_i N}{N^2}
    \right]
    \label{eq:tbyconsern}
\end{align}
written out here for a generically off-shell configuration, in terms of the
\((D-2)+1\) decomposition \labelcref{eq:metricg} of \(g_{ab}\) and the extrinsic
curvature \labelcref{eq:extrcurvadm} of constant \((\rho,\tau)\) slices. Again,
\(d_i\) is the covariant derivative associated to \(h_{ij}\). The first two
terms of \cref{eq:tbyconsern} generically diverge in the \(\rho\to 0\) limit
while the latter terms are \(\order{\rho^0}\). To be explicit, let us write
\begin{align}
  h_{ij}(\rho,\tau,y)
  &= h^0_{ij}(y)
    + h'_{ij}(\rho,\tau,y)
    \;,
    &
    \lim_{\rho\to 0}
      h'_{ij}(\rho,\tau,y)
    &= 0
    \;,
\end{align}
and use \((h^0)^{ij}\) to denote the inverse of \(h^0_{ij}\) (against the
prevailing convention of raising indices using \(h^{ij}\)). Let us suppose that
\(h_{ij}'\) generically vanishes slower than \(\rho^2\) as \(\rho\to 0\), so
that the latter two terms in \cref{eq:tbyconsern} can be neglected in the
following. Then, we find
\begin{align}
  \begin{split}
    \MoveEqLeft n_a\, D_b\, (T_{\mathrm{BY}})^{ab}
    \\
    &= \frac{1}{8\pi G_{\mathrm{N}}}\Bigg\{\frac{\partial_\rho N}{2N^2}\,
      (h^0)^{ij}\,
      \mathcal{L}_{\vec{N}} h^0_{ij}
    - \frac{\partial_\rho N}{2N^2}\left[
    (h^{0})^{ij}
    (\partial_\tau - \mathcal{L}_{\vec{N}})
    h'_{ij}
    + (h^0)^{ik}(h^0)^{j\ell}\,
      h'_{k\ell} \,
      \mathcal{L}_{\vec{N}} h^0_{ij}
    \right]
    \\
  &\phantom{{}={}}
    +\frac{1}{2N}\left[
    (h^0)^{ij}
    \partial_\rho(\partial_\tau-\mathcal{L}_{\vec{N}})
    h'_{ij}
    + \frac{1}{2}
    (h^0)^{ik} (h^0)^{j\ell}\,
    \partial_\rho h'_{k\ell} \,
    \mathcal{L}_{\vec{N}} h^0_{ij}
    \right]
    +\cdots
    \Bigg\}
    \;.
  \end{split}
  \label{eq:tbyconsernexpan}
\end{align}
The first term, of order \(\rho^{-2}\), is dominant over all other terms at
small \(\rho\). The remaining terms on the RHS's first line and the displayed
terms on the last line are respectively of order \(\rho^{-2}\,h'_{ij}\)
and \(\rho^{-1}\,\partial_\rho h'_{ij}\); which of the two is larger is
contingent on the \(\rho\)-dependence of \(h'_{ij}\), but at least one of
the two is larger than the omission \(\cdots\). The vanishing of
\cref{eq:tbyconsernexpan} to leading order implies that
the helical shift \(N^i \sim v^i\) of \(\gamma\) is divergence-free:
\begin{align}
  0
  &= \frac{1}{2} (h^0)^{ij}\mathcal{L}_{v} h^0_{ij}
    = d_i\, v^i
    \;.
\end{align}

Let us also consider for a moment \(v^i=0\), \ie{} non-helical singularities,
and suppose that \(h_{ij}\) has a small \(\rho\) expansion with first subleading
term \(h'_{ij}= f_1(\rho,y) h^1_{ij}(\tau,y) + \cdots\) for some \(f_1\) and
\(h^1_{ij}\). (\Eg{} for a power law expansion, as considered in
ref.~\cite{Dong:2019piw}, \(f_1=\rho^{s}\) for some \(s>0\) possibly dependent
on the strength of the conical singularity.) Then, we see that, the vanishing of
\cref{eq:tbyconsernexpan} generically\footnote{In the special smooth case,
  \(\partial_\rho N \sim 1\) and \(f_1(\rho)=\rho\), so the relevant terms in
  \cref{eq:tbyconsernexpan} automatically cancel.} implies that \((h^0)^{ij}
\partial_\tau h^1_{ij}=0\), which reproduces the well known extremality
condition for the area of a conically singular surface
\(\gamma\).\footnote{\Cf{} ref.~\cite{Dong:2019piw}'s eqs.~(A.64) and (A.65).}
More generally, in the presence of a nontrivial helical singularity, at least
just from examining the component \labelcref{eq:tbyconsern} alone, it appears
that the helical shift \(v^i\) will muddle this would-be extremality condition.

We leave for future work a thorough analysis of conditions on \(\gamma\) implied
by other bulk equations of motion.

\subsubsection{Surface terms under unconstrained variations}
\label{sec:surftermsunconstr}

Let us move on to the saddle-point conditions implied by the other terms in
\cref{eq:varaction}. Firstly, we note that the lapse term
\begin{align}
\int_{-\partial\mathscr{N}_\varepsilon}
  \dd{\tau} \wedge e_{\mathrm{BY}} \delta N
  \sim& 0
\end{align}
in \cref{eq:varaction} vanishes as \(\varepsilon\to 0\) automatically as a
consequence of \(N=\order{\rho}\) and \(h_{ij}=\order{\rho^0}\).
The remaining terms in \cref{eq:varaction} are
\begin{align}
  - \frac{\delta\mathrm{Area}(\gamma)}{4 G_{\mathrm{N}}}
  &= - \frac{1}{8 G_{\mathrm{N}}}
    \int_\gamma \tensor[^{(D-2)}]{\epsilon}{}\, h^{ij} \delta h_{ij}
    \;,
    \label{eq:vararea}
\end{align}
\begin{align}
  -\int_{-\partial\mathscr{N}_\varepsilon}
  \dd{\tau}
  \wedge
  \delta(p_{\mathrm{BY}})_i\, N^i
  &\sim
    -2\pi
    \int_\gamma \delta(p_{\mathrm{BY}})_i\, v^i
    \;,
    \label{eq:varhelterm}
\end{align}
\begin{align}
  - \frac{1}{2}
  \int_{-\partial\mathscr{N}_\varepsilon}
  u^A \epsilon_A\, (T_{\mathrm{BY}})^{ij} \delta h_{ij}
  &\sim \frac{1}{16\pi G_{\mathrm{N}}}
    \int_{-\partial\mathscr{N}_\varepsilon}
    u^A \epsilon_A\,
    \tensor{(K_u)}{^c_c} h^{ij}\delta h_{ij}
  \\
  &\sim
    -\frac{1}{16\pi G_{\mathrm{N}}}
    \int_{-\partial\mathscr{N}_\varepsilon}
    u^A \epsilon_A\,
    n^a n^b \tensor{(K_u)}{_{ab}}
    h^{ij}\delta h_{ij}
  \\
  &=
    -\frac{1}{16\pi G_{\mathrm{N}}}
  \int_{-\partial\mathscr{N}_\varepsilon}
    u^A \epsilon_A\,
    \frac{\partial_\rho N}{N}
    h^{ij}\delta h_{ij}
  \\
  &\sim
    \frac{1}{8 G_{\mathrm{N}}}
  \int_\gamma
    \tensor[^{(D-2)}]{\epsilon}{}\,
    \kappa\,
    h^{ij}\delta h_{ij}
    \;,
    \label{eq:vartbyterm}
\end{align}
and
\begin{align}
  - \frac{1}{g_{\mathrm{M}}^2}
  \int_{-\partial\mathscr{N}_\varepsilon}
  A \wedge \delta(* F)
  &\sim - \frac{2\pi}{g_{\mathrm{M}}^2}
    \int_\gamma \mu\, \delta(* F)
    \;.
    \label{eq:varholterm}
\end{align}
For unconstrained variations \(\delta h_{ij}\), \(\delta(p_{\mathrm{BY}})_i\),
and \(\delta(*F)\), requiring the terms
\labelcref{eq:vararea}+\labelcref{eq:vartbyterm}, \labelcref{eq:varhelterm},
and \labelcref{eq:varholterm} to vanish respectively lead to the saddle point
conditions
\begin{align}
  \kappa &= 1 \;,
  &
    v^i &= 0 \;,
  &
    \mu &= 0 \;.
          \label{eq:smoothsaddle}
\end{align}
Thus, unsurprisingly, unconstrained saddles are smooth at \(\gamma\).

\subsubsection{Constrained saddles}
\label{sec:constrsaddle}

However, we are also interested in constrained saddles where certain quantities,
which we will denote by \(\mathcal{C}^{(\alpha)}\), near \(\gamma\) are fixed.
For such saddles, the variation of the action \(\delta I\) need only vanish for
constrained variations preserving the fixed quantities,
\(\delta\mathcal{C}^{(\alpha)}=0\). Equivalently, for each constrained saddle,
there exist constant Lagrange multipliers \(\lambda_{(\alpha)}\), such that all
(unconstrained) variations from the saddle satisfy
\begin{align}
  \delta I + \lambda_{(\alpha)} \delta \mathcal{C}^{(\alpha)}
  &= 0
    \;.
    \label{eq:constrsaddle}
\end{align}

One quantity which we would like to fix is the area of \(\gamma\),
\begin{align}
  \mathcal{C}^{(1)}
  &= \mathrm{Area}(\gamma)
    \;,
    \label{eq:fixarea}
\end{align}
whose unconstrained variation is given by \cref{eq:vararea}.

We would also like to fix some notion of (angular) momentum on \(\gamma\). To
construct such a quantity from \((p_{\mathrm{BY}})_i\) defined in \cref{eq:pby2},
we must choose a vector with which to contract this dual-vector density. In
\(D=3\) spacetime dimensions, a natural choice is the unit-norm vector
\begin{align}
  \chi^i
  &= (\tensor[^{(D-2)}]{\epsilon}{})^i
    = h^{ij}\,(\tensor[^{(D-2)}]{\epsilon}{})_j
    \;.
    \label{eq:unitvec}
\end{align}
which serves also as the einbein on \(\gamma\),
\begin{align}
  h_{ij}
  &= \chi_i \chi_j
    \;.
    \label{eq:einbein}
\end{align}
Using \((p_{BY})_i\) and \(\chi^i\), we construct the quantity\footnote{Note
  that \((p_{\mathrm{BY}})_i \, \chi^i\) is a top form on \(\gamma\) --- see \cref{eq:pby2}.}
\begin{align}
  \mathcal{C}^{(2)}
  &= \int_\gamma (p_{\mathrm{BY}})_i \, \chi^i
    \;,
    \label{eq:fixmom}
\end{align}
which we would like to fix on \(\gamma\). The unconstrained variation of this
quantity is
\begin{align}
  \delta \mathcal{C}^{(2)}
  &= \int_\gamma \left[
    \delta(p_{\mathrm{BY}})_i \, \chi^i
    -\frac{1}{2} (p_{\mathrm{BY}})_i \, \chi^i \,
    h^{jk} \delta h_{jk}
    \right]
    \;.
\end{align}
In higher dimensions, it is less clear what vector(s) we should contract with
\((p_{\mathrm{BY}})_i\) to construct fixed quantities that lead to desirable
saddle-point conditions, \eg{} as described below. Let us therefore restrict all
discussion of fixed (angular) momentum to \(D=3\), at least for now.

Lastly, for the Maxwell field, we would also like to fix the electric charge
enclosed by \(\gamma\), as measured by Gauss's law,
\begin{align}
  \mathcal{C}^{(3)}
  &= \int_\gamma *F
    \;.
    \label{eq:fixcharge}
\end{align}

From \cref{eq:constrsaddle}, we see that fixing the quantities
\labelcref{eq:fixarea,eq:fixmom,eq:fixcharge} weakens the saddle-point
conditions at \(\gamma\) from \cref{eq:smoothsaddle} to now
\begin{align}
  0 &= \frac{\kappa-1}{8 G_{\mathrm{N}}}
      + \frac{\lambda_{(1)}}{2}
      -\frac{\lambda_{(2)}}{2}\,
      \iota_\chi(p_{\mathrm{BY}})_i \, \chi^i
      \;,
  \\
  0 &= -2\pi
      v^i + \lambda_{(2)} \chi^i
      \;,
  \\
  0 &= - 2\pi\mu + \lambda_{(3)}
      \;,
\end{align}
for arbitrary constants \(\lambda_{(\alpha)}\). As previously, \(\iota\) denotes
an interior product, \eg{} for \(\chi\) which we have defined in \(D=3\),
\begin{align}
  \iota_{\chi} \left(
  \tensor[^{(D-2)}]{\epsilon}{}
  \right)
  &= \chi^i\, \chi_i
    = 1
    \;.
\end{align}
The second condition can be used to eliminate
\(\lambda_{(2)}\) from the first. Also making use of
\cref{eq:unitvec,eq:einbein}, these saddle-point conditions can be succinctly
expressed as
\begin{align}
  \kappa
  - 8\pi G_{\mathrm{N}} \, \iota_\chi (p_{\mathrm{BY}})_i \, v^i
  &= \text{constant}
    \;,
    \label{eq:constrsaddle1}
  \\
  \chi_i\, v^i
  &= \text{constant}
    \;,
    \label{eq:constrsaddle2}
  \\
  \mu
  &= \text{constant}
    \label{eq:constrsaddle3}
    \;.
\end{align}
These saddle-point conditions will be satisfied by the constrained saddles to be
constructed in \cref{sec:bhsaddles} from black holes.

\section{Thermal partition function}
\label{sec:partfunc}

As described in the introduction in \cref{sec:intro}, a goal of this paper is to
better understand the gravitational path integral, in particular, as applied to
calculate the analogue of a thermal partition function \(Z(\beta,\Omega,\Phi)\).
More precisely, if the gravitational theory is holographically dual to a theory
on its boundary, then in this boundary theory,
\begin{align}
  Z(\beta,\Omega,\Phi)
  &= \tr\left(e^{-\beta\,H_{\xi,\Phi}}\right)
    \label{eq:partfunc}
\end{align}
is the grand canonical partition, where
\begin{align}
    H_{\xi,\Phi}
  &= H - \Omega\, J - \Phi\, Q
    \;,
\end{align}
\(H\) is the Hamiltonian generating evolution in a static time direction
\(\zeta\), \(J\) is the (angular) momentum generating translation or rotation in
a spatial direction \(\varphi\), and \(Q\) is electric charge.

In \cref{sec:pathintegral}, we will describe a naively Euclidean path integral
representation of \(Z(\beta,\Omega,\Phi)\). Of course, the integral over
Euclidean metrics is plagued by the conformal factor problem. Remaining
agnostic to what the precise contour of integration might be, we will use
this Euclidean setup merely to practice constructing (constrained) saddles, in
preparation for the better defined Lorentzian calculation in
\cref{sec:lorpartfunc}. Specifically, we will practice evaluating the path
integral in \cref{sec:saddles} by first fixing then later integrating quantities
\((\mathcal{S},\mathcal{J},\mathcal{Q})\) proportional to the area, integrated
momentum density, and electric flux on a generically singular codimension-two
surface \(\gamma\). We will construct conically, helically, and holonomically
singular constrained saddles for the intermediate integral at fixed
\((\mathcal{S},\mathcal{J},\mathcal{Q})\), allowing us to evaluate the path
integral by saddle-point methods. Many of the techniques used in this
\namecref{sec:pathintegral} will reappear in the Lorentzian calculation of
\cref{sec:lorpartfunc}.

\subsection{Specifying the Euclidean path integral}
\label{sec:pathintegral}

To give a partial description of the configurations which we would like to
include in our gravitational path integral, we might say the following. At the
spacetime boundary, \(\partial\mathscr{M}\), the configurations will be required
to satisfy certain boundary conditions, to be specified below. Within the
interior of the spacetime \(\mathscr{M}\), we integrate over configurations
which are smooth apart from a (possibly disconnected) codimension-two surface
\(\gamma\). On \(\gamma\), we allow singularities of the types introduced in
\cref{sec:singconfigs} for Euclidean signature. But, are we really integrating
over all such real Euclidean spacetimes \(\mathscr{M}\)? Unfortunately, such an
integral is ill-defined because of the conformal factor problem
\cite{Gibbons:1978ac}: the Euclidean action is unbounded below on this choice of
integration contour. We will leave the precise contours of the path integral
unspecified for now, but revisit this question in
\cref{sec:lorentzian,sec:discussion}.

Turning a temporary blind eye to this issue, let us continue by describing the
boundary conditions at the spacetime boundary \(\partial\mathscr{M}\) relevant
for the grand canonical partition function \cref{eq:partfunc}. The induced
geometry on \(\partial\mathscr{M}\) is fixed and possesses at least two Killing
vectors \(-i\,\zeta\) and \(\varphi\) interpreted as generating Euclidean time
evolution and spatial translation (or rotation). (To be consistent with later
notation, we have introduced \(\zeta\) as a Lorentzian time Killing vector,
hence the factor of \(-i\).) Moreover, we require the boundary conditions of all
fields to be invariant under these Killing symmetries. The parameters \(\beta\)
and \(\Omega\) do not influence the local rigid structure of
\(\partial\mathscr{M}\) but instead enter as follows. We require
\(\partial\mathscr{M}\) to have topology \(Y\times S^1\), where the constant
sections \(Y\) and fibres \(S^1\) need not be metric-orthogonal. The Killing
vector \(\varphi\) is tangent to constant sections \(Y\) while the \(S^1\)
fibres are the orbits of
\begin{align}
  -i\,\xi
  &= -i\,\zeta - i\,\Omega\, \varphi
    \;,
    \label{eq:corotvec}
\end{align}
such that \(e^{-i\,\beta\, \xi}\sim 1\) (acting on integer-spin fields) completes
one orbit. To obtain boundary conditions for a real Euclidean metric,
\(\Omega\) can be taken to be imaginary; on the other hand, real \(\Omega\)
describes a physical angular velocity, \eg{} for a physical Lorentzian black
hole or to give \(Z(\beta,\Omega,\Phi)\) a standard thermodynamic
interpretation. It will be useful to introduce a time coordinate \(\hat{\tau}
\sim \hat{\tau}+\beta\) starting at zero on a given constant section \(Y\) of
\(\partial\mathscr{M}\) and evolving as the Killing parameter of \(-i\,\zeta\) (and
thus also \(-i\,\xi\)). We will further denote the orbital period of \(\varphi\) by
\(\mathrm{Period}(\varphi)\), so that \(e^{\mathrm{Period}(\varphi)\, \varphi}
\sim 1\) (acting on integer-spin fields). Then we say that \(\varphi\) generates a
rotation if such a finite \(\mathrm{Period}(\varphi)\) exists; otherwise
\(\varphi\) generates translation.

The Maxwell field \(A\) will also be subject to boundary conditions on
\(\partial\mathscr{M}\). Sidestepping the possibly nontrivial analysis of field
asymptotics\footnote{See \eg{} \cite{Marolf:2006nd} for some analysis in
  asymptotically AdS spacetimes.} near \(\partial\mathscr{M}\), we will simply
assume boundary conditions which lead to a good variational principle in the
sense of \cref{eq:varbdy}. We require that the boundary conditions for \(A\) be
parametrized by a number \(\Phi\), with the interpretation of an electric
potential that is fixed on \(\partial\mathscr{M}\). We demand, for example, that
a trivial configuration \(A=0\) near \(\partial\mathscr{M}\) satisfies these
boundary conditions with \(\Phi=0\) and that shifting the time component of the
Maxwell field \(A\mapsto A + \hat{\mu} \dd{\hat{\tau}}\) leads to a
configuration satisfying boundary conditions with a shifted potential \(\Phi
\mapsto \Phi + i \hat{\mu}\). Moreover, we will also assume that \(\Phi\) enters
into the value of the boundary Hamiltonian in the manner befitting a fixed
potential ensemble, \eg{} as in \cref{eq:bdyhamval} below. It will be convenient
below to write the size of the Maxwell gauge group \(G_{\mathrm{M}}\) as
\(\mathrm{Period}(G_{\mathrm{M}})\), \eg{}
\begin{align}
  \mathrm{Period}(G_{\mathrm{M}})
  &= \begin{cases}
    2\pi & \text{if }G_{\mathrm{M}} = U(1)
    \\
    +\infty & \text{if }G_{\mathrm{M}} = \mathbb{R}
  \end{cases}
    \;.
\end{align}

Are boundary conditions specified by different values of \(\Omega\) or \(\Phi\)
inequivalent? For translation or non-compact \(G_{\mathrm{M}}\cong\mathbb{R}\),
yes seems to be the most natural answer. However, when \(\varphi\) generates
rotation, one might expect that inequivalent boundary conditions are specified
by \(\Omega\) only up to shifts by \(\frac{2\pi i}{\beta\, \Delta_J}\) for some
\(\Delta_J\).\footnote{Boundary conditions that are equivalent under this
  relation with \(\Delta_J\) given by \cref{eq:deltajint,eq:deltajhalfint}
  respectively are related by the \(T\) and \(T^2\) modular transformation of
  the boundary torus \(\partial\mathscr{M}\).} For example, if states of the
theory are all invariant under one full rotation
\(e^{\mathrm{Period}(\varphi)\,\varphi}\), then one would expect
\begin{align}
  \Delta_J
  &= \frac{2\pi}{\mathrm{Period}(\varphi)}
    \;.
    \label{eq:deltajint}
\end{align}
If two full rotations \(e^{2\mathrm{Period}(\varphi)\, \varphi}\) are required,
then one would instead expect
\begin{align}
  \Delta_J
  &= \frac{1}{2}\,\frac{2\pi}{\mathrm{Period}(\varphi)}
    \;.
    \label{eq:deltajhalfint}
\end{align}
This would be the case, for example, if the gravitational theory has a dual
boundary description in which fermions have anti-periodic Neveu-Schwarz
identification around orbits of \(\varphi\).\footnote{The partition function
  \cref{eq:partfunc}, without any fermion-parity \((-1)^F\) insertion,
  implicitly has Neveu-Schwarz conditions for fermions around the orbits
  generated by \(-i\,\xi\).}

For compact \(G_{\mathrm{M}}\cong U(1)\), we might similarly consider
inequivalent boundary conditions to be specified by \(\Phi \mod{\frac{2\pi i}{\beta\, \Delta_Q}}\), where one might most naturally expect
\begin{align}
  \Delta_Q
  &= \frac{2\pi}{\mathrm{Period}(G_{\mathrm{M}})}
    \;.
\end{align}
We will regard the equivalences
\begin{align}
  \Omega
  &\sim \Omega + \frac{2\pi i\, m}{\beta\,\Delta_J}\,
  &
    (m\in\mathbb{Z})
    \label{eq:periodangvel}
  \\
  \Phi &\sim \Phi + \frac{2\pi i\, n}{\beta\,\Delta_Q}
  &
    (n\in\mathbb{Z})
    \label{eq:periodelepot}
\end{align}
between boundary conditions as assumptions about how the path integral should be
formulated. (Depending on whether one wishes to ignore or make these
assumptions, the notation used later can be interpreted with \(m\) or \(n\) set
to zero or taking arbitrary integer values.) Assuming the equivalence(s), we
will conversely find that the path integral quantizes angular momentum and/or
charge with spacing \(\Delta_J\) and \(\Delta_Q\) respectively.

For simplicity, we focus on pure Einstein-Maxwell theory with a cosmological
constant. In \cref{sec:discussion}, we will touch on some of the challenges
associated to the inclusion of other matter fields, for the most part leaving
these extensions for future work.

\subsection{Saddle-point evaluation}
\label{sec:saddles}

We now construct saddles of the path integral which, as described in
\cref{sec:saddle}, extremize the action, possibly subject to constraints
on \(\gamma\). We do not claim that the list of saddles below is exhaustive.
Moreover, we will refrain from arguing whether the saddles actually contribute
to a saddle-point approximation of the path integral, leaving such an analysis
to \cref{sec:lorentzian,sec:discussion}. For now, we simply assume they do and
study their potential contributions to the grand canonical partition function.

\subsubsection{Empty thermal saddle}
\label{sec:emptysaddle}

One somewhat trivial saddle is given by an empty stationary background,
periodically identified along a stationary flow to produce the nowhere
degenerate \(S^1\). The Maxwell background \(A = -i\Phi \dd{\hat{\tau}}\) is
flat, with the time coordinate \(\hat{\tau}\) extended into the
bulk. The surface \(\gamma=\varnothing\) is trivial for this saddle.

The action for this stationary background is proportional to \(\beta\) and
independent of \(\Omega\) and \(\Phi\). The leading, classical contribution of
this saddle to the grand canonical partition function is therefore
\(e^{-\beta\,E}\) for some \(E\), interpreted as the background energy of this
saddle. Perturbative fluctuations around this saddle describe the thermal QFT
grand canonical ensemble on this empty background. Writing the QFT grand
canonical partition function, \eg{} including gravitons to one loop, as
\(Z_{\mathrm{th}}^{\mathrm{QFT}}(\beta,\Omega,\Phi)\), the corrected
gravitational partition function is
\begin{align}
  Z_{\mathrm{th}}(\beta,\Omega,\Phi)
  &= e^{-\beta E} Z_{\mathrm{th}}^{\mathrm{QFT}}(\beta,\Omega,\Phi) \;.
    \label{eq:thpartfunc}
\end{align}

\subsubsection{Black hole saddles}
\label{sec:bhsaddles}

As is well known, charged rotating black holes are possible smooth saddles when
considering the path integral of the grand canonical partition function
\(Z(\beta,\Omega,\Phi)\). However, it is instructive to dissect the calculation
of the path integral slightly. In particular, let us imagine first integrating
over everything with the exception of certain fixed quantities
\((\mathcal{S},\mathcal{J},\mathcal{Q})\) on a singular codimension-two surface
\(\gamma\) and, only at the end, integrating over those initially fixed
quantities. That is, we write
\begin{align}
  Z(\beta,\Omega,\Phi)
  &= \int
    \dd{\mathcal{S}}\dd{\mathcal{J}}\dd{\mathcal{Q}}
    Z(\beta,\Omega,\Phi;\mathcal{S},\mathcal{J},\mathcal{Q})
    \;.
    \label{eq:finalints}
\end{align}
where \(Z(\beta,\Omega,\Phi;\mathcal{S},\mathcal{J},\mathcal{Q})\) is the
integral over a codimension-three subcontour of fixed
\((\mathcal{S},\mathcal{J},\mathcal{Q})\) in the full contour of the path
integral \(Z(\beta,\Omega,\Phi)\). The quantities whose integrals we would like
to leave until the end are
\begin{align}
  \mathcal{S}
  &= \frac{\mathrm{Area}(\gamma)}{4 G_{\mathrm{N}}}
    \;,
  \\
  \mathcal{J}
  &= -i\,\frac{\mathrm{Area}(\gamma)}{\mathrm{Period}(\varphi)}
    \int_\gamma (p_{\mathrm{BY}})_i \, \chi^i
    \;,
    \label{eq:mom}
  \\
  \mathcal{Q}
  &= -\frac{i}{g_{\mathrm{M}}^2}\int_\gamma *F
    \;,
    \label{eq:charge}
\end{align}
which are related to the \(\mathcal{C}^{(\alpha)}\), introduced in
\cref{eq:fixarea,eq:fixmom,eq:fixcharge}, by rescaling so that they can be
equated to a black hole's Bekenstein-Hawking entropy, (angular) momentum, and
charge below. For definiteness, we have taken \(\varphi\) and \(\chi\) to point
in the same direction.\footnote{This statement is meaningful when \(\gamma\) can
  be homologously related to a boundary time slice \(Y\) by a preferred class of continuous
  deformations. We anticipate that this will be true in Lorentz signature for
  certain cases where \(\gamma\) is homologously related to a boundary time
  slice \(Y\) by spatial surfaces of bulk codimension one.} The factors of
\(-i\) in \cref{eq:mom,eq:charge} ensure that the RHSs coincide with the
standard real values of angular momentum and charge when evaluated on a
Lorentzian black hole with real angular velocity \(\Omega\) and potential
\(\Phi\). (The Lorentzian expressions \labelcref{eq:lormom,eq:lorcharge} for
\(\mathcal{J}\) and \(\mathcal{Q}\) are related to \cref{eq:mom,eq:charge} above
by the Wick rotations \labelcref{eq:hodgewick,eq:pbywick} of quantities on the
RHSs.)

We may now look for potential saddles for the subcontour path integral
\(Z(\beta,\Omega,\Phi;\mathcal{S},\mathcal{J},\mathcal{Q})\). As described in
\cref{sec:saddle}, such constrained saddles must satisfy
\cref{eq:constrsaddle1,eq:constrsaddle2,eq:constrsaddle3} on \(\gamma\) in
addition to equations of motion away from \(\gamma\). We can construct such
constrained saddles as follows.

First, we start with a smooth stationary black hole solution with
Bekenstein-Hawking entropy (\ie{} horizon area), (angular) momentum, and
electric charge set to the prescribed values of
\((\mathcal{S},\mathcal{J},\mathcal{Q})\) fixed in the integral
\(Z(\beta,\Omega,\Phi;\mathcal{S},\mathcal{J},\mathcal{Q})\). While (angular)
momentum and charge are ordinarily evaluated on \(\partial\mathscr{M}\), it is
straightforward to see that evaluating \cref{eq:mom,eq:charge} on the
bifurcation surfaces of stationary black holes reproduces the values obtained
from standard definitions on \(\partial\mathscr{M}\).\footnote{We have in mind
  here situations where \(\gamma\) is homologously related to a boundary time
  slice \(Y\) by a surface of bulk codimension one, which is preserved by the
  Killing vector \(\varphi\) extended into the bulk and coincides with a surface
  of constant \(\tau\) (or \(t\) in Lorentz signature, as will be considered in
  \cref{sec:lorentzian}) near \(\gamma\). In more general situations, \eg{} for
  arbitrary \(\mathrm{SL}(2,\mathbb{Z})\) black holes \cite{Maloney:2007ud}
  discussed in \cref{sec:bhsums}, \cref{eq:mom,eq:charge,eq:lormom,eq:lorcharge}
  might not correspond precisely to the standard definitions of (angular)
  momentum and charge on \(\partial\mathscr{M}\). \label{foot:momchargebh}} We
require this initial configuration to satisfy boundary conditions on
\(\partial\mathscr{M}\) with some parameters \((\beta_0,\Omega_0,\Phi_0)\)
possibly differing from the desired values \((\beta,\Omega,\Phi)\). We will now
correct this mismatch.

Let \(\hat{\tau}\) be the bulk stationary time extending the boundary time
introduced in \cref{sec:pathintegral}. Then, to satisfy the periodicity
prescribed by \(\beta\), we set by hand \(\hat{\tau} \sim \hat{\tau} + \beta\).
This induces a conical singularity with opening angle
\begin{align}
  2\pi\,\kappa
  &= 2\pi \frac{\beta}{\beta_0}
    \label{eq:bhcon}
\end{align}
around the bifurcation surface \(\gamma\).

\begin{figure}
  \centering
  \begin{subfigure}[t]{0.225\textwidth}
    \centering
    \includegraphics{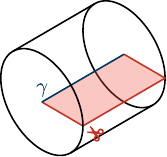}
    \caption{First, we cut open the spacetime along a stationary time slice
      which is preserved by the bulk extension of the boundary Killing vector
      \(\varphi\) generating spatial translation or rotation. For rotation, the
      front and back faces of the solid cylinder are identified.}
    \label{fig:helicalbh1}
  \end{subfigure}
  \hfill
  \begin{subfigure}[t]{0.225\textwidth}
    \centering
    \includegraphics{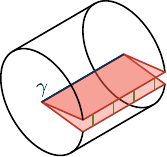}
    \caption{The top and bottom faces of the cut are initially identified in the
      trivial manner, as indicated by the green threads connecting identified
      points of the two red faces.}
    \label{fig:helicalbh2}
  \end{subfigure}
  \hfill
  \begin{subfigure}[t]{0.45\textwidth}
    \centering
    \includegraphics{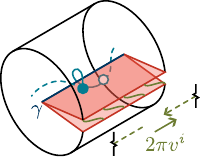}
    \caption{However, we can instead resew the top and bottom faces with a shift
      \(2\pi\,v^i\). The solid teal curve, drawn as winding helically around
      \(\gamma\), is now a closed curve, because the empty and filled endpoints
      are identified. The strength of the helical singularity at the bifurcation
      surface \(\gamma\) is given by \(v^i\). When the front and back faces of
      the solid cylinder are identified, different \(v^i\), \eg{} corresponding
      to the shifts shown as solid and dashed arrows, can lead to configurations
      that agree away from \(\gamma\). However, when regularized as described in
      \cref{sec:regsing}, these configurations become distinct in a
      neighbourhood of \(\gamma\). In particular, they have different
      contractible cycles, as illustrated by the solid and dashed teal curves.}
    \label{fig:helicalbh3}
  \end{subfigure}
  \caption{A cut, shift, and resew procedure on a stationary spacetime with a
    bifurcation surface \(\gamma\). By an appropriate choice of the shift, it is
    possible to ensure that the boundary co-rotating Killing vector \(-i\,\xi\)
    in \cref{eq:corotvec} has closed orbits on the spacetime boundary
    \(\partial\mathscr{M}\) (and the bulk extension of \(-i\,\xi\) also has
    closed orbits, \eg{} the solid teal curve in (\subref{fig:helicalbh3})).}
  \label{fig:helicalbh}
\end{figure}

To produce the shift prescribed by \(\Omega\), we can perform a cut, shift, and
resew procedure illustrated in \cref{fig:helicalbh}. First, we cut open the
spacetime along a stationary time slice, then relatively shift the two faces of
the cut, and finally reidentify the two faces so that closed orbits are
generated with period \(\beta\) by the desired co-rotating vector \(-i\,\xi\), given
by \labelcref{eq:corotvec}, on \(\partial\mathscr{M}\). This introduces a
helical singularity on \(\gamma\) with shift \eg{}
\begin{align}
  v^i
  &= \frac{\beta}{2\pi i} (\Omega-\Omega_0)\, \varphi^i
    \;,
\end{align}
where we have extended the Killing vector \(\varphi\), previously introduced on
\(\partial\mathscr{M}\) in \cref{sec:pathintegral}, now into the bulk. When
boundary conditions \labelcref{eq:periodangvel} are treated as
equivalent for \(m\in\mathbb{Z}\), one can make other choices for the shift,
turning on any
\begin{align}
  v^i
  &=  \left[ \frac{\beta}{2\pi i}
    (\Omega-\Omega_0)
    + \frac{m}{\Delta_J}
    \right] \varphi^i
    \;.
  &
    (m \in \mathbb{Z})
    \label{eq:bhhel}
\end{align}
These different shifts lead to configurations which are diffeomorphic to each
other away from \(\gamma\) --- see \cref{fig:helicalbh3} --- but have helical
singularities on \(\gamma\) which are physically distinct, as evident upon
regularization --- see \cref{sec:regsing}.

Finally, to match the prescribed electric potential \(\Phi\) on
\(\partial\mathscr{M}\), we can shift the Maxwell field
\(A\mapsto A-i\,(\Phi-\Phi_0)\dd{\hat{\tau}}\). Doing so turns on turns on a holonomic
singularity with strength
\begin{align}
  \mu
  &= \frac{\beta}{2\pi i}(\Phi - \Phi_0) \;.
\end{align}
If boundary conditions \labelcref{eq:periodelepot} are equivalent for
\(n\in\mathbb{Z}\), we can further shift \(A\) by arbitrary integer multiples of
\(\frac{2\pi}{\beta\,\Delta_Q} \dd{\hat{\tau}}\), giving
\begin{align}
  \mu
  &= \frac{\beta}{2\pi i}(\Phi - \Phi_0) + \frac{n}{\Delta_Q}
    \;.
  & (n\in\mathbb{Z})
    \label{eq:bhhol}
\end{align}

At last, we have some configuration(s) satisfying the desired boundary
conditions specified by \((\beta,\Omega,\Phi)\) at \(\partial\mathscr{M}\) and
reproducing the fixed values of \((\mathcal{S},\mathcal{J},\mathcal{Q})\) on
\(\gamma\). Equations of motion are satisfied away from \(\gamma\), because the
original smooth back hole is a solution. Moreover, we see from
\cref{eq:bhcon,eq:bhhel,eq:bhhol} and the bulk Killing symmetry \(\varphi\) that
\labelcref{eq:constrsaddle1,eq:constrsaddle2,eq:constrsaddle3} on \(\gamma\) are
satisfied. Thus, we have indeed constructed constrained saddle(s) which are
saddles for the integral
\(Z(\beta,\Omega,\Phi;\mathcal{S},\mathcal{J},\mathcal{Q})\), or rather many
constrained saddles labelled by integer \(m\) and/or \(n\) if the equivalences
\labelcref{eq:periodangvel} and/or \labelcref{eq:periodelepot} between boundary
conditions are assumed.

To evaluate the saddle-point contribution to this integral, we must calculate
the action \labelcref{eq:action} on the constrained saddle. It is helpful to
first consider the original smooth black hole with parameters
\((\beta_0,\Omega_0,\Phi_0)\), where the combination of terms
\begin{align}
  \int_{\mathscr{M}\setminus\mathscr{N}_\varepsilon} (L_{\mathrm{EH}} + L_{\mathrm{M}})
  + \int_{\partial\mathscr{M}} L_{\partial \mathscr{M}}
  + \int_{-\partial\mathscr{N}_\varepsilon} L_{\mathrm{GH}}
  &\sim \beta_0\, E_{\zeta + \Omega_0 \varphi,\Phi_0}(\mathcal{S},\mathcal{J},\mathcal{Q})
    \label{eq:actionsmoothbh}
\end{align}
is given by the value
\begin{align}
  E_{\zeta + \Omega_0 \varphi,\Phi_0}(\mathcal{S},\mathcal{J},\mathcal{Q})
  &= E(\mathcal{S},\mathcal{J},\mathcal{Q})
    - \Omega_0\, \mathcal{J} - \Phi_0\, \mathcal{Q}
    \label{eq:bdyhamval}
\end{align}
of the boundary Hamiltonian generating time evolution and translation or
rotation with (angular) velocity \(\Omega_0\) in the presence of an electric
potential \(\Phi_0\). Here, \(E(\mathcal{S},\mathcal{J},\mathcal{Q})\) is the
energy of the system with respect to just the time evolution. The
modifications undergone from this smooth black hole to construct our constrained
saddle should not locally modify the Lagrangian densities appearing in
\cref{eq:actionsmoothbh}. The only change in the value to
\cref{eq:actionsmoothbh} results from the adjustment of the time period
\(\beta_0\mapsto\beta\). Thus, for our constrained saddle,
\begin{align}
  \int_{\mathscr{M}\setminus\mathscr{N}_\varepsilon} (L_{\mathrm{EH}} + L_{\mathrm{M}})
  + \int_{\partial\mathscr{M}} L_{\partial \mathscr{M}}
  + \int_{-\partial\mathscr{N}_\varepsilon} L_{\mathrm{GH}}
  &\sim \beta\, E_{\zeta + \Omega_0 \varphi,\Phi_0}(\mathcal{S},\mathcal{J},\mathcal{Q})
    \;.
    \label{eq:bulkghsingbh}
\end{align}

With \cref{eq:bhhel,eq:bhhol}, we can also evaluate other terms in the action
\labelcref{eq:action}:
\begin{align}
  \int_{-\partial\mathscr{N}_\varepsilon} L_{\mathrm{hel}}
  &\sim -2\pi \int_\gamma v^i\,(p_{\mathrm{BY}})_i
    = \left[
    \beta\,(\Omega_0-\Omega)
    - \frac{2\pi i\, m}{\Delta_J}
    \right] \mathcal{J}
  \\
  \int_{-\partial\mathscr{N}_\varepsilon} L_{\mathrm{hol}}
  &\sim - \frac{2\pi}{g_{\mathrm{M}}^2} \int_\gamma \mu *F
    = \left[
    \beta\, (\Phi_0 - \Phi)
    - \frac{2\pi i\, n}{\Delta_Q}
    \right] \mathcal{Q}
    \;.
\end{align}
In the first line, we have used \cref{eq:laghel} and
\begin{align}
  \varphi^i
  &= \frac{\mathrm{Area}(\gamma)}{\mathrm{Period}(\varphi)} \, \chi^i
\end{align}
on \(\gamma\). Altogether, the action \labelcref{eq:action} of our constrained
saddle evaluates to
\begin{align}
  I
  &=  \beta\, E_{\xi,\Phi}(\mathcal{S},\mathcal{J},\mathcal{Q})
    -\mathcal{S}
    - 2\pi i\, m\, \frac{\mathcal{J}}{\Delta_J}
    - 2\pi i\, n\, \frac{\mathcal{Q}}{\Delta_Q}
    \;.
\end{align}
where
\begin{align}
  E_{\xi,\Phi}(\mathcal{S},\mathcal{J},\mathcal{Q})
  &= E(\mathcal{S},\mathcal{J},\mathcal{Q})
    - \Omega\, \mathcal{J} - \Phi\, \mathcal{Q}
    \;.
\end{align}

The contribution of each such saddle to the subcontour path integral
\(Z(\beta,\Omega,\Phi;\mathcal{S},\mathcal{J},\mathcal{Q})\) is perturbatively
given by \(e^{-I}\), with the above value of \(I\), corrected multiplicatively
by the partition function of perturbative quantum fluctuations, which we will
write as
\begin{align}
  Z_{\mathrm{BH}}^{\mathrm{QFT}}\left(
  \beta,
  \Omega+\frac{2\pi i\, m}{\beta\, \Delta_J},
  \Phi+\frac{2\pi i\, n}{\beta\, \Delta_Q};
  \mathcal{S},\mathcal{J},\mathcal{Q}
  \right)
  \;.
\end{align}
Including the final integrals over \((\mathcal{S},\mathcal{J},\mathcal{Q})\), and summing over \(m\) and \(n\)
as needed, the contribution to the full integral \labelcref{eq:finalints}
becomes
\begin{align}
  \sum_{m,n}
  Z_{\mathrm{BH}}
  \left(
  \beta,
  \Omega+\frac{2\pi i\, m}{\beta\, \Delta_J},
  \Phi+\frac{2\pi i\, n}{\beta\, \Delta_Q}
  \right)
  \;,
  \label{eq:allbhpartfunc}
\end{align}
where
\begin{align}
  Z_{\mathrm{BH}}(\beta,\Omega,\Phi)
  &= \int \dd{\mathcal{S}}\dd{\mathcal{J}}\dd{\mathcal{Q}}
    e^{\mathcal{S} - \beta\, E_{\xi,\Phi}(\mathcal{S},\mathcal{J},\mathcal{Q})}
    Z_{\mathrm{BH}}^{\mathrm{QFT}}\left(
    \beta, \Omega, \Phi
    ;\mathcal{S},\mathcal{J},\mathcal{Q}
    \right)
    \;.
    \label{eq:bhpartfunc}
\end{align}
The latter takes the expected form for a grand canonical ensemble of states with
energies \(E\), (angular) momenta \(\mathcal{J}\), electric charges \(\mathcal{Q}\), and density
\(e^{\mathcal{S}}\), with corrections from quantum fluctuations.\footnote{Recall that \(E\) can be
  regarded as a function of \((\mathcal{S},\mathcal{J},\mathcal{Q})\); alternatively, one can regard \(\mathcal{S}\) as
  a function of \((E,\mathcal{J},\mathcal{Q})\).}

When \(\varphi\) generates translation, the sum \(\sum_m\) has only the trivial
\(m=0\) term. For rotation, if all \(m\in\mathbb{Z}\) saddles are included, then
the sum over these saddles leads to a discrete angular momentum spectrum. To see
this, note that the summed contribution \labelcref{eq:allbhpartfunc} to the
partition function is periodic in imaginary values of \(\Omega\) as required by
\cref{eq:periodangvel}. Therefore, taking the inverse Laplace transform in
\(-\Omega\) (\ie{} inverse Fourier transform in \(-i\Omega\)), we conclude that
the thermodynamically conjugate variable, angular momentum, must have a discrete
spectrum with spacing \(\Delta_J\). To see this more directly in a toy
calculation, let us naively neglect quantum fluctuations by dropping
\(Z_{\mathrm{BH}}^{\mathrm{QFT}}\). Then, the sum over \(m\) becomes a Dirac
comb
\begin{align}
  \sum_{m=-\infty}^\infty  e^{2\pi i m\,\mathcal{J}/\Delta_J}
  &=
    \sum_{j=-\infty}^\infty \delta\left(
    \frac{\mathcal{J}}{\Delta_J} - j
    \right)
    \;,
    \label{eq:diraccomb}
\end{align}
selecting discrete values of \(\mathcal{J}\in\Delta_J\,\mathbb{Z}\).\footnote{Of
  course, this is only a toy calculation since
  \(Z_{\mathrm{BH}}^{\mathrm{QFT}}\) is a multiplicative correction to each
  saddle, so dropping this factor to perform the sum \(\sum_m\) is unjustified;
  moreover, \(\mathcal{J}\) captures only the angular momentum of the
  background. We expect the more general argument following from the periodicity
  in \(\Omega\) to hold even when the quantum corrections
  \(Z_{\mathrm{BH}}^{\mathrm{QFT}}\) are included.}

If the Maxwell gauge group \(G_{\mathrm{M}}\) is compact and we sum over all
\(n\in\mathbb{Z}\), then charge is similarly quantize as expected with spacing
\(\Delta_Q\).

\section{Lorentzian formulation}
\label{sec:lorentzian}

In this section, we turn now to configurations in Lorentzian signature. Our goal
will be twofold. Firstly, we would like to characterize and assign an action to
conical, helical, and holonomic singularities on a codimension-two surface
\(\gamma\) in a Lorentzian spacetime. In \cref{sec:lorsing}, we will be
partially successful in this respect, leaving for future work some open
challenges introduced by the presence of lightcones.

We will nonetheless be able to progress towards our second goal in
\cref{sec:lorpartfunc}, evaluating the gravitational thermal partition function
\(Z(\beta,\Omega,\Phi)\) using a Lorentzian path integral. This calculation, as
already sketched in \cref{sec:newstuff}, closely parallels the calculation of
ref.~\cite{Marolf:2022ybi}, but, like the naive Euclidean calculation of
\cref{sec:partfunc}, employs constrained saddles containing the more general
types of codimension-two singularities studied in this paper. The final answer
for black hole contributions to \(Z(\beta,\Omega,\Phi)\) is again
\cref{eq:allbhpartfunc,eq:bhpartfunc}, but derived now from a Lorentzian path
integral free from the conformal factor problem. This result and its advantages
over the previous analysis of \cite{Marolf:2022ybi} will be discussed in
\cref{sec:discussion}.

\subsection{Codimension-two singularities}
\label{sec:lorsing}

In Lorentzian signature, the singular codimension-two surface \(\gamma\) can be
timelike or spacelike.\footnote{The possibility of a null \(\gamma\) is perhaps
  more subtle.} In the former case, we do not expect any novelties relative to
our preceding Euclidean discussion, apart from the fact that \(h_{ij}\) in
\cref{eq:metricg} is now Lorentzian as opposed to Euclidean. (See, \eg{}
ref.~\cite{Tod:1994iop}, which treats special cases of time-like conical and
helical singularities constructed by cutting and gluing flat spacetime.) From
here on, we will try to understand the situation where \(\gamma\) is a spacelike
surface.

The primary challenges that arise when studying spacelike \(\gamma\) occur when
the spacetime possesses lightcones that emanate from \(\gamma\). Thus, it will
be easier to start in \cref{sec:nolightcones} to consider spacetimes where such
lightcones do not exist --- see \cref{fig:nolightcones}. In fact, this will be
the situation directly relevant to our construction of Lorentzian constrained
saddles for the thermal partition function in \cref{sec:lorpartfunc}. In
\cref{sec:lightcones}, we will address some of the subtleties that arise in the
presence of lightcones; however, these issues do not arise in the saddles used
in \cref{sec:lorpartfunc} so uninterested readers are free to skip
\cref{sec:lightcones} directly to \cref{sec:lorpartfunc}.

\subsubsection{Spacelike \(\gamma\) with no lightcones}
\label{sec:nolightcones}

\begin{figure}
  \centering
  \begin{subfigure}[t]{0.475\textwidth}
    \centering
    \includegraphics{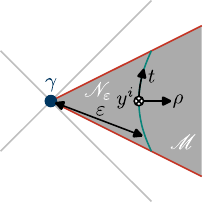}
    \caption{\(\gamma\) spacelike separated from nearby points in
      the spacetime \(\mathscr{M}\).}
    \label{fig:nolightconesspacelike}
  \end{subfigure}
  \hfill
  \begin{subfigure}[t]{0.475\textwidth}
    \centering
    \includegraphics{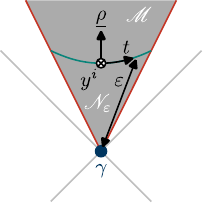}
    \caption{\(\gamma\) timelike separated from nearby points.}
    \label{fig:nolightconestimelike}
  \end{subfigure}
  \caption{The spacetime \(\mathscr{M}\) (shaded) near a spacelike singularity
    \(\gamma\) with no lightcones. The red surfaces are identified. The would-be
    lightcones (grey lines) lie beyond the spacetime \(\mathscr{M}\). Similar to
    \cref{fig:coordinates}, \(\rho\) or \(\underline{\rho}\) is a proper radial
    coordinate coming out from \(\gamma\), \(t\) is a boost angle coordinate
    with some period, say \(2\pi\), and \(y^i\) are coordinates along the directions of
    \(\gamma\). Additional cases are obtained by rotating
    (\subref{fig:nolightconesspacelike}) and (\subref{fig:nolightconestimelike})
    by \(\pi\).}
  \label{fig:nolightcones}
\end{figure}

When no lightcones emanate from the singular codimension-two surface \(\gamma\),
the singularity can still be characterized in almost the same manner as in
\cref{sec:singconfigs}, except that \(\tau\) in \cref{eq:metricg} should be
traded for a boost angle \(t\). Pictorially, the spacetime near
\(\gamma\) can be illustrated as in \cref{fig:nolightcones}, as we now describe.

The fact that the spacetime possesses no lightcones emanating from \(\gamma\)
indicates that the spacetime near \(\gamma\) does not span infinite proper boost
angles. Thus, the (possibly improper) boost angle coordinate \(t\) can be taken
to be periodically identified with some constant period, say \(2\pi\), much like
the Euclidean angle \(\tau\) of \cref{sec:singconfigs}. Topologically, a
neighbourhood of \(\gamma\) is thus again given by \(\gamma\) times a disk, as
illustrated in \cref{fig:coordinates}. However, the causal structure is better
exhibited by the Penrose diagrams in \cref{fig:nolightcones}. Note, in
particular, that closed curves are parametrized by \(t\) at fixed proper
spacelike separation \(\rho\) or timelike separation \(\underline{\rho}\) from
\(\gamma\) and fixed coordinates \(y^i\) along the directions of \(\gamma\).
Moreover, these circles degenerate at \(\gamma\), where the causal structure
becomes singular.

For the purposes of translating equations from previous sections to Lorentzian
signature, the boost angle \(t\) can roughly be viewed\footnote{This is not to
  say that an arbitrary real Euclidean spacetime necessarily becomes a real
  Lorentzian spacetime under this wick rotation. This is true for only certain
  (\eg{} static) spacetimes or in conjunction with the analytic continuation of
  other parameters (\eg{} \cref{eq:wicknormallapseshift} for stationary spacetimes).} as a
Wick rotation of the Euclidean angle \(\tau\):
\begin{align}
  \tau
  &= i\, t
    \label{eq:wickrot}
\end{align}
It will be convenient to refer to some Lorentzian quantities by reusing the same
symbols previously introduced in Euclidean signature. For example, the relation
between the Euclidean (\(\mathrm{E}\)) and Lorentzian (\(\mathrm{L}\)) unit normals,
lapses, and shifts with respect to \(\tau\) and \(t\) are:
\begin{align}
  n_{\mathrm{E}}
  &= N_{\mathrm{E}} \dd{\tau}
    = i\,N_{\mathrm{L}} \dd{t}
    = i\,n_{\mathrm{L}}
    \;,
  &
    N_{\mathrm{E}}
  &= N_{\mathrm{L}}
    \;,
    &
    N_{\mathrm{E}}^i
  &= -i\, N_{\mathrm{L}}^i
    \;,
    \label{eq:wicknormallapseshift}
\end{align}
From here on, we will work primarily with Lorentzian objects and omit subscripts
except when explicitly comparing Lorentzian and Euclidean quantities. In case
\(\gamma\) is timelike separated from nearby points as in
\cref{fig:nolightconestimelike}, we also rotate the proper radial coordinate
\begin{align}
  \rho
  &= i\, \underline{\rho} \;.
    \label{eq:radialwickrot}
\end{align}
The unit normals respectively to surfaces of constant \(\underline{\rho}\) and \(t\), and the lapse with
respect to \(t\) will be similarly be underlined in case \(\gamma\) is timelike
separated from nearby points as in \cref{fig:nolightconesspacelike}:
\begin{align}
  u
  &= \dd{\rho}
    = i \dd{\underline{\rho}}
    = i\, \underline{u}
    \;,
  &
    n
  &= N\, \dd{t}
    = i\, \underline{N}\, \dd{t}
    = i\, \underline{n}
    \;,
    &
    N
  &= i\, \underline{N} \;.
    \label{eq:wicknormallapse}
\end{align}

As in \cref{sec:conhelsings}, we can characterize the conical and helical nature
of the singularity \(\gamma\) by examining metric \labelcref{eq:metricG} near
\(\gamma\) in radially static gauge,
\begin{align}
  G_{AB} \dd{X^A} \dd{X^B}
  &= \dd{\rho}^2 + g_{ab} \dd{x}^a \dd{x}^b
    = -\dd{\underline{\rho}}^2 + g_{ab} \dd{x}^a \dd{x}^b
    \;.
    \label{eq:lormetricG}
\end{align}
The induced metric \(g_{ab}\) on surfaces of constant proper radius
\(\rho\) or \(\underline{\rho}\) now reads
\begin{align}
  g_{ab} \dd{x}^a \dd{x}^b
  &= -N^2 \dd{t}^2
    + h_{ij}\,
    (\dd{y}^i + N^i \dd{t})
    (\dd{y}^j + N^j \dd{t})
    \label{eq:lormetricg1}
  \\
  &= \underline{N}^2 \dd{t}^2
    + h_{ij}\,
    (\dd{y}^i + N^i \dd{t})
    (\dd{y}^j + N^j \dd{t})
    \;.
    \label{eq:lormetricg2}
\end{align}
Again, we can consider small \(\rho\) or \(\underline{\rho}\) behaviour of the lapse
\(N\) or \(\underline{N}\) and shift \(N^i\) as in \cref{eq:shiftlapsesing}:
\begin{align}
 \begin{split}
  N
  &= \kappa\, \rho
    + (\text{subleading in }\rho)
    \;,
   \\
   \underline{N}
  &= \kappa\, \underline{\rho}
    + (\text{subleading in }\underline{\rho})
    \;,
 \end{split}
  &
    N^i
  &= v^i + \order{\rho^2}
    = v^i + \order{\underline{\rho}^2}
    \;.
    \label{eq:lorlapseshift}
\end{align}
Similar to in \cref{sec:conhelsings}, the conical nature of the singularity is
determined by \(\kappa\) which describes the passage of proper hyperbolic angle
around \(\gamma\) per coordinate time \(t\). If constant, \(2\pi\,\kappa\) is the
total opening angle when the period of \(t\) is taken to be \(2\pi\). More
generally, we will allow \(\kappa\) to depend on \(t\) or \(y^i\) below.
Meanwhile, the strength of the helical singularity is parametrized by \(v^i\) as
before, which also might depend on \(t\) and/or \(y^i\). (In
\cref{sec:lightcones}, we will explain why we might want to relax some of the
assumptions described in \cref{sec:stationarity} forbidding \(t\)-dependence.)

As previously mentioned in \cref{sec:conhelsings}, the degenerate limit
\(\varepsilon\to 0\) of a closed circle \(\mathscr{C}_{\varepsilon}\), at fixed
\(\rho,\underline{\rho}=\varepsilon\) and \(y^i\), can retain a finite length,
because the metric registers the path \(\mathscr{C}_\varepsilon\) as moving in a
direction along \(\gamma\) (even though, topologically, it is not). Since this
is a spacelike length even in the current Lorentzian context, there is a sort of
ergo region near \(\gamma\), where \(\partial_t\) is spacelike.
(\Cref{fig:nolightcones} may be misleading in this respect, because
\(\partial_t\) has a finite projection \(N^i\sim v^i\) onto the directions
orthogonal to the page.)

The holonomic singularity of the Maxwell field \(A\) is again characterized by
the leading behaviour of the electric potential near \(\gamma\) as in
\cref{eq:holsing},
\begin{align}
  [(\partial_t)^a - N^a]\, A_a
  &= \mu + \order{\rho^2}
    \;.
    \label{eq:lorholsing}
\end{align}
The holonomic singularity strength \(\mu\), possibly a function of \(t\) and/or
\(y^i\), again enters into the Wilson loop \(\int_{\mathscr{C}_\varepsilon} A\)
on a small circle \(\mathscr{C}_\varepsilon\) now parametrized by \(t\).

Let us now consider the action for such singular configurations. Because the
causal structure breaks down at \(\gamma\), the strategy used in
\cref{sec:curvature,sec:actionprop} to derive the action by smoothing out the
singularity would now likely involve complex regulated configurations. Rather
than deriving the Lorentzian action from scratch in this way, we will instead
simply analytically continue the Euclidean answer \labelcref{eq:action}.

For clarity, let us state some additional conventions for Wick rotation. We will
adopt the conventional relations between Euclidean and Lorentzian Lagrangian
densities and actions,
\begin{align}
  L_{\mathrm{E}}
  &= -i\, L_{\mathrm{L}}
    \;,
  &
    I_{\mathrm{E}}
  &= -i\, I_{\mathrm{L}}
    \;,
    \label{eq:lagactwick}
\end{align}
where the factor of \(i\) comes from the continuation \labelcref{eq:wickrot} of
the volume form,
\begin{align}
    \epsilon_{\mathrm{E}}
  &=
    N\dd{\tau} \wedge \dd{\rho} \wedge \tensor[^{(D-2)}]{\epsilon}{}
    =
    i\,N\dd{t} \wedge \dd{\rho} \wedge \tensor[^{(D-2)}]{\epsilon}{}
    = i\, \epsilon_{\mathrm{L}}
    \;.
\end{align}
Also because of this Wick rotation of the volume form,
\begin{align}
  *_{\mathrm{E}}
  &= i *_{\mathrm{L}}
    \;.
    \label{eq:hodgewick}
\end{align}
Due to the signs in \cref{eq:wickrot}, the Euclidean and Lorentzian stress
tensors, \eg{} the Brown-York stress tensors on surfaces of constant \(\rho\),
are related by\footnote{To understand equations such as \cref{eq:stresswick}, it
  is important to be consistent with the tensor indices. For example,
  \((T_{\mathrm{EBY}})^{\tau\tau}=-(T_{\mathrm{LBY}})^{\tau\tau}=(T_{\mathrm{LBY}})^{tt}\).}
\begin{align}
  (T_{\mathrm{EBY}})^{ab}
  &= -(T_{\mathrm{LBY}})^{ab}
    \;,
    \label{eq:stresswick}
\end{align}
so that, in Lorentzian signature, an extra sign appears in
\cref{eq:stressby}\footnote{Recall our convention for orientations as described
  in \cref{foot:stresssign}. This carries over to our Lorentzian setting as
  described below \cref{eq:lorlaghol}.}
\begin{align}
  (T_{\mathrm{BY}})^{ab}
  &= \frac{1}{8\pi G_{\mathrm{N}}} \left[
    (K_u)^{ab} - g^{ab} \tensor{(K_u)}{^c_c}
    \right]
  = \frac{i}{8\pi G_{\mathrm{N}}} \left[
    (K_{\underline{u}})^{ab} - g^{ab} \tensor{(K_{\underline{u}})}{^c_c}
    \right]
    \;.
    \label{eq:lorstressby}
\end{align}
Due to the Wick rotation \labelcref{eq:wicknormallapse}, this is imaginary for
spacelike surfaces of constant \(\underline{\rho}\) as indicated in the last
expression. However, the momentum density on surfaces of constant \((\rho,t)\)
or \((\underline{\rho},t)\),
\begin{align}
  (p_{\mathrm{BY}})^i
  &= \tensor[^{(D-2)}]{\epsilon}{} \,
    (T_{\mathrm{BY}})^{ab} n_a \tensor{h}{_b^i}
    =
    \frac{1}{8\pi G_{\mathrm{N}}}\,
    \tensor[^{(D-2)}]{\epsilon}{}\,
    \tensor{(K_u)}{^{ab}}
    n_a \tensor{h}{_b^i}
  \\
  &= \tensor[^{(D-2)}]{\epsilon}{} \,
    (T_{\mathrm{BY}})^{ab}\,i\,\underline{n}_a \tensor{h}{_b^i}
    =
    -\frac{1}{8\pi G_{\mathrm{N}}}\,
    \tensor[^{(D-2)}]{\epsilon}{}\,
    \tensor{(K_{\underline{u}})}{^{ab}}
    \underline{n}_a \tensor{h}{_b^i}
    \;,
\end{align}
is always real. From \cref{eq:wicknormallapseshift,eq:stresswick}, we see that
the relation between the Euclidean momentum density \labelcref{eq:pby2} and its
Lorentzian counterpart (to which we implicitly refer above and elsewhere in this
section) is
\begin{align}
  (p_{\mathrm{EBY}})^i
  &= -i\, (p_{\mathrm{LBY}})^i
    \;.
    \label{eq:pbywick}
\end{align}

With these conventions, the Lorentzian action is given by
\begin{align}
  I
  &= \int_{\mathscr{M}\setminus\mathscr{N}_\varepsilon} (L_{\mathrm{EH}} + L_{\mathrm{M}})
    + \int_{\partial\mathscr{M}} L_{\partial \mathscr{M}}
    - i\, \frac{\mathrm{Area}(\gamma)}{4 G_{\mathrm{N}}}
    + \int_{-\partial\mathscr{N}_\varepsilon}(
    L_{\mathrm{GH}} + L_{\mathrm{hel}} + L_{\mathrm{hol}}
    )
    \;,
    \label{eq:loraction}
\end{align}
where now
\begin{align}
  L_{\mathrm{EH}}
  &= \frac{1}{16\pi G_{\mathrm{N}}}
    (R - 2\Lambda)\, \epsilon
    \;,
  &
    L_{\mathrm{M}}
  &= -\frac{1}{2 g_{\mathrm{M}}^2} F\wedge * F
    \;,
\end{align}
and
\begin{align}
  L_{\mathrm{GH}}
  &= \frac{1}{8\pi G_{\mathrm{N}}}\,
    u^A \epsilon_A\,
    \, g^{ab} \tensor{(K_{u})}{_{ab}}
  = -\frac{1}{8\pi G_{\mathrm{N}}}\,
    \underline{u}^A \epsilon_A\,
    \, g^{ab} \tensor{(K_{\underline{u}})}{_{ab}}
    \;,
    \label{eq:lorlaggh}
  \\
  L_{\mathrm{hel}}
  &= \frac{1}{8\pi G_\mathrm{N}}
    u^A \epsilon_A\,
    \, (\dd{t})_a\, g^{ab} \, N^i\,
    (K_{u})_{b i}
  = -\frac{1}{8\pi G_\mathrm{N}}
    \underline{u}^A \epsilon_A\,
    \, (\dd{t})_a\, g^{ab} \, N^i\,
    (K_{\underline{u}})_{b i}
    \;,
    \label{eq:lorlaghel}
  \\
  &= -\dd{t} \wedge (p_{\mathrm{BY}})_i\, N^i
  \\
  L_{\mathrm{hol}}
  &= \frac{1}{g_{\mathrm{M}}^2} A \wedge * F
    \;.
    \label{eq:lorlaghol}
\end{align}
As in the Euclidean case, \(\mathscr{N}_\varepsilon\) is a neighbourhood of
\(\gamma\) such that \(\partial\mathscr{N}_\varepsilon\) is a surface of
constant proper radius \(\rho\) or \(\underline{\rho}\) of order \(\varepsilon\)
from \(\gamma\) --- see \cref{fig:nolightcones}. Note that the \emph{vectors}
\(u^A\) and \(\underline{u}^A\) point respectively outward from and into
\(\mathscr{N}_\varepsilon\), \ie{} into and outward from
\(\mathscr{M}\setminus\mathscr{N}_\varepsilon\). The orientation of
\(\partial\mathscr{N}_\varepsilon\) is therefore given by \(u^A \epsilon_A\) and
\(-\underline{u}^A \epsilon_A\); the opposite orientation applies to
\(\partial(\mathscr{M}\setminus\mathscr{N}_\varepsilon)=-\partial\mathscr{N}_\varepsilon\)
appearing in \cref{eq:loraction}.

More generally, as described in \cref{sec:shapeindep}, there are certain cases
where we can alternatively work with other choices of the excised neighbourhood
\(\mathscr{N}_\varepsilon\) and cutoff surface
\(\partial\tilde{\mathscr{N}}_\varepsilon\). In particular, if the singularity
is nonhelical, \ie{} \(v^i=0\), then there is full freedom to choose the profile
of \(\mathscr{N}_\varepsilon\) to be given by \(\rho=\tilde{\rho}\,e^{f(x)}\) or
\(\underline{\rho}=\underline{\tilde{\rho}}\,e^{f(t)}\) for fixed \(\tilde{\rho}\) and
\(\underline{\tilde{\rho}}\) of order \(\varepsilon\) and any function \(f\) of
\(x^a=(t,y^i)\). If the singularity is helical, then, around each connected
component of \(\gamma\), we can still choose an arbitrary function
\(f(t)\) but only of \(t\), provided \(\lim_{\rho,\underline{\rho}\to0}h_{ij}\),
\(\kappa\), and \(v^i\) are \(t\)-independent or provided
\(\lim_{\rho,\underline{\rho}\to0}h_{ij}\) is \(t\)-independent and \(\kappa\)
is \(y^i\)-independent. The cutoff surface terms in the action
\labelcref{eq:action} is then given by
\(\int_{-\partial\tilde{\mathscr{N}}_\varepsilon}(L_{\mathrm{GH}}+L_{\mathrm{hel}}+L_{\mathrm{hol}})\)
where the Gibbons-Hawking and helical Lagrangian densities are
\begin{align}
  \tilde{L}_{\mathrm{GH}}
  &= \frac{1}{8\pi G_{\mathrm{N}}}
    \tilde{u}^A \epsilon_A\,
    \, \tilde{g}^{ab} \tensor{(\tilde{K}_{\tilde{u}})}{_{ab}}
    = - \frac{1}{8\pi G_{\mathrm{N}}}
    \underline{\tilde{u}}^A \epsilon_A\,
    \, \tilde{g}^{ab} \tensor{(\tilde{K}_{\underline{\tilde{u}}})}{_{ab}}
    \;,
    \label{eq:lorlagghwiggly}
  \\
  \tilde{L}_{\mathrm{hel}}
  &= \frac{1}{8\pi G_\mathrm{N}}
    \tilde{u}^A \epsilon_A\,
    \, (\dd{t})_a\, \tilde{g}^{ab} \, \tilde{N}^i\,
    (\tilde{K}_{\tilde{u}})_{b i}
    = - \frac{1}{8\pi G_\mathrm{N}}
    \underline{\tilde{u}}^A \epsilon_A\,
    \, (\dd{t})_a\, \tilde{g}^{ab} \, \tilde{N}^i\,
    (\tilde{K}_{\underline{\tilde{u}}})_{b i}
    \;,
    \label{eq:lorlaghelwiggly}
\end{align}
while \(L_{\mathrm{hol}}\) remains the same as in \cref{eq:lorlaghol}. As in
\cref{sec:shapeindep}, the induced metric \(\tilde{g}_{ab}\), its inverse
\(\tilde{g}^{ab}\), and the shift vector \(\tilde{N}^i\) in \cref{eq:lorlaghel}
are understood as being defined on the chosen surface
\(\partial\tilde{\mathscr{N}}_\varepsilon\). We have also written the spacelike
and timelike unit normals of \(\partial\tilde{\mathscr{N}}_\varepsilon\)
respectively as \(\tilde{u}_A=\underline{\tilde{u}}_A\) and
\(\underline{\tilde{u}}_A\).

\subsubsection{Lightcones and resulting complications}
\label{sec:lightcones}

\begin{figure}
  \centering
  \begin{subfigure}[t]{0.475\textwidth}
    \centering
    \includegraphics{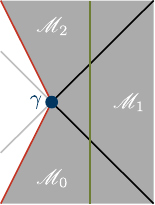}
    \caption{Spacetime \(\mathscr{M}\) (shaded) constructed by gluing together
      wedges \(\mathscr{M}_i\). The red surfaces are identified.}
    \label{fig:gluing}
  \end{subfigure}
  \hfill
  \begin{subfigure}[t]{0.475\textwidth}
    \centering
    \includegraphics{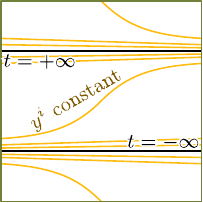}
    \caption{A timelike surface (green in previous panel) passing through
      lightcones near \(\gamma\). This illustration depicts angles measured by
      the metric at least approximately. Surfaces of constant \(t\) are exactly
      horizontal. A surface of constant \(y^i\) is drawn for the case of a
      constant nontrivial helical shift \(v^i\) and periodically identified
      \(y^i\).}
    \label{fig:dragging}
  \end{subfigure}
  \caption{A spacetime with lightcones emanating from a codimension-two singular
    surface \(\gamma\).}
  \label{fig:lightcones}
\end{figure}

Let us now consider more general configurations that contain lightcones for the
singular codimension-two surface \(\gamma\). Below, we will describe a procedure
for constructing such configurations in a neighbourhood of each connected component
of \(\gamma\). We then have in mind that the full spacetime \(\mathscr{M}\) is
an extension of these neighbourhoods which is smooth with the exception of
\(\gamma\) (and possibly its lightcones, as we will later describe).

In a neighbourhood of a given connected component of \(\gamma\), there can be
any even number \(\mathcal{N}\ge 0\) of lightcone components meeting at
\(\gamma\). The case of \(\mathcal{N}=0\) has already been treated in
\cref{sec:nolightcones}. Configurations with \(\mathcal{N}>0\) in this
neighbourhood can be constructed by a gluing procedure similar to that described
in ref.~\cite{Marolf:2022ybi}. However, the pieces of spacetime that we sew
together are \((\mathcal{N}+1)\)-many wedges \(\mathscr{M}_i\), each of which is
the decompactified limit of a spacetime described in \cref{sec:nolightcones}.
That is, in each \(\mathscr{M}_i\), we relax the periodic identification of
\(t\). Instead, in \(\mathscr{M}_0\), the range of \(t\) is taken to be a
semi-infinite line from a finite value \(t_0\) to \(+\infty\); in each
\(\mathscr{M}_{0<i<\mathcal{N}}\), \(t\) takes values on an infinite line;
finally, in \(\mathscr{M}_{\mathcal{N}}\), \(t\) again lies on a semi-infinite
line from \(-\infty\) to a finite value \(t_{\mathcal{N}}\).\footnote{As we have
  implicitly assumed before, \(\kappa\) is required to be finite, so infinite
  proper boost angles are only reached at infinite \(t\).} These wedges are then
sewn together along the lightcones where \(\Re t = \pm\infty\); as we will later
see, at least in certain situations, it is helpful to view the semi-infinite or
infinite lines of \(t\) for different wedges to have imaginary offsets relative
to each other. Finally, we identify the surface \(t=t_0\) in \(\mathscr{M}_0\)
with \(t=t_{\mathcal{N}}\) in \(\mathscr{M}_{\mathcal{N}}\). In this procedure,
\(\gamma\) must be spacelike and timelike separated from nearby points
respectively in \(\mathscr{M}_{\text{even }i}\) and \(\mathscr{M}_{\text{odd
  }i}\), or vice versa. This in particular implies that \(\mathcal{N}\) must be
even. In the resulting spacetime within a neighbourhood of \(\gamma\), the
number of connected lightcone components meeting at \(\gamma\) is
\(\mathcal{N}\). A case with \(\mathcal{N}=2\) is illustrated in
\cref{fig:gluing}.

The possible outputs of the above procedure describe the configurations we
consider only in a neighbourhood of a each component of \(\gamma\). As already
mentioned, we envision the configuration on the full spacetime \(\mathscr{M}\)
to be an extension of these neighbourhoods. In principle, we see no reason to
require \(\mathcal{N}\) to be identical for all connected components of
\(\gamma\).\footnote{As constructed through our procedure, the spacetimes we
  consider are real and have Lorentzian signature, with the possible exception
  of a codimension-two surface \(\gamma\) where the causal structure can break
  down. However, if one further allows complex geometries, then there also does
  not seem to be a reason to forbid \(\mathcal{N}\) from varying within a
  connected component of \(\gamma\).} We also see no reason to rule out
nontrivial connections and identifications between the wedges associated to
various connected components of \(\gamma\) through the full spacetime
\(\mathscr{M}\).

\paragraph{Lightcone singularities.}

As previously hinted, the treatment of such configurations can be subtle due to
complications that arise from the lightcones emanating from \(\gamma\). One
issue is whether these spacetimes are singular on the lightcones. For example,
the presence of a helical shift \(v^i\) can pose a threat at \(\Re t =
\pm\infty\) if the configuration, \eg{} the induced metric \(h_{ij}\) on
surfaces of constant \(\rho\) or \(\underline{\rho}\) and \(t\), near \(\gamma\)
is not invariant under the shift, \eg{} \(\mathcal{L}_{v}h_{ij}\ne0\). To see
the problem, suppose the metric \(\lim_{\rho,\underline{\rho}\to0}h_{ij}\) on
\(\gamma\) is not shift-invariant due to some inhomogeneity near some value of
\(y^i\). On nearby surfaces of fixed \(\rho\) or \(\underline{\rho}\) and
\(t\), the induced metric \(h_{ij}\) should have a similar inhomogeneity
whose motion follows a worldline of fixed \(y^i\). Over time \(\delta t\), the
metric will register this worldline as moving by some amount \(v^i \delta t\) in
the directions \(y^i\) along \(\gamma\). Near a lightcone, \(t\) grows
arbitrarily large, so the worldline runs infinitely rapidly along the \(y^i\)
directions if \(v^i\) is not turned off --- see \cref{fig:dragging}. Due to this
effect, the metric near the lightcone can vary arbitrarily rapidly. To avoid
such singularities, one might therefore want to consider \(t\)-dependent helical
shifts \(v^i\) that turn off sufficiently quickly as \(\Re t\to\pm\infty\). On
the other hand, one might allow these lightcone singularities, in which case one
might expect a nontrivial contribution to the action from the lightcones.

A somewhat similar effect occurs for the Maxwell field. Consider a path normal
to surfaces of constant \(t\) passing near \(\gamma\), \eg{} a vertical line in
\cref{fig:dragging}. Close to \(\gamma\), parallel transport on the principal
bundle along this path includes a motion \(\int_t^{t+\delta t}A=\mu\,\delta t\)
in the fibre direction for every increment \(\delta t\) of boost time. However,
this does \emph{not} give rise to a physical singularity on the lightcone. One
way to see this is simply by changing to a gauge in which the Wilson line \(\int A\)
between different points along the path remains finite even as one crosses the
lightcone. Perhaps more conceptually, the fibre direction is homogeneous just
like the gauge group. So, in analogy to the previous paragraph, there is no
``inhomogeneity'' that is dragged infinitely rapidly in the fibre direction as
we cross the lightcone.

\paragraph{The action: a first pass.}

Let us now turn to a discussion of the action for a configuration constructed
from the gluing procedure described above. We expect the action for such
configurations to again be given by \cref{eq:loraction}, where now
\(\mathscr{M}\) is the full spacetime resulting from the gluing and extension
procedure. A rough\footnote{However, we will argue later that there are ``pole''
  contributions to the cutoff surface \(\partial\mathscr{N}_\varepsilon\) terms
  of the action, which can be viewed as a consequence of the gluing between the
  wedges.} argument for this is that the action near each connected component of
\(\gamma\) should be a sum of the actions for the wedges \(\mathscr{M}_i\), with
the exception of the area term. We keep only one copy of the area term even as
more wedges are sewn together, because, in a sense, it represents the \(2\pi\)
opening angle around \(\gamma\) of a \emph{smooth} configuration. This notion is
made precise in \cref{eq:concontact1} when considering configurations where the
would-be singularity on \(\gamma\) has been regulated. Several subtleties remain
however, all related to the presence of lightcones.

Firstly, if the lightcones emanating from \(\gamma\) are singular as described
previously, then there may be contributions to the action associated to these
lightcone singularities. Secondly, if we take as before the cutoff surface
\(\partial\mathscr{N}_\varepsilon\) in the action \cref{eq:loraction} to be a
surface of constant proper radius \(\rho\) and \(\underline{\rho}\), then this
cutoff surface will run far away from \(\gamma\) in terms of affine distance
along the lightcones. In particular, the lightcones are part of the excised
neighbourhood \(\mathscr{N}_\varepsilon\) of \(\gamma\).

Rather than a bug, we suggest viewing this as a possible feature of the action
which accounts for possible contributions from lightcone singularities. This
suggestion is motivated by the fact that the Euclidean action, which
analytically continues to our Lorentzian action \labelcref{eq:loraction}, was
derived in \cref{sec:curvature,sec:action} by considering the regulated
configurations described in \cref{sec:regsing}. Our procedure for constructing
these regulated configuration involves interpolating various field components,
\eg{} the shift \(N^i\), from values matching the unregulated singular
configuration on \(\partial\mathscr{N}_\varepsilon\) to a behaviour at
\(\rho=0\) consistent with a smooth configuration. Now, in our current
Lorentzian context, the locus of \(\rho=0\) includes not only \(\gamma\) but
also the lightcones. Thus, it is natural to guess that the area and cutoff
surface \(\partial\mathscr{N}_\varepsilon\) terms in the action
\cref{eq:loraction}, roughly speaking, account for the action
\(\int_{\mathscr{N}_\varepsilon}(L_{\mathrm{EH}}+L_{\mathrm{M}})_{\mathrm{reg}(\varepsilon)}\)
of a configuration that is regulated to be smooth on the locus of \(\rho=0\),
which includes both \(\gamma\) and the lightcones in the singular configuration.
(The possibly singular causal structure at \(\gamma\) means the full regulation
procedure likely involves complex geometries; however, to regulate just the
helical shift which is responsible for the type of lightcone singularity
described in reference to \cref{fig:dragging}, a real interpolation of the shift
\(N^i\) as a function of \(\rho\) would seem to suffice.) We leave a more
careful analysis of this intuition for future work.

Setting aside this issue, in the remainder of this section, we would like to
highlight the fact that the cutoff surface terms in the action receive an
imaginary ``pole'' contribution for every lightcone component emanating from
\(\gamma\). This is true even if the lightcones themselves are smooth.

\paragraph{Crossing a lightcone.}

\begin{figure}
  \centering
  \includegraphics{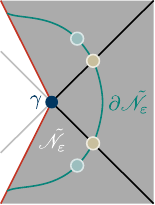}
  \caption{A neighbourhood \(\tilde{\mathscr{N}}_\varepsilon\) of \(\gamma\) in
    the spacetime constructed by gluing in \cref{fig:gluing}. The cutoff surface
    \(\partial\tilde{\mathscr{N}}_\varepsilon\) remains everywhere affinely close to
    \(\gamma\). Highlighted by dots on \(\partial\tilde{\mathscr{N}}_\varepsilon\) are
    surfaces that play important roles in the cutoff surface terms of the action: the
    light blue dots are surfaces where
    \(\partial\tilde{\mathscr{N}}_\varepsilon\) becomes null; the beige
    dots are the intersection of \(\partial\tilde{\mathscr{N}}_\varepsilon\)
    with lightcones emanating from \(\gamma\), where \(\Re t=\pm\infty\).}
  \label{fig:neighbourhood}
\end{figure}

To make progress, we will find it helpful to restrict our attention to certain
configurations in which we have the freedom, as described around
\cref{eq:lorlagghwiggly,eq:lorlaghelwiggly}, to choose a cutoff surface
\(\partial\tilde{\mathscr{N}}_\varepsilon\) which is everywhere affinely close
to \(\gamma\) --- see \cref{fig:neighbourhood}. We will now argue that, given
certain simplifying assumptions, such a choice is possible; in the process, we
will also better understand how the boost time contours of neighbouring wedges
are joined together.

From here on, we will firstly assume that the lightcones are in fact smooth,
\eg{} because of symmetries or because the shift \(v^i\) turns off sufficiently
quickly at \(\Re t\to \pm \infty\) as discussed previously. Secondly, we will require
\(\kappa\) for each connected component of \(\gamma\) to only be a function of
\(t\) and not \(y^i\). Finally, we will assume that that
\(N/\rho=\underline{N}/\underline{\rho}\) quickly approaches \(\kappa\) as we
approach any lightcone (near but finitely separated from \(\gamma\)).\footnote{It is perhaps sufficient to merely require
  \(N/\rho=\underline{N}/\underline{\rho}\) to quickly become independent of
  \(y^i\) as we approach the lightcone. Then, one can consider the curvature of
  the effective two-dimensional metric \cref{eq:effmetric},
  \begin{align}
    \tensor[^{(2)}]{R}{}
    &= - \frac{2}{N}\, \partial_\rho^2 N
      = \frac{2}{\underline{N}}\, \partial_{\underline{\rho}}^2 \underline{N}
      \;.
      \label{eq:effcurv}
  \end{align}
  As we approach a smooth lightcone, given that \(N\to 0\) but the above should
  remain finite, we then expect \(N-\kappa\,\rho\) to vanishes no slower than
  \(\rho^2 N\), \ie{} \(\rho^3\). In particular, the ``subleading'' terms in
  \cref{eq:lorlapseshift} are truly smaller than the \(\kappa\) terms as we
  approach a lightcone, even though one might have initially feared that the
  ``subleading'' terms could be enhanced by \(\Re t\to\pm\infty\).
  \label{foot:effcurv}}

As argued in \cref{sec:shapeindep} and reviewed around
\cref{eq:lorlagghwiggly,eq:lorlaghelwiggly}, the profile of each connected
component of the cutoff surface \(\partial\tilde{\mathscr{N}}_\varepsilon\) can
be taken to be given by \(\rho=\tilde{\rho}\,e^{f(t)}\) or
\(\underline{\rho}=\underline{\tilde{\rho}}\,e^{f(t)}\) for fixed
\(\tilde{\rho}\) and \(\underline{\tilde{\rho}}\) of order \(\varepsilon\) and
an arbitrary function \(f(t)\). It is helpful to view this surface as a
worldline moving in an effective two-dimensional geometry,
\begin{align}
  \dd{s}_\perp^2
  &= \dd{\rho}^2 - N^2 \dd{\tau}^2
  = -\dd{\underline{\rho}}^2 + \underline{N}^2 \dd{\tau}^2
    \;.
    \label{eq:effmetric}
\end{align}
In particular, the lapse \(\tilde{N}\) on
\(\partial\tilde{\mathscr{N}}_\varepsilon\) is the einbein of a worldline in
this effective geometry. Near any given lightcone, by assumption,
\(N/\rho=\underline{N}/\underline{\rho}\) and thus the above effective metric
quickly become independent of \(y^i\), so the problem truly becomes
two-dimensional.

To cross lightcones affinely near (but finitely separated from) a given
connected component of \(\gamma\), it is helpful to consider Kruskal coordinates
\begin{align}
  z
  &= \rho\, e^{i\,\vartheta}
    \;,
  &
    \bar{z}
  &= \rho\, e^{-i\,\vartheta}
    \;,
    \label{eq:kruskal1}
\end{align}
where the proper Euclidean angle \(\vartheta\), or equivalently the proper hyperbolic angle
\(-i\,\vartheta\), is defined (up to an integration constant) by
\begin{align}
  \dd{\vartheta}
  &= \kappa \dd{\tau}
    = i\,\kappa \dd{t}
    \;.
\end{align}
Using these coordinates, \cref{eq:effmetric} can be expressed as
\begin{align}
  \dd{s}_\perp^2
  &= \dd{z}\dd{\bar{z}}
    + \left[
    (\kappa\, \rho)^2
    - N^2
    \right] \dd{t}^2
    \;.
\end{align}
The first term is manifestly smooth, while \((\kappa\,\rho)^2-N^2\)
vanishes quickly as we approach the lightcone such that the second term is also
smooth.\footnote{From \cref{foot:effcurv}, we expect
  \((\kappa\,\rho)^2-N^2\) to vanish no slower than \(\rho^4\) as we
  approach the lightcone. Meanwhile,
  \(\kappa^2\,\rho^2\dd{t}=z\dd{\bar{z}}-\bar{z}\dd{z}\).}

To cross a lightcone at \(\Re t=+\infty\) or \(\Re t=-\infty\) respectively, we
therefore want to take \(z\to0\) while keeping \(\bar{z}\) finite or take
\(\bar{z}\to0\) while keeping \(z\) finite. This can be achieved for a connected
component of \(\partial\tilde{\mathscr{N}}_\varepsilon\) by choosing its time
profile \(f(t)\) to behave like \(\pm i\,\vartheta = \mp \int^t\dd{t'}\kappa(t')\) asymptotically
at \(\Re t=\pm\infty\).

\begin{figure}
  \centering
  \includegraphics{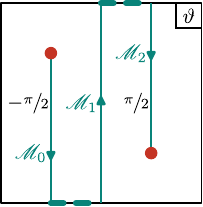}
  \caption{The contour traced out by the proper Euclidean angle \(\vartheta\) or
    the proper hyperbolic angle \(-i\,\vartheta=\int^t\dd{t'}\kappa(t')\), for
    the spacetime in \cref{fig:gluing}. The red points correspond to the
    identified red surfaces in \cref{fig:gluing}. The dashed lines indicate that
    the contour is completed by segments taken to infinity.}
  \label{fig:timecontour}
\end{figure}

Before proceeding, let us remark that the Kruskal coordinates
\labelcref{eq:kruskal1},
\begin{align}
  z
  &= \rho\, e^{i\,\vartheta}
    = i\underline{\rho}\, e^{i\,\vartheta}
    \;,
  &
    \bar{z}
  &= \rho\, e^{-i\,\vartheta}
    = i\underline{\rho}\, e^{-i\,\vartheta}
    \;,
    \label{eq:kruskal2}
\end{align}
also tell us how to relate the contours of \(t\) or \(\vartheta\) in
neighbouring wedges. In particular, \(\vartheta\) jumps by \(\pi/2\) when moving
from spacelike to timelike separation from \(\gamma\) across a lightcone at
\(\Re t=+\infty\) and when moving from timelike to spacelike separation from
\(\gamma\) across a lightcone at \(\Re t=-\infty\). The contour traced out by
\(\vartheta\) when going around near a connected component of \(\gamma\) in the
full spacetime \(\mathscr{M}\) thus consists of semi-infinite and infinite lines
for the wedges \(\mathscr{M}_i\), joined by segments at \(\Im
\vartheta=\pm\infty\), \ie{} \(\Re t=\pm\infty\). This is illustrated in
\cref{fig:timecontour} for the spacetime of \cref{fig:gluing}.

\paragraph{The action: surface terms revisited.}

We have argued above that, for configurations satisfying certain simplifying
assumptions\footnote{We will continue to assume that the geometry to be smooth
  away from \(\gamma\), even on the lightcones, and \(\kappa\) is a function
  only of \(t\) near each connected component of \(\gamma\).}, the cutoff
surface \(\partial\tilde{\mathscr{N}}_\varepsilon\) can be chosen to be affinely
close to \(\gamma\), as illustrated in \cref{fig:neighbourhood}. With this
picture in mind, let us revisit the terms
\(\int_{-\partial\tilde{\mathscr{N}}_\varepsilon}(L_{\mathrm{GH}}+L_{\mathrm{hel}}+L_{\mathrm{hol}})\)
in the action integrated over this cutoff surface.

A first comment is that the Gibbons-Hawking Lagrangian density
\labelcref{eq:lorlagghwiggly} pulled back to
\(\partial\tilde{\mathscr{N}}_\varepsilon\) has a pole wherever
\(\partial\tilde{\mathscr{N}}_\varepsilon\) becomes null --- see
\cref{fig:neighbourhood}. While the volume form on
\(-\partial\tilde{\mathscr{N}}_\varepsilon\) vanishes here, the trace of the
extrinsic curvature \(\tilde{g}^{ab} \tensor{(\tilde{K}_{\tilde{u}})}{_{ab}}\)
diverges even more strongly. This can be understood as the result of the
extrinsic curvature \((\tilde{K}_{\tilde{u}})_{ab}\) being defined in terms of a
unit-normalized normal \(\tilde{u}^A\). By rewriting the extrinsic curvature and
induced volume form on \(-\partial\tilde{\mathscr{N}}_\varepsilon\) in terms of
a smoothly varying but non-normalized normal, it can be shown that the residue
of the pole is proportional to the volume form on the codimension-one section of
\(\partial\tilde{\mathscr{N}}_\varepsilon\) where
\(\partial\tilde{\mathscr{N}}_\varepsilon\) is null. A prescription for going
around the pole can then be obtained by rotating the contour of Lorentzian time
slightly towards the negative imaginary direction.\footnote{Very roughly
  speaking, if matter fields are placed on the spacetime, then their evolution
  with respect to a Lorentzian time \(u\) is generated by the contour-ordered
  exponential \(\exp\left(-i\int_{\mathfrak{T}}\dd{u}
    H^{\mathrm{QFT}}(u)\right)\) where \(\hat{H}^{\mathrm{QFT}}\) is the matter
  Hamiltonian and \(\mathfrak{T}\) is the time contour. Whereas ordinary
  Lorentzian time evolution involves \(\mathfrak{T}\) along the real line, our
  prescription slightly rotates \(\mathfrak{T}\) so that \(\dd{u}\) along it is
  slightly negative imaginary. This effectively adds a mild exponential damping
  to high energy fluctuations. In contrast, the opposite prescription gives a
  dangerous exponential enhancement of such fluctuations.} This analysis has
been carried out previously, \eg{} as described in
refs.~\cite{Neiman:2013ap,Colin-Ellerin:2020mva,Marolf:2022ybi}, and we will
refrain from repeating it.

The upshot is that the Gibbons-Hawking action should be understood as a
principal value (PV) integral plus a (half) residue contribution from the pole:
\begin{align}
  \int_{-\partial\tilde{\mathscr{N}}_\varepsilon}
  \tilde{L}_{\mathrm{GH}}
  &\sim \mathrm{PV}
    \int_{-\partial\tilde{\mathscr{N}}_\varepsilon}
    \tilde{L}_{\mathrm{GH}}
    + \frac{i}{16 G_{\mathrm{N}}}
    \sum_{\text{connected }\gamma_\ell \subset \gamma}
    \mathcal{N}_\ell\, \mathrm{Area}(\gamma_\ell)
    \;,
    \label{eq:ghpole}
\end{align}
where the sum is over connected components \(\gamma_\ell\) of \(\gamma\). Above,
in the \(\varepsilon\to 0\) limit intended in the relation \(\sim\), we have
equated the area element along sections of \(\partial\tilde{\mathscr{N}}_\varepsilon\)
where \(\partial\tilde{\mathscr{N}}_\varepsilon\) is null to the area element along
\(\gamma\). From the discussion in \cref{sec:singdisambig},\footnote{The function
  \(f\) in \cref{sec:singdisambig} should not be confused with the function
  \(f\) in \cref{sec:shapeindep} and mentioned below.} we expect this to be
valid in the presence of helical singularities if (each connected component of)
the section of \(\partial\tilde{\mathscr{N}}_\varepsilon\) tends to a surface of
constant \(t\). This is indeed true in the current case where (near each
connected component of \(\gamma\)) \(\kappa(t)\) depends only on \(t\) and the
profile of \(\partial\tilde{\mathscr{N}}_\varepsilon\) is specified by a function
\(f(t)\) also only of \(t\) --- see \cref{eq:effmetric}.

The aforementioned pole on sections where
\(\partial\tilde{\mathscr{N}}_\varepsilon\) becomes null appears only when
considering the component of the extrinsic curvature normal to said section.
Other components, \eg{} the ones appearing in the helical term
\labelcref{eq:lorlaghelwiggly}, diverge, if at all, no faster than the induced
volume form vanishes on sections where
\(\partial\tilde{\mathscr{N}}_\varepsilon\) becomes null.

However, there are somewhat similar poles in the helical and holonomic terms
\labelcref{eq:lorlaghelwiggly,eq:lorlaghol}, located on the intersection of
\(\partial\tilde{\mathscr{N}}_\varepsilon\) and the lightcones --- see
\cref{fig:neighbourhood}. With \(\Re t\to\pm\infty\) here, the helical and
holonomic Lagrangian densities diverge because of the appearance of \(\dd{t}\)
explicit in \cref{eq:lorlaghelwiggly} and implicit in the Maxwell potential \(A\) as
described by \cref{eq:lorholsing}. The proper angle contour illustrated in
\cref{fig:timecontour} provides a prescription for going around these poles.
Specifically, the vertical and horizontal segments of \cref{fig:timecontour}
give the principal value and pole contributions to the integrals of the helical
and holonomic surface terms \labelcref{eq:lorlaghel,eq:lorlaghol} over
\(-\partial\tilde{\mathscr{N}}_\varepsilon\).

In fact, the proper angle contour gives an integration prescription for all the
surface terms on \(\partial\tilde{\mathscr{N}}_\varepsilon\), including
reproducing \cref{eq:ghpole}. Continuing \cref{eq:wigglyall} to Lorentz
signature, we find that the pullback of
\(\tilde{L}_{\mathrm{GH}}+\tilde{L}_{\mathrm{hel}}+L_{\mathrm{hol}}\) to
\(\partial\tilde{\mathscr{N}}_\varepsilon\) in the \(\varepsilon\to 0\) limit is
equivalent, up to a total derivative, to a density on \(S^1\times\gamma\) given
by
\begin{align}
  \begin{split}
    \tilde{\phi}^*_{\varepsilon\to 0}\left(
  \tilde{L}_{\mathrm{GH}}
  +\tilde{L}_{\mathrm{hel}}
  +L_{\mathrm{hol}}
  \right)
    &=
      \dd{t}\wedge\left[
      -\frac{\kappa}{8\pi G_{\mathrm{N}}}\,
      \tensor[^{(D-2)}]{\epsilon}{}
      - v^i\,(p_{\mathrm{BY}})_i
      + \frac{1}{g_{\mathrm{M}}^2} \mu \,* F
      \right]
    \\
  &\phantom{{}={}}+ (\text{total derivative})
  \end{split}
  \\
  \begin{split}
    &= i\dd{\vartheta}\wedge\left[
    \frac{1}{8\pi G_{\mathrm{N}}}\,
     \tensor[^{(D-2)}]{\epsilon}{}
      + \frac{v^i}{\kappa}\, (p_{\mathrm{BY}})_i
      - \frac{1}{g_{\mathrm{M}}^2}
      \frac{\mu}{\kappa}\, * F
      \right]
    \\
  &\phantom{{}={}}+ (\text{total derivative})
    \;.
  \end{split}
\end{align}
The combinations \(v^i/\kappa\) and \(\mu/\kappa\) give the helical and
holonomic shifts per proper hyperbolic angle. The \(\vartheta\) integral over
the \(S^1\) is to be performed over the contour illustrated in
\cref{fig:timecontour}. We see that the horizontal segments of the contour,
where the pullback of \(\dd\vartheta\) is real, gives an imaginary contribution
to the cutoff surface terms in the Lorentzian action:
\begin{multline}
  \Im\int_{-\partial\tilde{\mathscr{N}}_\varepsilon}(
  \tilde{L}_{\mathrm{GH}} + \tilde{L}_{\mathrm{hel}} + L_{\mathrm{hol}}
  )
  \\
  \sim \sum_{\text{connected }\gamma_\ell \subset \gamma}
    \left\{
      \frac{\mathcal{N}_\ell\, \mathrm{Area}(\gamma_\ell)}{16 G_{\mathrm{N}}}
      + \frac{\pi}{2}\sum_{i=1}^{\mathcal{N}_\ell}
    \int_{\gamma_\ell^i}
    \left[
      \frac{v^i}{\kappa} (p_{\mathrm{BY}})_i
      - \frac{1}{g_{\mathrm{M}}^2} \frac{\mu}{\kappa}\, * F
    \right]
    \right\}
    \;,
    \label{eq:lorcutoffimag}
\end{multline}
where \(\gamma_\ell^i\) is the section of \(S^1\times\gamma_\ell\) corresponding
to the \(i\)-th lightcone component emanating from \(\gamma_\ell\). (If the
terms in the square brackets are \(t\)- or \(\vartheta\)-independent, as we
sometimes assumed in our Euclidean calculations, then the sum just gives a
multiplicative factor of \(\mathcal{N}_\ell\) like for the area term.)

In summary, we see that the cutoff surface terms in the action acquire imaginary
contributions \labelcref{eq:lorcutoffimag}. To deduce \cref{eq:lorcutoffimag},
we focused on configurations which are smooth on the lightcones of \(\gamma\)
and satisfy some simplifying assumptions which allowed us to consider a cutoff
surface \(\partial\tilde{\mathscr{N}}_\varepsilon\) which is everywhere affinely
close to \(\gamma\), as illustrated in \cref{fig:neighbourhood}. Because this
cutoff surface itself crosses over the lightcones, it is natural to include the
horizontal real (imaginary) segments of the (hyperbolic) angle contour
\cref{fig:timecontour} which give rise to the pole contributions.

However, we expect that, even when working with a cutoff surface
\(\partial\mathscr{N}_\varepsilon\) which asymptotes to but naively never
crosses the lightcones, the correct prescription for evaluating the Lorentzian
action should nonetheless implicitly include similar pole contributions. After
all, in the special cases where we were allowed to choose cutoff surfaces
\(\partial\tilde{\mathscr{N}}_\varepsilon\) which do cross lightcones, we ought
to be able to get the same answer by treating the original cutoff surface
\(\partial\mathscr{N}_\varepsilon\) with a consistent prescription. Another
reason is that, the Lorentzian action for a smooth configuration with
\(\mathcal{N}=4\) ought to be purely real with no special contributions near
\(\gamma\). In order to cancel the explicitly imaginary area term in the action
\labelcref{eq:loraction}, the cutoff surface terms on
\(\partial\mathscr{N}_\varepsilon\) must also have imaginary contributions. We
will leave to future work a more thorough analysis of imaginary contributions to
the Lorentzian action.

\subsection{Thermal partition function}
\label{sec:lorpartfunc}

As described in the introduction in \cref{sec:intro} and briefly mentioned in
\cref{sec:pathintegral}, an obstacle of the Euclidean analysis in
\cref{sec:partfunc} is the conformal factor problem \cite{Gibbons:1978ac}. Given
the sickness of the gravitational path integral contour over purely Euclidean
configurations, this raises the question of what the right integration contour
should be. We will adopt the perspective that the fundamental starting point
should be the Lorentzian gravitational path integral with a contour over real
Lorentzian configurations. Using such a Lorentzian path integral in this
section, we will reevaluate the grand canonical partition function
\(Z(\beta,\Omega,\Phi)\).

As emphasized in ref.~\cite{Marolf:2022ybi}, to recover the expected
thermodynamic contributions from the analogue of Euclidean black holes, it is
important to include configurations in the Lorentzian path integral which are at
least conically singular on codimension-two surface \(\gamma\). We will extend
this analysis to include helical and holonomic of singularities. Thus, we
include in our path integral all Lorentzian configurations which are smooth
apart from \(\gamma\) (and possibly the lightcones of \(\gamma\)), on which we
allow singularities of the kind described in \cref{sec:lorsing}. Including this
broader class of singularities leads to a richer set of constrained saddles, as
we have already seen in \cref{sec:partfunc} in Euclidean signature.
Consequently, we will find a larger set of stability conditions for a given
saddle to contribute to the partition function \(Z(\beta,\Omega,\Phi)\), beyond
the positivity of specific heat found in ref.~\cite{Marolf:2022ybi}. (In
\cref{sec:discussion}, we will provide a Morse theory explanation for why these
additional conditions were not visible from the analysis of
ref.~\cite{Marolf:2022ybi}.)

Let us briefly review our strategy for evaluating \(Z(\beta,\Omega,\Phi)\),
which was already explained in \cref{sec:intro} and closely parallels
ref.~\cite{Marolf:2022ybi}. First, using saddle-point methods, we will evaluate
a Lorentzian path integral \(Z_{\mathrm{L}}(T,\Omega,\Phi)\) with real
Lorentzian boundary conditions specified by parameters \((T,\Omega,\Phi)\) as we
elaborate in \cref{sec:lorpathint}. Such an integral has the quantum mechanical
interpretation of a trace
\begin{align}
  Z_{\mathrm{L}}(T,\Omega,\Phi)
  &= \tr \exp\left(-i\, T\, H_{\zeta + \Omega\, \varphi, \Phi}\right)
    \;,
    \label{eq:lorpartfunc}
\end{align}
Here, \(H_{\zeta + \Omega\, \varphi, \Phi}\) is the same Hamiltonian as in
\cref{eq:partfunc}. Naively, the thermal partition function
\(Z(\beta,\Omega,\Phi)\) is simply an analytic continuation of
\(Z_{\mathrm{L}}(T,\Omega,\Phi)\). Unfortunately, the trace
\labelcref{eq:lorpartfunc} does not converge to a function, but rather gives a
distribution in \(T\). However, we can nonetheless recover
\(Z(\beta,\Omega,\Phi)\) through an integral transform of
\(Z_{\mathrm{L}}(T,\Omega,\Phi)\),
\begin{align}
  Z(\beta,\Omega,\Phi)
  &= \int_{-\infty}^\infty \dd{T} f_\beta(T)\, Z_{\mathrm{L}}(T,\Omega,\Phi)
    \;.
    \label{eq:inttrans}
\end{align}
The kernel \(f_\beta(T)\) is chosen such that
\begin{align}
  e^{-\beta\,E} g_{E_0}(E)
  &= \int_{-\infty}^\infty \dd{T} f_\beta(T)\, e^{-i\, T\, E}
    \;.
    \label{eq:intkernelprop}
\end{align}
for some function \(g_{E_0}(E)\) evaluating to \(1\) for \(E\ge E_0\), where
\(E_0\) is a lower bound on the eigenvalues of
\(H_{\zeta+\Omega\,\varphi,\Phi}\). For example, we can take\footnote{The sign
in \cref{eq:intkernelexample} is due to the fact that the real integration
contour in \cref{eq:intkernelprop}, completed at infinity in the lower half
plane, forms a \emph{clockwise} contour around the pole at \(T=-i\,\beta\).}
\begin{align}
  f_\beta(T)
  &= - \frac{1}{2\pi\, i}
    \frac{
    e^{E_0\,(-\beta + i\, T)}
    }{
    T + i\,\beta
    }
    \;.
  \label{eq:intkernelexample}
\end{align}
In \cref{sec:discussion}, we will discuss
the validity of the integral transform \labelcref{eq:inttrans} in conjunction
with saddle point methods applied to approximate the path integral.

\subsubsection{Boundary conditions}
\label{sec:lorpathint}

The Lorentzian path integral \(Z_{\mathrm{L}}(T,\Omega,\Phi)\) is a function of
three parameters \((T,\Omega,\Phi)\) specifying boundary conditions at the
spacetime boundary \(\partial\mathscr{M}\) essentially identical to those
described in \cref{sec:pathintegral} up to Wick rotation.

Again, we require the induced geometry on \(\partial\mathscr{M}\) to be fixed
and to possess two Killing vectors \(\zeta\) and \(\varphi\), under which all
boundary conditions are required to be invariant. The main difference relative
to \cref{sec:pathintegral} is that the Killing vector \(\zeta\) here is timelike and
related by Wick rotation to its Euclidean counterpart \(-i\,\zeta\). As before,
we require \(\partial\mathscr{M}\) to have topology \(Y\times S^1\), where the
spacial constant sections \(Y\) need not be metric-orthogonal to the \(S^1\)
fibres. The constant sections \(Y\) are preserved by \(\varphi\) while the
fibres \(S^1\) are the orbits of the Killing vector
\begin{align}
  \xi
  &= \zeta + \Omega\,\varphi
    \;,
    \label{eq:lorcorotvec}
\end{align}
such that \(e^{T\, \xi}\) (acting on integer-spin fields) completes one orbit.
Notice that a real Lorentzian boundary geometry is obtained from taking real
\(T\) and \(\Omega\), in contrast to the Euclidean case where \(\Omega\) was
required to be imaginary to specify a real boundary. Let us refer to the
boundary Killing time parameter of \(\zeta\) as \(\hat{t}\), which is constant
on constant sections \(Y\) and is periodically identified
\(\hat{t}\sim\hat{t}+T\).

The boundary conditions for the Maxwell field \(A\) are the same as described in
\cref{sec:pathintegral}. In particular, they are parametrized by an electric
potential \(\Phi\) which has the following properties: \(\Phi=0\) for
configurations which are trivial \(A=0\) near \(\partial\mathscr{M}\); a shift
\(A\mapsto A+\hat{\mu}\dd{\hat{t}}\) in the gauge potential shifts the electric
potential \(\Phi\mapsto\Phi+\hat{\mu}\); and \(\Phi\) enters into the boundary
Hamiltonian in the expected manner for a fixed potential ensemble as we will
later specify.

As in \cref{eq:periodangvel,eq:periodelepot}, when \(\varphi\) generates
rotation or the Maxwel gauge group is compact, we expect that values for the
angular velocity \(\Omega\) or electric potential \(\Phi\) related by
\begin{align}
  \Omega
  &\sim \Omega + \frac{2\pi\, m}{T\,\Delta_J}\,
  &
    (m\in\mathbb{Z})
    \label{eq:lorperiodangvel}
  \\
  \Phi &\sim \Phi + \frac{2\pi\, n}{T\,\Delta_Q}
  &
    (n\in\mathbb{Z})
    \label{eq:lorperiodelepot}
\end{align}
lead to equivalent boundary conditions. (Otherwise, refraining from imposing
these equivalence relations, \(m\) and \(n\) can be interpreted as being set to
zero in the following.)

\subsubsection{Saddle-point evaluation}
\label{sec:lorsaddles}

Let us now consider saddles for the path integral
\(Z_{\mathrm{L}}(T,\Omega,\Phi)\) and constrained saddles for its subcontour
integrals.

An example of a saddle with a trivial singular surface \(\gamma=\varnothing\) is
the empty thermal saddle described in \cref{sec:emptysaddle}, Wick rotated to
Lorentzian signature in the obvious way. Analogous to \cref{eq:thpartfunc}, we
expect a perturbative contribution of
\begin{align}
  Z_{\mathrm{Lth}}(T,\Omega,\Phi)
  &= e^{-i\,T\,E}
    Z_{\mathrm{Lth}}^{\mathrm{QFT}}(T,\Omega,\Phi)
\end{align}
from this saddle, for some constant \(E\) and corrections
\(Z_{\mathrm{Lth}}^{\mathrm{QFT}}(T,\Omega,\Phi)\) from quantum fluctuations,
\eg{} one-loop gravitons. Performing the integral transform
\labelcref{eq:inttrans} on the above, we expect to recover the contribution,
which we called \(Z_{\mathrm{th}}(\beta,\Omega,\Phi)\) in \cref{eq:thpartfunc},
to the thermal partition function \(Z(\beta,\Omega,\Phi)\). This is clear by
virtue of \cref{eq:intkernelprop}, if we ignore the quantum corrections
\(Z_{\mathrm{Lth}}^{\mathrm{QFT}}(T,\Omega,\Phi)\) and
\(Z_{\mathrm{th}}^{\mathrm{QFT}}(\beta,\Omega,\Phi)\), or given that
\(Z_{\mathrm{Lth}}^{\mathrm{QFT}}(T,\Omega,\Phi)\) itself takes the form of an
oscillatory trace \labelcref{eq:lorpartfunc}, \eg{} now over a QFT Hilbert
space.

More interesting are the Lorentzian analogues of Euclidean black holes. As
emphasized by ref.~\cite{Marolf:2022ybi}, in the absence of singularities
\(\gamma\) where the causal structure breaks down, it is hard to imagine how one
would obtain contributions to the Lorentzian path integral analogous to
Euclidean black holes that have contractible boundary time circles. However, as
described in \cref{eq:smoothsaddle}, saddles that are stationary under all
unconstrained variations are not expected to possess a nontrivial singular
surface \(\gamma\). Thus, as in ref.~\cite{Marolf:2022ybi}, we are again led to
consider \emph{constrained} saddles analogous to those constructed in
\cref{sec:bhsaddles}.\footnote{As in \cref{sec:bhsaddles}, we will in fact find
  a family of constrained saddles labelled by integers \(m\) and \(n\). While
  the boundary time circle is contractible in constrained saddles with \(m=0\),
  we will find other constrained saddles with \(m\ne 0\) where the time circle
  is instead homologous to some number of copies of \(\gamma\).}

Analogous to the Euclidean case described there, these constrained saddles are
saddles for path integrals
\(Z_{\mathrm{L}}(T,\Omega,\Phi;\mathcal{S},\mathcal{J},\mathcal{Q})\) over
subcontours of fixed
\begin{align}
  \mathcal{S}
  &= \frac{\mathrm{Area}(\gamma)}{4 G_{\mathrm{N}}}
    \;,
  \\
  \mathcal{J}
  &= -\frac{\mathrm{Area}(\gamma)}{\mathrm{Period}(\varphi)}
    \int_\gamma (p_{\mathrm{BY}})_i \, \chi^i
    \;,
    \label{eq:lormom}
  \\
  \mathcal{Q}
  &= \frac{1}{g_{\mathrm{M}}^2}\int_\gamma *F
    \;.
    \label{eq:lorcharge}
\end{align}
These quantities will be equated to the Bekenstein-Hawking entropy, (angular)
momentum, and electric charge of a black hole shortly.\footnote{The sign in
  \cref{eq:lormom} results from the fact that the surface (angular) momentum
  density \((p_{\mathrm{BY}})_i\) was defined on
  \(-\partial\mathscr{N}_\varepsilon\) while viewing this surface as an internal
  boundary of the spacetime \(\mathscr{M}\setminus\mathscr{N}_\varepsilon\). In
  contrast, for black hole solutions, we would like to equate \(\mathcal{J}\) to
  the usual notion of (angular) momentum defined on the outer boundary
  \(\partial\mathscr{M}\). While the surface
  \(-\partial\mathscr{N}_\varepsilon\) has orientation \(-u^A \epsilon_A\), the
  surface \(\partial\mathscr{M}\) instead has orientation
  \(\tilde{u}^A\epsilon_A\), where \(\tilde{u}\) is the outward normal to
  \(\partial\mathscr{M}\).} The path integral \(Z_{\mathrm{L}}(T,\Omega,\Phi)\)
over the full contour of Lorentzian configurations is obtained by formally
including the integrals over \((\mathcal{S},\mathcal{J},\mathcal{Q})\):
\begin{align}
  Z_{\mathrm{L}}(T,\Omega,\Phi)
  &= \int
    \dd{\mathcal{S}}\dd{\mathcal{J}}\dd{\mathcal{Q}}
    Z_{\mathrm{L}}(T,\Omega,\Phi;\mathcal{S},\mathcal{J},\mathcal{Q})
    \;.
    \label{eq:lorfinalints}
\end{align}
However, as written above, \(Z_{\mathrm{L}}(T,\Omega,\Phi)\) fails to converge
to a function, which is consistent with the observation noted below
\cref{eq:lorpartfunc} that \(Z_{\mathrm{L}}(T,\Omega,\Phi)\) instead has the
interpretation of a distribution with respect to \(T\). Practically, this means that any
integral over \(T\), \eg{} the integral transform \labelcref{eq:inttrans}, should
be performed before the integrals over
\((\mathcal{S},\mathcal{J},\mathcal{Q})\).

A brief technical comment is that, as in our Euclidean analysis, it will be
sufficient, for our discussion of the partition function, to include or consider
in our path integral only configurations where the quantities
\((\kappa,v^i,\mu)\) and
\(\lim_{\rho\to0}(h_{ij},(p_{\mathrm{BY}})_i,(*F)_{i_1\cdots i_{D-2}})\),
describing the configuration near singularity \(\gamma\), are constant with
respect to hyperbolic time \(t\). In particular, the constrained saddles we will
construct shortly are highly symmetric and include no lightcones for \(\gamma\)
where one might want to turn off the helical shift to avoid lightcone
singularities as described in \cref{sec:lightcones}. More generally, however,
one may wish to allow \((\kappa,v^i,\mu)\) and
\(\lim_{\rho\to0}((p_{\mathrm{BY}})_i,(*F)_{i_1\cdots i_{D-2}})\) to vary with
respect to \(t\). When including such configurations in the path integral, an
extra average over \(t\) should be included in certain equations, \eg{}
\cref{eq:fixmom,eq:fixcharge,eq:constrsaddle1,eq:lormom,eq:lorcharge}, without
significantly altering our results. (In particular, the configurations
constructed below will continue to be constrained saddles.)

Let us now construct constrained saddles which are saddles for the subcontour
integral \(Z_{\mathrm{L}}(T,\Omega,\Phi;\mathcal{S},\mathcal{J},\mathcal{Q})\).
To contribute to this subcontour integral, these configurations must satisfy the
boundary conditions parametrized by \((T,\Omega,\Phi)\) on the spacetime
boundary \(\partial\mathscr{M}\) and have the prescribed values of
\((\mathcal{S},\mathcal{J},\mathcal{Q})\) on \(\gamma\). Moreover, as described
in \cref{sec:saddle}, constrained saddles must satisfy the bulk equations
of motion away from \(\gamma\) and also
\cref{eq:constrsaddle1,eq:constrsaddle2,eq:constrsaddle3} appropriately
Wick-rotated on \(\gamma\).

\begin{figure}
  \centering
  \includegraphics{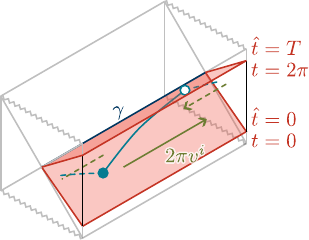}
  \caption{A conically and helically singular constrained saddle. Illustrated in
    grey is a stationary black hole spacetime. When the spatial Killing vector
    \(\varphi\) generates rotation (as opposed to translation), the front and
    back faces (where the Penrose diagram is drawn) are identified. The
    spacetime of the constrained saddle resides between the two red surfaces of
    constant time. These surfaces are identified with a relative shift
    \(2\pi v^i\), \eg{} represented by the solid green arrow. In particular, the points
    marked by empty and filled circles are identified, so the solid teal curve
    connecting these points is closed. When the helical shift \(v^i\) of
    \(\gamma\) is as illustrated by the solid arrow, the solid teal curve is
    contractible when the singularity is regularized --- see \cref{sec:regsing}.
    Other choices of helical shift, \eg{} the dashed arrow, can give
    configurations that are diffeomorphic to this one away from \(\gamma\)
    before regularization, but, upon regularization, have different
    contractible cycles, \eg{} the dashed teal curve.}
  \label{fig:lorhelicalbh}
\end{figure}

Similar to the Euclidean construction of \cref{sec:bhsaddles}, the starting
point is a smooth Lorentzian stationary black hole solution whose
Bekenstein-Hawking entropy, (angular) momentum, and electric charge are set to
the given values of \((\mathcal{S},\mathcal{J},\mathcal{Q})\). Again, for a
stationary black hole, it is straightforward to see that evaluating
\cref{eq:lormom,eq:lorcharge} on the bifurcation surface reproduces the values
of (angular) momentum and electric charge obtained from standard definitions on
\(\partial\mathscr{M}\).\footnote{See \cref{foot:momchargebh}.} We further
require this initial configuration to satisfy boundary conditions on one of its
boundaries \(Y\times \mathbb{R}\) which are locally similar to those imposed by
the path integral \(Z_{\mathrm{L}}(T,\Omega,\Phi)\) --- for example, we require
the black hole to have boundary Killing vectors \(\zeta\) and \(\varphi\) which
we then expect to extend into the bulk. However, in this black hole solution,
the periodic identification \(\hat{t}\sim\hat{t}+T\) is relaxed, the boundary
limit of the horizon generating Killing vector is given by
\(\zeta+\Omega_0\,\varphi\) for some \(\Omega_0\) possibly differing from
\(\Omega\), and the electric potential \(\Phi_0\) might differ from \(\Phi\). We
will now correct these mismatches by making the bifurcation surface \(\gamma\)
singular, as illustrated in \cref{fig:lorhelicalbh}

Firstly, extending the boundary stationary time \(\hat{t}\) into the bulk, we
can simply impose by hand \(\hat{t}\sim\hat{t}+T\) and consider a fundamental
domain of this identification in one exterior of the black hole. This produces a
spacetime of the kind illustrated in \cref{fig:nolightconesspacelike} with no
lightcones for \(\gamma\). The metric near \(\gamma\) is given by
\cref{eq:lormetricg1} where, taking \(t\sim t+2\pi\) proportional to
\(\hat{t}\), the leading coefficient \(\kappa=\kappa_0\, T/2\pi\) of the lapse
\labelcref{eq:lorlapseshift} is determined by the horizon surface gravity
\(\kappa_0\) of the black hole (as conventionally defined with respect to
\(\hat{t}\)). The hyperbolic opening angle of the now conically singular surface
\(\gamma\) is then given by
\begin{align}
  2\pi\, \kappa
  &= T\, \kappa_0
    \;,
    \label{eq:lorbhcon}
\end{align}
analogous to \cref{eq:bhcon}.

When we identify the bulk surfaces \(\hat{t}=0\) and \(\hat{t}=T\), there is a
freedom to relatively shift the surfaces along the Killing vector \(\varphi\)
extended into the bulk. We can use this freedom to ensure that the Killing
vector \(\xi\), given in \cref{eq:lorcorotvec} for the prescribed value of
\(\Omega\) or any of its representatives under the equivalence relation
\labelcref{eq:lorperiodangvel}, has closed orbits with period \(T\). This
introduces a helical singularity on \(\gamma\), such that the helical shift in
\cref{eq:lorlapseshift} is given by
\begin{align}
  v^i
  &=  \left[ \frac{T}{2\pi}
    (\Omega-\Omega_0)
    + \frac{m}{\Delta_J}
    \right] \varphi^i
    \;,
  &
    (m \in \mathbb{Z})
    \label{eq:lorbhhel}
\end{align}
analogous to \cref{eq:bhhel}. As illustrated in \cref{fig:lorhelicalbh}, these
different \(v^i\) lead to configurations which are diffeormorphic to each other
away from \(\gamma\), but are physically distinct in the way they are regulated
on \(\gamma\) --- see \cref{sec:regsing}.

Finally, to attain the electric potential \(\Phi\) prescribed by the boundary
conditions or a representative under \labelcref{eq:lorperiodelepot}, we can
shift the Maxwell field,
\begin{align}
  A
  &\mapsto
    A+
    \left(
    \Phi-\Phi_0
    +\frac{2\pi\,n}{T\,\Delta_Q}
    \right)
    \dd{\hat{t}}
  \;.
  & (n\in\mathbb{Z})
\end{align}
This turns on a holonomic singularity with strength
\begin{align}
  \mu
  &= \frac{T}{2\pi}(\Phi - \Phi_0) + \frac{n}{\Delta_Q}
    \;.
  & (n\in\mathbb{Z})
    \label{eq:lorbhhol}
\end{align}
analogous to \cref{eq:bhhol}.

We now have some configuration(s) satisfying the boundary conditions prescribed
by \((\beta,\Omega,\Phi)\) at \(\partial\mathscr{M}\) and having the fixed
values of \((\mathcal{S},\mathcal{J},\mathcal{Q})\) on \(\gamma\). Equations of
motion are satisfied away from \(\gamma\), as in the original black hole
solution, and the constrained saddle-point conditions
\labelcref{eq:constrsaddle1,eq:constrsaddle2,eq:constrsaddle3} are satisfied on \(\gamma\), as can be seen from
\cref{eq:lorbhcon,eq:lorbhhel,eq:lorbhhol} together with the bulk Killing
symmetry \(\varphi\). Thus, we have successfully constructed constrained
saddle(s) which are saddles for the subcontour integral
\(Z(T,\Omega,\Phi;\mathcal{S},\mathcal{J},\mathcal{Q})\), completely analogous
to the Euclidean constrained saddles obtained in \cref{sec:bhsaddles}.

The action \labelcref{eq:loraction} for these Lorentzian constrained saddles is
calculated in nearly identical fashion to that section. The bulk and
Gibbons-Hawking terms evaluate to
\begin{align}
  \int_{\mathscr{M}\setminus\mathscr{N}_\varepsilon} (L_{\mathrm{EH}} + L_{\mathrm{M}})
  + \int_{\partial\mathscr{M}} L_{\partial \mathscr{M}}
  + \int_{-\partial\mathscr{N}_\varepsilon} L_{\mathrm{GH}}
  &\sim -T\, (E - \Omega_0\, \mathcal{J} - \Phi_0\, \mathcal{Q})
    \;,
    \label{eq:lorbulkghsingbh}
\end{align}
where
\begin{align}
  E_{\zeta + \Omega_0 \varphi,\Phi_0}(\mathcal{S},\mathcal{J},\mathcal{Q})
  &= E(\mathcal{S},\mathcal{J},\mathcal{Q})
    - \Omega_0\, \mathcal{J}
    - \Phi_0\, \mathcal{Q}
\end{align}
is the value of the boundary Hamiltonian generating the evolution
\(\zeta+\Omega_0\,\varphi\) with an electric potential \(\Phi_0\) in the
original black hole solution. The helical and holonomic singularities on
\(\gamma\) as described by \cref{eq:lorbhhel,eq:lorbhhol} do not affect
\cref{eq:lorbhcon}, but instead enter into the action through the helical and
holonomic terms:
\begin{align}
  \int_{-\partial\mathscr{N}_\varepsilon} L_{\mathrm{hel}}
  &\sim -2\pi \int_\gamma v^i\,(p_{\mathrm{BY}})_i
    = \left[
    T\,(\Omega-\Omega_0)
    + \frac{2\pi\, m}{\Delta_J}
    \right] \mathcal{J}
    \;,
  \\
  \int_{-\partial\mathscr{N}_\varepsilon} L_{\mathrm{hol}}
  &\sim \frac{2\pi}{g_{\mathrm{M}}^2} \int_\gamma \mu *F
    = \left[
    T\, (\Phi - \Phi_0)
    + \frac{2\pi\, n}{\Delta_Q}
    \right] \mathcal{Q}
    \;,
\end{align}
where we have used \cref{eq:lormom,eq:lorcharge}. The action
\labelcref{eq:loraction} evaluated for our constrained saddles is thus
\begin{align}
  I
  &=  -T\, E_{\xi,\Phi}(\mathcal{S},\mathcal{J},\mathcal{Q})
    -i\,\mathcal{S}
    + 2\pi\, m\, \frac{\mathcal{J}}{\Delta_J}
    + 2\pi\, n\, \frac{\mathcal{Q}}{\Delta_Q}
    \;.
\end{align}
where
\begin{align}
  E_{\xi,\Phi}(\mathcal{S},\mathcal{J},\mathcal{Q})
  &= E(\mathcal{S},\mathcal{J},\mathcal{Q})
    - \Omega\, \mathcal{J}
    - \Phi\, \mathcal{Q}
\end{align}

Summing over \(m\) and \(n\) as needed (or picking only \(m=0\) or \(n=0\)), the
contribution of these constrained saddle(s) to the Lorentzian path integral
\(Z_{\mathrm{L}}(T,\Omega,\Phi)\) is given by the analogue of
\cref{eq:allbhpartfunc},
\begin{align}
  \sum_{m,n}
  Z_{\mathrm{LBH}}
  \left(
  T,
  \Omega+\frac{2\pi\, m}{T\, \Delta_J},
  \Phi+\frac{2\pi\, n}{T\, \Delta_Q}
  \right)
  \;,
  \label{eq:lorallbhpartfunc}
\end{align}
where
\begin{align}
  Z_{\mathrm{LBH}}(T,\Omega,\Phi)
  &= \int \dd{\mathcal{S}}\dd{\mathcal{J}}\dd{\mathcal{Q}}
    e^{\mathcal{S} - i\,T\, E_{\xi,\Phi}(\mathcal{S},\mathcal{J},\mathcal{Q})}
    Z_{\mathrm{LBH}}^{\mathrm{QFT}}\left(
    T, \Omega, \Phi
    ;\mathcal{S},\mathcal{J},\mathcal{Q}
    \right)
    \;,
    \label{eq:lorbhpartfunc}
\end{align}
and
\(Z_{\mathrm{LBH}}^{\mathrm{QFT}}(T,\Omega,\Phi;\mathcal{S},\mathcal{J},\mathcal{Q})\)
denotes the quantum corrections to the saddle-point evaluation of
\(Z_{\mathrm{BH}}(T,\Omega,\Phi;\mathcal{S},\mathcal{J},\mathcal{Q})\). The
integral \labelcref{eq:lorbhpartfunc} does not converge, because of the
exponentially enhanced integrand at large \(\mathcal{S}\), just as the
oscillatory trace \labelcref{eq:lorpartfunc} over infinitely many states does
not converge. Rather, as anticipated below
\cref{eq:lorpartfunc,eq:lorfinalints}, \cref{eq:lorbhpartfunc} gives a
distribution over \(T\) and the integrals displayed here should really be
performed last, \eg{} after the integral transform \labelcref{eq:inttrans}.
Similar to the empty thermal saddle, we expect that the integral transform of
\cref{eq:lorbhpartfunc} will recover the Euclidean result \cref{eq:bhpartfunc}
for the black hole contribution to the grand canonical partition function
\(Z(\beta,\Omega,\Phi)\).

\section{Discussion}
\label{sec:discussion}

\subsection{Summary}
\label{sec:summary}

In this paper, we have considered a class of singularities on codimension-two
surfaces \(\gamma\), in Einstein-Maxwell theory possibly with a cosmological
constant, which generalizes conical singularities. In particular, as described
in \cref{sec:sing}, the helical and holonomic types of singularities involve
shifts along \(\gamma\) and along the fibres of the Maxwell principal bundle as
one winds around \(\gamma\) in a metric-orthogonal and connection-horizontal
manner. Having given a prescription for regulating singularities with smooth
configurations, we studied the curvature contributions of these singularities
and subsequently, in \cref{sec:action}, proposed an action for singular
configurations. We then studied constrained saddles, which have stationary
action under variations that fix, on a codimension-two surface \(\gamma\), area
and quantities associated to (angular) momentum and electric charge. The upshot
is that these constrained saddles can possess nontrivial conical, helical, and
holonomic singularities on \(\gamma\).

A motivation for considering such constrained saddles is for the purpose of
evaluating the gravitational partition function. The grand canonical partition
function \(Z(\beta,\Omega,\Phi)\) is most often evaluated by a path integral in
Euclidean signature, where smooth black holes are well-known saddles.
Alternatively, one may reorganize the path integral as we have in
\cref{sec:partfunc} so as to leave for last the integrals over the
Bekenstein-Hawking entropy \(\mathcal{S}\), (angular) momentum \(\mathcal{J}\),
and electric flux \(\mathcal{Q}\) evaluated locally on a codimension-two surface
\(\gamma\). The aforementioned constrained saddles are then saddles for the
initial integral over subcontours of fixed
\((\mathcal{S},\mathcal{Q},\mathcal{J})\) and can be constructed by modifying
smooth black holes to make their bifurcation surfaces conically, helically, and
holonomically singular. The resulting constrained saddle-point contribution to
the grand canonical partition function takes the form given by
\cref{eq:allbhpartfunc,eq:bhpartfunc}. Although the evaluation of the
gravitational partition function is often associated with Euclidean signature,
due to the conformal mode, the Euclidean gravitational action is unbounded from
below on the integration contour over real Euclidean geometries
\cite{Gibbons:1978ac}. The path integral over this contour is therefore
manifestly divergent and ill-defined.

Turning to Lorentz signature, where integrals are instead oscillatory and have a
distributional meaning, the need to consider singular configurations in the path
integral becomes more pronounced. In this context, the grand canonical partition
function \(Z(\beta,\Omega,\Phi)\) is given by an integral transform
\labelcref{eq:inttrans} on the time period \(T\) of a Lorentzian path integral
\(Z_{\mathrm{L}}(T,\Omega,\Phi)\). As previously emphasized by
ref.~\cite{Marolf:2022ybi}, in order to receive contributions analogous to
Euclidean black holes, it is necessary to allow at least conically singular
surfaces \(\gamma\) in the Lorentzian configurations where the time circle can
contract to a point. In \cref{sec:lorentzian}, we generalized
ref.~\cite{Marolf:2022ybi}'s Lorentzian construction to include also helical and
holonomic singularities on \(\gamma\). The resulting saddle-point evaluation of
the Lorentzian path integral \(Z_{\mathrm{L}}(T,\Omega,\Phi)\) indeed receives
contributions, given in \cref{eq:lorallbhpartfunc,eq:lorbhpartfunc}, from
singular constrained saddles constructed from black holes. Performing the
integral transform from \(T\) to \(\beta\), one again recovers the expected
black hole contributions to the thermal partition function
\(Z(\beta,\Omega,\Phi)\), given in \labelcref{eq:allbhpartfunc,eq:bhpartfunc}.

\subsection{Contributing saddles and the stability of black holes}
\label{sec:bhstability}

What have we gained by including helical and holonomic singularities in our
analysis? One result of our more inclusive analysis is a better understanding of
which saddles are relevant for the gravitational thermal partition function. In
particular, we will address below some puzzles left open by the study of purely
conical singularities in ref.~\cite{Marolf:2022ybi}, regarding the thermodynamic
stability of relevant black hole saddles, with respect to variations in (angular)
momentum and charge.

A direct result of including more general types of singularities is a richer set
of constrained saddles. In particular, we saw explicitly in our evaluation of
the Lorentzian path integral \(Z_{\mathrm{L}}(T,\Omega,\Phi)\) how helical and
holonomic, as well as conical, singularities appear in constrained saddles where
(angular) momentum \(\mathcal{J}\) and electric charge \(\mathcal{Q}\), as well
as area \(\mathcal{S}\), are fixed to arbitrary values. The integrals over these
quantities can therefore all be saved for last while the constrained saddles
provide a saddle-point evaluation of the Lorentzian path integral
\(Z_{\mathrm{L}}(T,\Omega,\Phi;\mathcal{S},\mathcal{J},\mathcal{Q})\) over
subcontours of fixed \((\mathcal{S},\mathcal{J},\mathcal{Q})\) --- see
\cref{eq:lorfinalints}. (As we will see, the order of integration can play an
important role when analyzing the thermodynamic stability of saddles
relevant to the final thermal partition function.)

However, the mere existence of a saddle, \eg{} a constrained saddle constructed
from black holes in \cref{sec:lorsaddles}, does not guarantee that it
contributes with nonzero weight to a path integral, \eg{}
\(Z_{\mathrm{L}}(T,\Omega,\Phi;\mathcal{S},\mathcal{J},\mathcal{Q})\). Having
chosen the integration contour to be over real Lorentzian configurations, one
can in principle determine the weight with which a saddle contributes using
Morse theory, as reviewed in ref.~\cite{Witten:2010cx,Marolf:2022ybi}. Let us
summarize some of the pertinent results in this regard.

In general, let us consider an integral \(\int_{\mathfrak{X}}\dd{x}e^{i\,I(x)}\)
over a middle-(real-)dimensional contour \(\mathfrak{X}\) in a complex manifold
\(X\). The idea is then to deform \(\mathfrak{X}= \sum_p \mathfrak{n}_p\,
\mathfrak{J}_p\) into some multiples of particular contours \(\mathfrak{J}_p\),
called Lefschetz thimbles, with coefficients \(\mathfrak{n}_p\), while leaving
the value of integral unchanged. For each
isolated\footnote{Ref.~\cite{Witten:2010cx} also considers cases with
  non-isolated critical points, \ie{} critical manifolds, with generalized
  Lefschetz thimbles \(\mathfrak{J}_p\) where the label \(p\) now enumerates the
  middle-(real-)dimensional cycles of critical manifolds.} critical point \(p\)
of the action \(I\), the corresponding Lefschetz thimble \(\mathfrak{J}_p\) is
simply the contour of steepest descent for the Morse function \(\Re(i\,I)\);
meanwhile, the coefficient \(\mathfrak{n}_p\) is given by the intersection
number of the original contour \(\mathfrak{X}\) with the contour
\(\mathfrak{K}_p\) of steepest ascent for \(\Re(i\,I)\). In this way, we may
decompose the integral
\begin{align}
  \int_{\mathfrak{X}}\dd{x}e^{i\,I(x)}
  &= \sum_p \mathfrak{n}_p \int_{\mathfrak{J}_p}
    \dd{x}
    e^{i\,I(x)}
    \label{eq:lefschetz}
\end{align}
into a weighted sum over Lefschetz thimbles. If one so wishes, the integral over
each Lefschetz thimble can then be perturbatively approximated as a Gaussian with
corrections.

A simplifying observation \cite{Marolf:2022ybi} is that, if
\(\Re(i\,I)\) is constant on \(\mathfrak{X}\), then \(\mathfrak{n}_p\) must take
values \(\pm 1\) for saddles \(p\) lying on \(\mathfrak{X}\). Indeed, if we
decompose the subcontour integral
\(Z_{\mathrm{L}}(T,\Omega,\Phi;\mathcal{S},\mathcal{J},\mathcal{Q})\) into
discrete sectors in which the number of lightcone components
\(\mathcal{N}_\ell\) for each connected component \(\gamma_\ell\) of \(\gamma\)
is fixed, then the imaginary part of the action \labelcref{eq:loraction} takes a
constant value \(-\mathcal{S}\) in the sector with no lightcones for \(\gamma\).
Moreover, the (constrained) saddles constructed in \cref{sec:lorsaddles} do lie
on the original contour of integration over Lorentzian configurations in the
lightcone-less sector. The upshot then is that all such saddles do contribute
with weight \(\mathfrak{n}_p=\pm 1\) to the subcontour integral
\(Z_{\mathrm{L}}(T,\Omega,\Phi;\mathcal{S},\mathcal{J},\mathcal{Q})\).

As described around \cref{eq:inttrans,eq:lorbhpartfunc}, to recover the grand
canonical partition function \(Z(\beta,\Omega,\Phi)\), we must perform the
integral transform \labelcref{eq:inttrans} from \(T\) to \(\beta\) and the final
integrals over \((\mathcal{S},\mathcal{J},\mathcal{Q})\). The final result for
the contribution of the constrained saddles built from Lorentzian black holes
should then agree with the answer \labelcref{eq:bhpartfunc} deduced from the
naive Euclidean analysis (provided the integral transform in \(T\) behaves in
the expected manner, as we will elaborate in \cref{sec:toyexample1}). However,
whereas the Euclidean path integral did not even have a good and natural choice
of integration contour to speak of, we have now argued that the natural choice
of Lorentzian contour can be deformed into appropriate Lefschetz thimbles so as
to recover the expected black hole contribution \labelcref{eq:bhpartfunc} to the
grand canonical partition function.

To make the connection to the more standard Euclidean analysis of purely smooth
configurations, let us consider the leading semiclassical approximation, where
we ignore \(Z_{\mathrm{BH}}^{\mathrm{QFT}}\) and take the saddle-point values
for the \((\mathcal{S},\mathcal{J},\mathcal{Q})\) integrals in
\cref{eq:bhpartfunc}. Then, the saddle-point values for
\((\mathcal{S},\mathcal{J},\mathcal{Q})\) are simply the on-shell values for the
Bekenstein-Hawking entropy, (angular) momentum, and charge for a black hole with
the values of \((\beta,\Omega,\Phi)\) prescribed by the grand canonical
ensemble; in particular, the Euclidean counterparts of these black holes are
smooth everywhere, including the bifurcation surface \(\gamma\). In this sense,
smooth Euclidean black holes are saddles for the thermal partition function, as
one might have guessed by considering a naive path integral over Euclidean
configurations. However, let us emphasize again that this naive path integral is
plagued by the conformal factor problem, so it was therefore important for us to
derive this result from a Lorentzian starting point.

From \cref{eq:bhpartfunc}, we can also see how the thermodynamic stability of a
given on-shell black hole comes into play.\footnote{In this discussion of
  thermodynamic stability, we have in mind saddles which lie on the undeformed
  real integration contour for \((\mathcal{S},\mathcal{J},\mathcal{Q})\), along
  which we consider real variations of these variables. It would be interesting
  to also consider saddles not on this integration contour which one might
  naturally expect, for example, at nonzero values of \(m\) or \(n\) in
  \cref{eq:allbhpartfunc} (which is the integral transform
  \labelcref{eq:inttrans} of the Lorentzian result
  \labelcref{eq:lorallbhpartfunc}).
  \label{foot:complexsaddlesmn}} In particular (again in the semiclassical
limit), local maxima of the integrand in \cref{eq:bhpartfunc}, correspond to
black holes minimizing the free energy
\begin{align}
  F(\beta,\Omega,\Phi;\mathcal{S},\mathcal{J},\mathcal{Q})
  &= E(\mathcal{S},\mathcal{J},\mathcal{Q}) - \Omega\, \mathcal{J} - \Phi\, \mathcal{Q}
    -\frac{1}{\beta}\,\mathcal{S}
    \;,
\end{align}
with respect to variations in \((\mathcal{S},\mathcal{J},\mathcal{Q})\), where
\(E(\mathcal{S},\mathcal{J},\mathcal{Q})\) is the energy of an on-shell black
hole with Bekenstein-Hawking entropy \(\mathcal{S}\), (angular) momentum
\(\mathcal{J}\), and electric charge \(\mathcal{Q}\). Equivalently,
\begin{align}
  F(\beta,\Omega,\Phi;E,\mathcal{J},\mathcal{Q})
  &= E - \Omega\, \mathcal{J} - \Phi\, \mathcal{Q}
    -\frac{1}{\beta}\,\mathcal{S}(E,\mathcal{J},\mathcal{Q})
\end{align}
is minimized with respect to the more standard set of independent quantities
\((E,\mathcal{J},\mathcal{Q})\), while
\(\mathcal{S}(E,\mathcal{J},\mathcal{Q})\) is a function giving the on-shell
Bekenstein-Hawking entropy of a black hole with prescribed energy \(E\). Black
hole solutions which are thermodynamically unstable in any of the directions in
the space of \((\mathcal{S},\mathcal{J},\mathcal{Q})\) or equivalently
\((E,\mathcal{J},\mathcal{Q})\) will extremize but not locally maximize the
integrand in \cref{eq:bhpartfunc}. The question of whether such unstable black
holes contribute to the partition function subtly depends on definitions, as we
will elaborate in \cref{sec:toyexample2}. Indeed, we will argue there that such
contributions can sometimes be poorly defined (not just subleading) unless the
integrals along the Lefshetz thimbles of thermodynamically stable black holes
are specified precisely.

To appreciate what we have gained from allowing helical and holonomic
singularities into our Lorentzian path integral, it is helpful to contrast the
above with the limited stability condition found by ref.~\cite{Marolf:2022ybi}.
The story told by ref.~\cite{Marolf:2022ybi}, summarized in
\cref{sec:marolfreview}, is similar to the above, except that only
\(\mathcal{S}\) is fixed then integrated last:
\begin{align}
  Z(\beta,\Omega,\Phi)
  &= \int \dd{\mathcal{S}} \dd{T}
    f_\beta(T)\,
    Z_{\mathrm{L}}(T,\Omega,\Phi;\mathcal{S})
    \;.
    \label{eq:fixareaonly}
\end{align}
Saddles for the Lorentzian fixed-\(\mathcal{S}\) path integral
\(Z_{\mathrm{L}}(T,\Omega,\Phi;\mathcal{S})\) have purely conical singularities
and can be constructed from Lorentzian black holes as in \cref{sec:bhsaddles}.
(Helical and holonomic singularities, even if they are allowed in the path
integral \(Z_{\mathrm{L}}(T,\Omega,\Phi;\mathcal{S})\), do not appear in saddles for this
path integral where \(\mathcal{J}\) and \(\mathcal{Q}\) are integrated over
and not fixed.) Going through the same argument as presented above, one finds
that the conically singular saddles \(p\) all contribute with weight
\(\mathfrak{n}_p=\pm 1\) to \(Z_{\mathrm{L}}(T,\Omega,\Phi;\mathcal{S})\).
Naively performing the integral transform from \(T\) to \(\beta\), one seems to
find that a sufficient condition for a black hole solution to correspond to a
local maximum for the \(\mathcal{S}\) integrand in \cref{eq:fixareaonly} is for
it to minimize
\begin{align}
  F(\beta,\Omega,\Phi;\mathcal{S})
  &= E(\Omega,\Phi;\mathcal{S})
    - \Omega\, J(\Omega,\Phi;\mathcal{S})
    - \Phi\, Q(\Omega,\Phi;\mathcal{S})
    - \frac{1}{\beta}\, \mathcal{S}
\end{align}
with respect to \(\mathcal{S}\), where \(E(\Omega,\Phi;\mathcal{S})\),
\(J(\Omega,\Phi;\mathcal{S})\), and \(Q(\Omega,\Phi;\mathcal{S})\) are functions
giving the energy, angular momentum, and charge of an on-shell black hole with a
prescribed (angular) velocity \(\Omega\), electric potential \(\Phi\), and
Bekenstein-Hawking entropy \(\mathcal{S}\). In other words, one seems to find
only a stability condition in the \(\mathcal{S}\) direction, which can be
equivalently rephrased again in terms of energy and corresponds to the
positivity of specific heat \(C\) as defined in \labelcref{eq:specheat}.

In contrast, by performing the integrals over
\((\mathcal{S},\mathcal{J},\mathcal{Q})\) after the integral transform in \(T\),
we have found a stronger set of conditions corresponding to a more complete
notion of thermodynamic stability. Though this is a desirable result, one is
led to wonder: how was it possible for us to obtain a stronger set of stability
conditions by simply reorganizing the path integral?

\subsection{Unstable saddles and the integral transform on Lorentzian time \(T\)}
\label{sec:toyexamples}

The emergence of the extra stability conditions we have found relative to
ref.~\cite{Marolf:2022ybi} clearly has something to do with the order of
integration in \(\mathcal{J}\) and \(\mathcal{Q}\) versus the integral transform
in \(T\). To understand this subtlety, it will be instructive to retreat to a
very simple toy example where we can fully dissect the calculation in
\cref{sec:toyexample1}. By considering a slightly more complete example in
\cref{sec:toyexample2}, we will also explain whether and in what sense an
unstable saddle can contribute to the thermal partition function.

\subsubsection{A simple Gaussian toy example}
\label{sec:toyexample1}

Let us first study an example that is perhaps overly simple but nonetheless
instructive for understanding how the integral transform \labelcref{eq:inttrans}
acts on contributions to the Lorentzian path integral
\(Z_{\mathrm{L}}(T,\Omega,\Phi)\) from the Lefschetz thimbles of
various saddles. Here, we consider a toy analogue for these contributions from
saddles which are ``stable'' (\(+\)) or ``unstable'' (\(-\)) --- in the sense to
be explained below --- with respect to \(\mathcal{J}\) or \(\mathcal{Q}\),
\begin{align}
  z_{\mathrm{L}}^\pm(T)
  &= \int
    \frac{\dd{x}}{\sqrt{\hbar}}\,
    e^{-\frac{i}{\hbar}\,T\,(H_0 \pm \omega\,x^2)}
    = \sqrt{\frac{\pi}{\pm i\,T\,\omega}}
    \,
    e^{-\frac{i}{\hbar}\,T\,H_0}
    \;,
    \label{eq:toysaddle}
\end{align}
where \(H_0\) is a constant, \(\omega>0\) is an energy scale introduced to make
\(x\) dimensionless, and we have restored \(\hbar\). For real \(T\), the
integration contour can be taken to be \(\mathbb{R}\), which is then deformable
to the Lefschetz thimble for the saddle at \(x=0\) so the integral converges. In
this analogy, \(x\) plays the role of \(\mathcal{S}\), \(\mathcal{J}\), or
\(\mathcal{Q}\), which are integrated over in \(Z_{\mathrm{L}}(T,\Omega,\Phi)\).

However, it should be noted that \(z_{\mathrm{L}}^\pm(T)\) is a purely
oscillatory integral, while the Lorentzian gravitational action has an imaginary
part given by \(-\mathcal{S}\) (in the no-lightcone sector). Relatedly,
\(z_{\mathrm{L}}^\pm(T)\) is a well-defined (branched) function of \(T\) unlike
the gravitational path integral \(Z_{\mathrm{L}}(T,\Omega,\Phi)\) which must be
interpreted as a distribution, as described below \cref{eq:lorpartfunc}. The
simplicity of our example therefore allows \(z_{\mathrm{L}}^\pm(T)\) to be
continued to complex values of \(T\) by inspection. The integral expression
\labelcref{eq:toysaddle} for \(z_{\mathrm{L}}^\pm(T)\) continues to complex
\(T\) as well, provided the integration contour rotates along with the Lefschetz
thimble of \(x=0\) to ensure convergence. By considering the defining integral
\labelcref{eq:toysaddle} at even just real \(T\), it is easy to see that the
branch cut of the analytic continuation lies respectively in the upper (\(+\))
or lower (\(-\)) half \(T\)-plane.

The analytic continuation of \(z_{\mathrm{L}}^\pm(T)\) to negative imaginary values of
\(T\) in particular,
\begin{align}
  z^\pm(\beta)
  &\equiv \int
    \frac{\dd{x}}{\sqrt{\hbar}}\,
    e^{
    -\frac{1}{\hbar}\,\beta\,
    (H_0 \pm \omega\,x^2)
    }
    = z_{\mathrm{L}}^\pm(-i\,\beta)
    = \sqrt{\frac{\pi}{\pm\beta\,\omega}}
    \,
    e^{-\frac{1}{\hbar}\,\beta\,H_0}
    \;,
    \label{eq:toysaddleeuc}
\end{align}
has the interpretation, in this toy example, as the corresponding contribution
to the thermal partition function \(Z(\beta,\Omega,\Phi)\). A few things are
noteworthy here. Firstly, we see that the sign \(\pm\) multiplies the piece of
the ``Euclidean action'' \(\beta\,(H_0\pm \omega\,x^2)\) or
the ``free energy'' \(H_0\pm\omega\,x^2\) quadratic in the fluctuation
away from the saddle \(x=0\). Thus, depending on the sign \(\pm\), one would
ordinarily refer to the saddle as stable (\(+\)) or unstable
(\(-\)).\footnote{This notion of stability comes from considering real
  perturbations of \(x\). Of course, the saddle is always a maximum for the
  magnitude of the integrand along the Lefschetz thimble of saddle. For the
  ``unstable'' case, the Lefschetz thimble is along the imaginary \(x\) axis. We
  will see in the more complete toy example of \cref{sec:toyexample2} how such
  thimbles can combine with those of other ``stable'' saddles to give an
  integration contour deformable to the real line.} Secondly, in the unstable
case, \(z^-(\beta)\) has a sign ambiguity due to the aforementioned branch cut.
Using a more complete example in \cref{sec:toyexample2}, we will explain how to
make sense of this sign-ambiguous contribution to the partition function in
terms of Stokes phenomena.

Instead of analytically continuing by inspection, let us now instead apply the
integral transform \labelcref{eq:inttrans} to \(z_{\mathrm{L}}^\pm(T)\), using
the integral kernel \(f_\beta(T)\) given in \cref{eq:intkernelexample}. As we
have emphasized in \cref{sec:lorpartfunc}, our prescription is always to perform
\emph{first} this \(T\) integral \emph{then} the remaining integrals over
\((\mathcal{S},\mathcal{J},\mathcal{Q})\). In this toy example, the correct
prescription is therefore to evaluate
\begin{align}
  \int \frac{\dd{x}}{\sqrt{\hbar}}\,
  \int_{-\infty}^\infty \dd{T}
  f_\beta(T)\,
  e^{
  -\frac{i}{\hbar}\,T\,
  (H_0 \pm \omega\,x^2)
  }
  &=
    \int \frac{\dd{x}}{\sqrt{\hbar}}\,
    e^{
    -\frac{1}{\hbar}\,\beta\,
    (H_0 \pm \omega\,x^2)
    }
    = z^\pm(\beta)
    \;,
    \label{eq:toyexample1inttrans}
\end{align}
where we have applied the residue theorem in the first equality, taking
advantage of the analyticity of \(e^{-\frac{i}{\hbar}\,T\,(H_0 \pm
  \omega\,x^2)}\) and the exponential suppression of
\(f_\beta(T)\,e^{-\frac{i}{\hbar}\,T\,(H_0 \pm \omega\,x^2)}\) in the lower half
\(T\) plane, provided the constant \(E_0\) in \(f_\beta(T)\) is chosen to be
less than \(H_0\).\footnote{Recall that, as \(T\) varies over complex values, we
  previously allowed the \(x\) contour of integration in \cref{eq:toysaddle} to
  rotate along with the Lefschetz thimble of the saddle \(x=0\). We should,
  however, select a \(T\)-independent choice of \(x\) contour in
  \cref{eq:toyexample1inttrans}, where the \(x\) integral is performed after the
  \(T\) integral transform. Here, we can consistently choose the \(x\) contour
  to be over real (\(\pm=+\)) and imaginary (\(\pm=-\)) values respectively.
  This choice of \(x\) contour is deformable to the Lefshetz thimble of the
  \(x=0\) saddle for the integral \labelcref{eq:toysaddle}, for any \(T\) in the
  lower half complex plane; moreover, this choice of \(x\) contour coincides
  with the Lefshetz thimble of the \(x=0\) saddle for the integral in the middle
  expression of \cref{eq:toyexample1inttrans}. Note that, on this \(x\) contour,
  \(H_0\pm \omega\,x^2\) is real and takes its minimal value \(H_0\) at
  \(x=0\).} As expressed in the second equality, this matches the analytic
continuation \labelcref{eq:toysaddleeuc}.

However, ref.~\cite{Marolf:2022ybi} implicitly takes the opposite ordering of
the \(\mathcal{J}\) and \(\mathcal{Q}\) integrals versus the \(T\) integral
transform. Let us see the consequences of this alternative ordering of the
\(T\) and \(x\) integrals in our example.

Similar to the above, as long as \(E_0<H_0\),
\(f_\beta(T)\,z_{\mathrm{L}}^\pm(T)\) is exponentially suppressed at large
negative imaginary \(T\). Moreover, \(z_{\mathrm{L}}^+(T)\) is analytic in the
lower half \(T\)-plane, so we can again use the residue theorem to
evaluate\footnote{\Cref{eq:inttransstable,eq:inttransunstable} are the only
  equations in this paper where we take seriously the written order of
  performing the integral transform in \(T\) after other, either implicit or
  explicit, integrals in the integrand. Elsewhere, \eg{} in \cref{eq:inttrans},
  our convention of performing the \((\mathcal{S},\mathcal{J},\mathcal{Q})\)
  integrals last, as explained below that equation, always prevails.}
\begin{align}
  \int_{-\infty}^\infty \dd{T} f_\beta(T)\, z_{\mathrm{L}}^+(T)
  &= z_{\mathrm{L}}^+(-i\,\beta)
    = z^+(\beta)
    \;,
    \label{eq:inttransstable}
\end{align}
which agrees with the analytic continuation \labelcref{eq:toysaddleeuc}. For
this stable case, the order of integration therefore does not matter in this
simple example.

In the unstable case, \(z_{\mathrm{L}}^-(T)\) possesses a branch cut
in the lower half plane. Choosing this branch cut to be along negative imaginary
\(T\) where the pole of \(f_\beta(T)\) also resides,
\begin{align}
  \int_{-\infty}^\infty \dd{T} f_\beta(T)\, z_{\mathrm{L}}^-(T)
  &=
    \frac{e^{-\frac{1}{\hbar}\,\beta\,E_0}}{\sqrt{\pi}}\,
    \,
    \mathrm{PV} \int_0^\infty \dd{\beta'}
    \frac{
    e^{-\frac{1}{\hbar}\,\beta'\,(H_0 - E_0)}
    }{
    \sqrt{\beta'\,\omega}\, (\beta - \beta')
    }
    \label{eq:inttransunstable}
  \\
  &=
    \frac{2\,e^{-\frac{1}{\hbar}\,\beta\,E_0}}{\sqrt{\beta\,\omega}}
    \,
    F\left(
    \sqrt{\frac{\beta\,(H_0-E_0)}{\hbar}}
    \right)
      \label{eq:inttransdawson}
\end{align}
can be expressed as a principal value (PV) integral along the branch cut. The
result can be written, as in the second equality, in terms of the Dawson \(F\)
function\footnote{Specifically, \(F(y)\) is given by \texttt{DawsonF[y]} in
  Mathematica.} and disagrees with the analytic continuation
\labelcref{eq:toysaddleeuc}. The Dawson \(F\) function has an
asymptotic\footnote{There is a non-perturbative correction \(\sim e^{-y^2}\) to
  the expansion \cref{eq:dawsonexpan} which becomes important when the argument
  has a large imaginary part.} power law expansion at large argument,
\begin{align}
  F(y)
  &= \frac{1}{2\,y}
    + \order{\frac{1}{y^3}}
    \;,
    \label{eq:dawsonexpan}
\end{align}
so the prefactor in \cref{eq:inttransdawson} determines the exponential scaling
in the semiclassical limit \(\hbar\to 0\). Unexpectedly, this is set by the
somewhat arbitrary choice of \(E_0\) rather than \(H_0\) as in
\cref{eq:toysaddleeuc}.

To summarize what we have learned from this example, we see that, for an
unstable saddle, the order of the integral transform \labelcref{eq:inttrans} in
\(T\) versus the integration over \(x\) --- the analogue of
\((\mathcal{S},\mathcal{J},\mathcal{Q})\) --- does matter. In particular, our
prescription of first performing the integral transform in \(T\) while saving
the \(x\)-integral for last leads to the expected analytic continuation
\labelcref{eq:toysaddleeuc} from Lorentzian to Euclidean time. This contribution
to the thermal partition function is weighted exponentially by minus the
``classical Euclidean action'' \(\beta\,H_0\) and has the expected ``one-loop
determinant'' \(\sqrt{\frac{\pi}{\pm\beta\,\omega}}\), though the ambiguous sign
from the choice of branch in the unstable (\(\pm=-\)) case requires a more
complete example to explain in \cref{sec:toyexample2}. In contrast, the opposite
prescription leads to a contribution to the thermal partition function which
nonsensically depends on the parameter \(E_0\) introduced in the integral kernel
\(f_\beta(T)\) in \cref{eq:intkernelexample}. This parameter has no bearing on
the physical system and so should drop out from any physical calculation.
Altogether, we view this toy example as illustrating what can go wrong if the
integral transform in \(T\) is performed before other integrals, over naively
``unstable'' variables in the path integral.

In particular, this explains why the analysis of ref.~\cite{Marolf:2022ybi}
found only the positivity of the specific heat to be relevant for evaluating the
significance of a black hole saddle to the thermal partition function. Instead,
by performing the integrals over \(\mathcal{J}\) and \(\mathcal{Q}\), as well as
\(\mathcal{S}\), after the integral transform in \(T\), we ensure that the
latter does not misbehave when acting on saddles which are unstable with respect
to \((\mathcal{S},\mathcal{J},\mathcal{Q})\). Locally maximizing the integrand
for the final integrals over \((\mathcal{S},\mathcal{J},\mathcal{Q})\) then
leads to a more complete thermodynamic stability condition as described in
\cref{sec:bhstability}.

Before moving on, let us mention that the analyticity of \(z_{\mathrm{L}}^+(T)\)
in the lower half \(T\)-plane in this example is not a general property shared
by integrals over the Lefschetz thimbles of all stable saddles in more
nontrivial examples. In particular, non-analyticity in \(T\) can occur if the
Lefshetz thimble jumps as \(T\) varies across a Stokes ray, as we will see in
\cref{sec:toyexample2}. Because these jumps occur at finite separation from the
saddle, we expect the non-analyticity to appear only beyond the perturbative
expansion around the stable saddle. Nonetheless, if (opposing our general
prescription) we perform the \(T\) integral transform \emph{after} the integral
over the Lefshetz thimbles of these stable saddles, the non-anaylticity in the
lower half \(T\)-plane can lead to unexpectedly large pathologies, similar to
what we saw above for unstable saddles.

In fact, as we will see in \cref{sec:toyexample2} using an example with multiple
saddles for a given integral, the non-analyticities of the contributions from
unstable and stable saddles can cancel against each other. Let us suppose we are
able to \emph{exactly} evaluate the integrals over the Lefshetz thimbles of all
relevant saddles. Then, analyticity in the lower half \(T\)-plane of the sum of
exact contributions would imply that performing the \(T\) integral transform
last still corresponds to taking the expected analytic continuation
\(T\to-i\beta\). However, if we keep only perturbative expansions around each
saddle or if we were to study the contribution of each saddle individually,
then this ordering of the integrals can become problematic, as in the example
studied above.

\subsubsection{A more complete toy example: do unstable saddles contribute?}
\label{sec:toyexample2}

The question of whether and how saddles that are unstable (in the sense
described above) contribute to an integral can subtly depend on definitions. We
will demonstrate this explicitly by studying a slightly more complete toy
example\footnote{Recently, while this paper was in preparation,
  ref.~\cite{Lee:2024hef} appeared, which studied the same kind of toy example
  and others in its appendices.} which includes both stable and unstable
saddles. We will see that ambiguities in specifying the contribution from an
unstable saddle stems from ambiguities in specifying the Lefshetz thimbles for
the stable saddles. Consequently, when these ambiguities are present, unless the
integrals along the latter thimbles are specified to non-perturbative accuracy,
it can sometimes be meaningless to include contributions from the unstable
saddles.

The example we will consider starts with the following ``Lorentzian'' integral,
\begin{align}
  z_{\mathrm{L}}(T)
  &= \int_{\mathfrak{X}}
    \frac{\dd{x}}{\sqrt{\hbar}}\,
    e^{
    i\, I(x)
    }
    \;,
  &
    I(x)
    &= -\frac{1}{\hbar}\,T\,\omega\,\left(
    -\frac{x^2}{2} + \frac{x^4}{4}
      \right)
      \;,
    \label{eq:toyexample2}
\end{align}
where \(\omega>0\) is again a constant introduced for dimensional consistency
and, for real \(T\), the contour of integration is \(\mathfrak{X}=\mathbb{R}\)
(very slightly rotated to ensure convergence). We view this integral as a toy
analogue for \(Z_{\mathrm{L}}(T,\Omega,\Phi)\) where again \(x\) can represent
any of \((\mathcal{S},\mathcal{J},\mathcal{Q})\).

\Cref{eq:toyexample2} is a slightly richer toy example compared to the Gaussian
example of \cref{sec:toyexample1}, because we now have three saddle points, at
\(x=-1,0,1\). But near each saddle, the integrand of \cref{eq:toyexample2} is
still approximated by that of the Gaussian example \labelcref{eq:toysaddle} (for
possibly differing \(\omega\)), so we can again classify the saddles by their
stability. In the language of \cref{sec:toyexample1}, the \(x=-1,1\) saddles
are stable while the \(x=0\) saddle is unstable --- again, this notion of
stability will become more intuitive once we pass from \(T\) to thermal time
\(\beta\) below in \cref{eq:toyexample2euc}.

The main question which we would like to understand with this toy example is
whether the saddles \(x=-1,0,1\) contribute to the integral
\(z_{\mathrm{L}}(T)\) and, more importantly, its ``Euclidean'' analogue
\labelcref{eq:toyexample2euc} below which represents the thermal partition
function. Let us therefore give a more precise definition for what counts as a
contributing saddle. As briefly reviewed around \cref{eq:lefschetz}, the contour
\(\mathfrak{X}\) chosen for an integral, \eg{} \(\mathfrak{X}=\mathbb{R}\) for
\cref{eq:toyexample2}, can be deformed into a linear combination
\(\sum_p\mathfrak{n}_p\,\mathfrak{J}_p\) of Lefschetz thimbles. We shall say
that a given saddle \(p\) contributes to the integral if the saddle's
corresponding Lefschetz thimble \(\mathfrak{J}_p\) appears in this linear
combination with nontrivial coefficient \(\mathfrak{n}_p\ne 0\).

\begin{figure}
  \centering
  \includegraphics{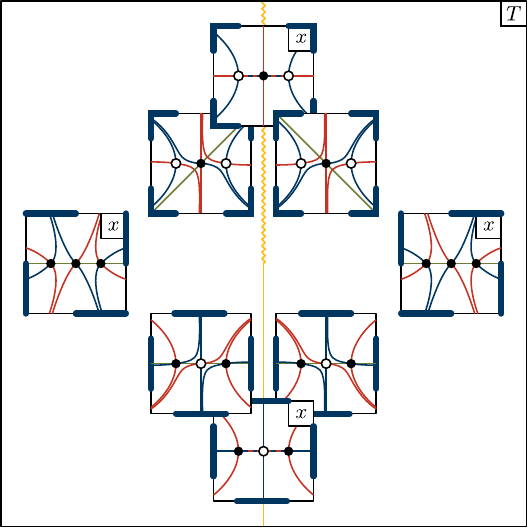}
  \caption{Pertinent contours for the integral \(z_{\mathrm{L}}(T)\) given in
    \cref{eq:toyexample2}. This figure illustrates the complex \(x\) plane
    ``fibred'' over the complex \(T\) plane: negative and positive \(T\)
    correspond to the left- and right-most panels while imaginary \(T\)
    correspond to the top- and bottom-most panels; additional panels showing
    limits approaching the imaginary \(T\)-axis from the left and right are
    positioned accordingly. \textbf{For the \(x\)-plane in each panel:} Saddle
    points for the action \(I(x)\) are illustrated by dots, which are solid if
    \(\Re(i\, I(x))\) is maximized among the three saddle points. Lefshetz
    thimbles (\ie{} contours of steepest descent) and contours of steepest
    ascent for the Morse function \(\Re(i\, I(x))\) are respectively illustrated
    by dark blue and red curves. All contours that give convergent integrals,
    \eg{} the Lefshetz thimbles, must be in the relative homology of the
    asymptotic regions indicated by thick dark blue lines where
    \(\Re(i\,I(x))\to-\infty\). The green line indicates the contour of
    integration \(\mathfrak{X}\) defining \(z_{\mathrm{L}}(T)\). \textbf{In the
      \(T\)-plane:} Stokes rays lie on the imaginary \(T\) axis, as highlighted
    in yellow. The function \(z_{\mathrm{L}}(T)\), with the aforementioned
    choice of integration contours, has a branch cut in the upper half
    \(T\)-plane, say along the positive imaginary axis as shown by a zigzag;
    otherwise, \(z_{\mathrm{L}}(T)\) is analytic in \(T\).}
  \label{fig:stokes}
\end{figure}

In the ``Lorentzian'' case of real \(T\), all three saddles contribute. This
follows from the fact that, at real \(T\), the integral
\labelcref{eq:toyexample2} defining \(z_{\mathrm{L}}(T)\) is purely oscillatory
on \(\mathfrak{X}=\mathbb{R}\), so the same argument given in the paragraph
following \cref{eq:lefschetz} applies and says that \(\mathfrak{X}\) has
intersection number \(\mathfrak{n}_p=\pm 1\) with the contours of steepest
ascent \(\mathfrak{K}_p\) for saddles \(p\in\mathfrak{X}\). For our relatively
simple example, we can be even more explicit and illustrate all pertinent
contours in \cref{fig:stokes}. The left- and right-most panels here are relevant
for the case of real \(T\) currently under consideration. It is clear that the
integration contour \(\mathfrak{X}=\mathbb{R}\) (green line) intersects each
ascent contour \(\mathfrak{K}_p\) (red curve) once and can be deformed into a
sum of all Lefshetz thimbles \(\mathfrak{J}_p\) (dark blue curves).\footnote{The
  sign of \(\mathfrak{n}_p\) just corresponds to how one chooses to define the
  orientations of \(\mathfrak{J}_p\) and \(\mathfrak{K}_p\).}

The analogue of a thermal partition function in this simple example is
\begin{align}
  z(\beta)
  &\equiv
    \int
    \frac{\dd{x}}{\sqrt{\hbar}}\,
    e^{
    -\frac{1}{\hbar}\,\beta\,
    \omega\,\left(
    -\frac{x^2}{2} + \frac{x^4}{4}
      \right)
    }
    = z_{\mathrm{L}}(-i\,\beta)
    \;,
    \label{eq:toyexample2euc}
\end{align}
obtained by analytically continuing the integrand in the defining integral
\labelcref{eq:toyexample2}. As illustrated in \cref{fig:stokes}, we can continue
to make the choice \(\mathfrak{X}=\mathbb{R}\) of integration contour everywhere
in the lower half \(T\)-plane --- in particular, the contour
\(\mathfrak{X}=\mathbb{R}\) connects asymptotic regions in the \(x\)-plane where
\(\Re(i\,I(x))\to -\infty\). Because \(z_{\mathrm{L}}(T)\) is merely a
(branched) function of \(T\), we also have the option to analytically continue
\(z_{\mathrm{L}}(T)\) to negative imaginary \(T\) after performing the
\(x\)-integral like in \cref{sec:toyexample1}.

In reality though, the Lorentzian gravitational path integral
\(Z_{\mathrm{L}}(T,\Omega,\Phi)\) is not a function but a distribution in real
\(T\) as described below \cref{eq:lorpartfunc}. The thermal partition function
\(Z_{\mathrm{L}}(T,\Omega,\Phi)\) is therefore obtained by an integral transform
\labelcref{eq:inttrans} rather by a naive analytic continuation. However,
granted we perform integrals in the correct order, we expect this integral
transform to amount to an analytic continuation of the \emph{integrand} before
the \((\mathcal{S},\mathcal{J},\mathcal{Q})\) integrals, \eg{} taking
\cref{eq:lorbhpartfunc} to \cref{eq:bhpartfunc}. This is certainly true in both
the simpler toy example of \cref{sec:toyexample1} and the current toy example:
applying the integral transform \labelcref{eq:inttrans} before the
\(x\)-integral, as per our prescription, simply has the effect of analytically
continuing the \(x\)-integrand to \(T=-i\,\beta\). It just so happens in these
toy examples that we can equivalently perform the analytic continuation after
the \(x\) integral as well.\footnote{In the current example, due to the
  analyticity of \(z_{\mathrm{L}}(T)\) in the lower half \(T\)-plane, it is
  possible also to perform the \(T\) integral transform after the \(x\)
  integral, as mentioned at the end of \cref{sec:toyexample1}. But again, this
  can be dangerous if the \(x\) integral is evaluated approximately using
  saddle-point methods.}

Let us now work towards answering the question of whether the saddles
\(x=-1,0,1\) contribute to \(z(\beta)\). To this end, we consider again the
Lefschetz thimbles \(\mathfrak{J}_p\) and steepest ascent contours
\(\mathfrak{K}_p\) for each saddle point \(p\). As illustrated in the
bottom-most panel of \cref{fig:stokes}, the Lefshetz thimble of the \(x=0\)
saddle lies on the imaginary \(x\) axis, while those of the \(x=-1,1\) saddles
include the negative and positive real \(x\) axes respectively. In particular,
the latter thimbles connect the \(x=-1,1\) saddles to the \(x=0\) saddle, but it
then becomes somewhat ambiguous whether and in which direction one should extend
these thimbles after that point. Correspondingly, the steepest ascent contour
\(\mathfrak{K}_{x=0}\) for \(x=0\) has an ambiguous intersection number with the
original integration contour \(\mathfrak{X}=\mathbb{R}\). This situation where
multiple saddle points are connected by Lefshetz thimbles is known as a Stokes
phenomenon\footnote{We refer the reader to \cite{Witten:2010cx} for an
  accessible review of Morse theory including a discussion of Stokes phenomena.}
and the aforementioned ambiguities are closely tied to the question of whether
and with what sign the ``unstable'' saddle \(x=0\) contributes to \(z(\beta)\).

To better appreciate this phenomenon, let us move slightly off the so-called
Stokes ray --- here, the negative imaginary \(T\) axis --- and consider limits
where we approach this ray from the left and right. This is illustrated in the
two panels closely straddling but not directly on the negative imaginary \(T\)
axis in \cref{fig:stokes}. Decomposing \(\mathfrak{X}=\mathbb{R}\) into the
Lefshetz thimbles, going between the two panels, it is clear that the
contribution to \(z(\beta)\) from the Lefschetz thimble of \(x=0\) flips sign,
as the direction of integration along the imaginary \(T\) axis reverses.
Additionally, we see that pieces of the Lefschetz thimbles for the \(x=-1,1\)
saddles jump discontinuously from positive to negative imaginary \(T\) and vice
versa. In total, the sign flip in the contribution from the \(x=0\) saddle and
the discontinuous jumps in the Lefshetz thimbles of the \(x=-1,1\) saddles
cancel to give a function \(z_{\mathrm{L}}(T)\) that is analytic across the
negative imaginary \(T\) axis, so the value of the thermal partition function
\(z(\beta)=z_{\mathrm{L}}(-i\,\beta)\) remains unambiguous.

We are now in a position to give an answer to the question of whether various
saddles contribute to the thermal partition function. This answer is based
purely on our simple toy example --- we leave it for future work to generalize
the qualitative lessons learned here to more realistic examples arising from the
gravitational path integral. ``Stable'' saddles, in this example \(x=-1,1\),
unambiguously contribute as their Lefshetz thimbles \(\mathfrak{J}_p\) are
included with unambiguous coefficients in the decomposition
\(\mathfrak{X}=\sum_p \mathfrak{n}_p\,\mathfrak{J}_p\) of the contour of
integration. There is nonetheless a question of whether one wants to extend
thimbles after they run into an ``unstable'' saddle, in this example \(x=0\),
when we have a Stokes phenomenon. If not, then we should not include the
Lefschetz thimble of the unstable saddle. If we choose to extend the Lefshetz
thimbles of stable saddles after they reach unstable saddles, then, as in this
example, there might be an ambiguity in which direction they should be extended.
However, these ambiguities are to be exactly cancelled by the equally ambiguous
contribution from the Lefshetz thimble of the unstable saddle.

In particular, in situations where we only have perturbative knowledge of the
integrals along the Lefschetz thimbles of the stable saddles, it might seem
pointless to include any contributions from the Lefshetz thimbles of the
unstable saddle. After all, it is not immediately obvious how the perturbative
expansion of the integral along the Lefshetz thimble of a stable saddle can know
about the choice of extension, if any, of this Lefshetz thimble at finite
separation from the saddle. Even if we wanted to include the non-perturbatively
small correction from the unstable saddle, would we even know what sign to
assign to it?

However, let us again emphasize that further study is required to generalize the
analysis carried out here for a very simple toy example, in particular involving
a \emph{one-dimensional} integral. For example, ref.~\cite{Lee:2024hef}
advocates in certain situations for the inclusion of saddles with an even number
of unstable variables. In particular, on a Stokes ray, ref.~\cite{Lee:2024hef}
suggests taking an average from both sides of the ray. With this prescription,
it might be possible to define an unambiguous nonzero coefficient
\(\mathfrak{n}_p\) for a saddle \(p\) with an even number of unstable variables.
Moreover, ref.~\cite{Lee:2024hef} provides examples where the contributions from
unstable saddles, unambiguously so defined, can provide sizable corrections to
the (potentially asymptotic, optimally truncated) perturbative expansions around
stable saddles. We will leave further study of these ideas for future work.

\subsection{Equivalent boundary conditions, inequivalent singularities, and
  black hole sums}
\label{sec:bhsums}

On the topic of saddles, another interesting feature worth discussing is the sum
over constrained saddles constructed from black holes --- see
\cref{eq:allbhpartfunc,eq:lorallbhpartfunc}. Around \cref{eq:diraccomb}, we
already explained how this sum gives rise to a discrete spectrum for angular
momentum and charge. Let us review how these sums came about from identifying
equivalent boundary conditions and distinguishing inequivalent ``internal''
structures of helical and holonomic singularities. Below, we will also relate
one of the integer sums in \cref{eq:allbhpartfunc} to the sum over an
integer-parameter subset of the \(\mathrm{SL}(2,\mathbb{Z})\) black holes in
AdS\(_3\) \cite{Maloney:2007ud}.

The Euclidean and Lorentzian path integrals studied in
\cref{sec:partfunc,sec:lorpartfunc} were in part specified by supplying grand
canonical boundary conditions at the spacetime boundary
\(\partial\mathscr{M}=Y_{\mathrm{space}}\times S^1_{\mathrm{time}}\),
parameterized by an inverse temperature \(\beta\) in the Euclidean case or a
Lorentzian time period \(T\), a(n angular) velocity \(\Omega\), and an electric
potential \(\Phi\).\footnote{Recall that the \(S^1_{\mathrm{time}}\) is the
  orbit of the co-rotating Killing vector given in \cref{eq:corotvec} or
  \cref{eq:lorcorotvec}, and is not metric-orthogonal to \(Y_{\mathrm{space}}\)
  when \(\Omega\ne 0\).} In particular, we noted that it is natural to identify
different values of \(\Omega\) and \(\Phi\) as equivalent if they related by
discrete increments which are finite for the case of rotation or a compact
Maxwell gauge group. This discrete identification of \(\Omega\), given in
\cref{eq:periodangvel,eq:lorperiodangvel}, arose from the observation that a
full rotation or a double rotation ought to act trivially --- if a dual boundary
theory exists, then the distinction corresponds to whether the dual theory has
fermions that are anti-periodically identified under the rotation. Similarly,
a full rotation around the Maxwell gauge group should be tantamount to the
identity, motivating the identification of \(\Phi\) given in
\cref{eq:periodelepot,eq:lorperiodelepot}.

One might wonder whether there are further identifications that should be made
between naively distinct boundary conditions. While we have been vague about the
precise spacetime asymptotics at \(\partial\mathscr{M}\) we are considering,
with an eye towards AdS/CFT, one setting we are particularly interested in is
when the cosmological constant is negative and the spacetime is asymptotically
locally AdS. In this case, it is natural to identify boundary conditions related
by all modular transformations for the conformal boundary (preserving spin
structure, if present). For the case of \(D=3\) bulk spacetime dimensions, a
(Euclidean\footnote{In relation to the parameters \(\beta\) and \(\Omega\)
  specifying Euclidean boundary conditions as described in
  \cref{sec:pathintegral},
  \begin{align}
    \tau=i\beta\,\left( \frac{1}{2\pi}-\frac{\Omega}{\mathrm{Period}(\varphi)} \right)
  \end{align}
  where we recall that \(\Omega\) is imaginary in order for the boundary to be
  Euclidean.}) boundary torus \(\partial\mathscr{M}=S^1_{\mathrm{space}}\times
S^1_{\mathrm{time}}\) can be specified in the usual way by identifying a complex
plane under shifts by \(1\) and by a modular parameter \(\tau\) in the upper
half plane --- specifically, the line segments from \(0\) to \(1\) and \(\tau\)
respectively correspond to the \(S^1_{\mathrm{space}}\) and
\(S^1_{\mathrm{time}}\) cycles. The modular transformations are then given by
\begin{align}
  \tau
  &\mapsto \frac{a\,\tau+b}{c\,\tau+d}
    \;,
  &
    \begin{pmatrix}
    a & b \\
    c & d
    \end{pmatrix}
  &\in \mathrm{PSL}(2,\mathbb{Z})
    \;,
    \label{eq:modularaction}
\end{align}
Our identification \labelcref{eq:periodangvel} on \(\Omega\) is generated
by\footnote{This is the usual ``\(T\)'' element of
  \(\mathrm{PSL}(2,\mathbb{Z})\) and \(\mathscr{S}\) in \cref{eq:btz} below is
  the usual ``\(S\)''; we use script symbols, because \(T\) already refers to a
  Lorentzian time period in this paper.}
\begin{align}
  \mathscr{T}
  &= \begin{pmatrix}
    1 & 1 \\
    0 & 1
  \end{pmatrix}
    \in \mathrm{PSL}(2,\mathbb{Z})
\end{align}
if states come back to themselves after one full rotation, or \(\mathscr{T}^2\) if two
rotations are required.

In \cref{sec:bhsaddles,sec:lorsaddles}, we saw how the identifications in
\(\Omega\) and \(\Phi\) naturally give rise to a discrete set of constrained
saddles, labelled by two integers \(m,n\in\mathbb{Z}\), at fixed \((\beta\text{
  or }T,\Omega,\Phi;\mathcal{S},\mathcal{J},\mathcal{Q})\). These constrained
saddles are actually diffeomorphic and gauge equivalent to each other \emph{if}
we cut out the singular surface \(\gamma\). However, as described in
\cref{sec:singdisambig}, the helical and holonomic singularities in these
constrained saddles are distinguished by the way in which a small neighbourhood
\(\mathscr{N}_\varepsilon\) of \(\gamma\) is smoothly filled in by the regulated
versions of these configurations. For example, the constrained saddles labelled
by different \(m\) have different combinations of the boundary cycles which are
contractible in the (regulated\footnote{The notion of contractibility might seem
  ambiguous in configurations where the helical singularity can prevent cycles
  from contracting to zero proper size. To precisely define contractibility, we
  therefore refer to regulated versions of such configurations where the helical
  singularity has been smoothed out over a small neighbourhood
  \(\mathscr{N}_\varepsilon\) --- see \cref{sec:regsing}.}) bulk. Specifically,
if we choose the \(m=0\) constrained saddle to be one in which
\(S^1_{\mathrm{time}}\) is contractible, then more general \(m\) will have
contractible cycles \(m\,S^1_{\mathrm{space}}+S^1_{\mathrm{time}}\) or
\(2m\,S^1_{\mathrm{space}}+S^1_{\mathrm{time}}\) respectively if boundaries
\(\partial\mathscr{M}\) related by \(\mathscr{T}\) or \(\mathscr{T}^2\) are
deemed equivalent. Thus, the internal structures of these helical singularities
are physically inequivalent and such distinct configurations should be summed
over in the path integral. Similar comments can be made about constrained
saddles with different \(n\) and different contractible cycles in the Maxwell
principal fibre bundle.

Let us now connect this discussion back to previous work \cite{Maloney:2007ud}
which considered sums over \emph{smooth} \(\mathrm{SL}(2,\mathbb{Z})\) black holes. As
explained in \cref{sec:bhstability}, smooth Euclidean black holes, in an
appropriate sense, appear as saddles for the final
\((\mathcal{S},\mathcal{J},\mathcal{Q})\) integrals in the thermal partition
function. Specifically, for a given Lorentzian or Euclidean constrained saddle
labelled by \(m\) and \(n\), the saddle-point values of
\((\mathcal{S},\mathcal{J},\mathcal{Q})\) (which can depend on \(m\) and \(n\))
are precisely those for which the helical and holonomic singularities, as well
as the Euclidean conical singularity, vanish. However, as we tune
\((\mathcal{S},\mathcal{J},\mathcal{Q})\) to their saddle-point values, the
topology of each constrained saddle does not change; in particular, the smooth
Euclidean saddles with different \(m\) must continue to have different boundary
cycles \(m\,S^1_{\mathrm{space}}+S^1_{\mathrm{time}}\) or
\(2m\,S^1_{\mathrm{space}}+S^1_{\mathrm{time}}\) that are contractible.

We can now identify these smooth Euclidean saddles as an integer-parameter
subset of the \(\mathrm{SL}(2,\mathbb{Z})\) black holes in AdS\(_3\)
\cite{Maloney:2007ud}. Let us recall that these latter geometries
\(\mathscr{M}_{c,d}\) are labelled by two coprime integers \(c\ge 0\) and \(d\).
In particular, \(\mathscr{M}_{0,1}(\tau)\) is the thermal but otherwise empty
AdS\(_3\) saddle, described in \cref{sec:emptysaddle}, with a contractible
\(S_{\mathrm{space}}^1\). The other \(\mathscr{M}_{c,d}(\tau)\) can be
constructed from \(\mathscr{M}_{0,1}\) using \cref{eq:modularaction},
\begin{align}
  \mathscr{M}_{c,d}(\tau)
  &= \mathscr{M}_{0,1}\left(\frac{a\,\tau+b}{c\,\tau+d}\right)
    \;.
\end{align}
(The RHS depends only on \((c,d)\), because we require \(ad-bc=1\) in
\cref{eq:modularaction} and because
\(\mathscr{M}_{0,1}(\tau)=\mathscr{M}_{0,1}(\mathscr{T}(\tau))\).) A notable
example is the BTZ black hole
\begin{align}
  \mathscr{M}_{1,0}(\tau)
  &=\mathscr{M}_{0,1}(\mathscr{S}(\tau))
    \;,
    &
    \mathscr{S}
  &= \begin{pmatrix}
    0 & -1 \\
    1 & 0
  \end{pmatrix}
    \in \mathrm{PSL}(2,\mathbb{Z})
    \label{eq:btz}
\end{align}
where the time circle \(S_{\mathrm{time}}^1\) is contractible. More generally,
the contractible cycle in \(\mathscr{M}_{c,d}(\tau)\) is given by
\(c\,S^1_{\mathrm{time}}+d\,S^1_{\mathrm{space}}\). We therefore expect our
Euclidean saddles labelled by \(m\) to correspond to the geometries
\begin{align}
  \mathscr{M}_{1,m}(\tau)
  &= \mathscr{M}_{0,1}\left(
    \mathscr{S}\circ\mathscr{T}^m(\tau)
    \right)
\end{align}
or \(\mathscr{M}_{1,2m}(\tau)\) respectively if boundaries
\(\partial\mathscr{M}\) related by \(\mathscr{T}\) or \(\mathscr{T}^2\) are
deemed equivalent.

Let us remark that our sum over inequivalent helical singularities bears close
resemblance to the sum over different ``KK instantons'' (which we call holonomic
singularities) in \(D=2\) dimensions in ref.~\cite{Maxfield:2020ale}. However,
whereas our helical singularities are fundamentally singular, the (single
insertions of) KK instantons are the dimensional reductions of the smooth
bifurcation surfaces in \(\mathrm{SL}(2,\mathbb{Z})\) black holes. As described
above, the sum over a subset of \(\mathrm{SL}(2,\mathbb{Z})\) black holes (and
thus over a subset of single KK instanton insertions) corresponds to taking the
on-shell values of \((\mathcal{S},\mathcal{J},\mathcal{Q})\) in our sum over
otherwise singular constrained saddles.

The above considerations naturally lead to some further open questions. Firstly,
how might the other \(\mathrm{SL}(2,\mathbb{Z})\) black holes arise as saddles
for the thermal partition function, starting from singular constrained saddles
of a fixed-\((\mathcal{S},\mathcal{J},\mathcal{Q})\) path integral? Relatedly,
one might also ask whether it is possible to recover the other
\(\mathrm{SL}(2,\mathbb{Z})\) black holes from a purely Lorentzian starting
point. Once we understand whether and how each of these
\(\mathrm{SL}(2,\mathbb{Z})\) black holes arise from a Lorentzian starting
point, we can turn to the question of whether the Lefschetz thimble of each
saddle really contributes to the path integral as described in
\cref{sec:bhstability,sec:toyexamples}.\footnote{See
  \cref{foot:complexsaddlesmn}.} We will leave these questions largely for
future study.

\begin{figure}
  \centering
  \begin{subfigure}[t]{0.475\textwidth}
    \centering
    \includegraphics{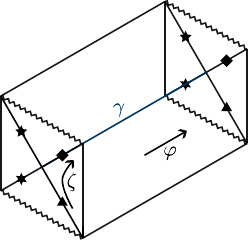}
    \caption{A BTZ black hole. Points indicated by like shapes are identified
      along the spatial circle direction running between the front and back
      faces (where the Penrose diagram is drawn). Time translation, \ie{} boosts
      around the bifurcation surface \(\gamma\), are generated by the
      \emph{non-rotating} Killing vector \(\zeta=\partial_{\hat{t}}\). Rotation
      along the spatial circle is generated by the Killing vector \(\varphi\).
      (These Killing vectors were originally defined in \cref{sec:lorpathint} on
      the spacetime boundary \(\partial\mathscr{M}\), but have obvious bulk
      extensions.) For simplicity let us take this BTZ black hole to be
      non-rotating, \(\Omega=0\).}
    \label{fig:btzbh}
  \end{subfigure}
  \hfill
  \begin{subfigure}[t]{0.475\textwidth}
    \centering
    \includegraphics{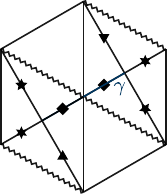}
    \caption{A quotient of the BTZ black hole gives the CRT-twisted black hole
      \cite{Harlow:2023hjb}. The halved depth of this figure relative to
      (\subref{fig:btzbh}) is intentional. Notice that the front and back faces
      are identified after a twist around the bifurcation
      surface \(\gamma\). Despite appearances, there is only one connected
      asymptotic region.}
    \label{fig:crtbh}
  \end{subfigure}
  \par\bigskip
  \begin{subfigure}[t]{\textwidth}
    \centering
    \includegraphics{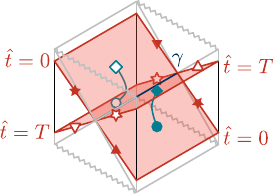}
    \caption{A constrained saddle with a conical singularity on the bifurcation
      surface \(\gamma\) resulting from fixing area, \ie{} the
      Bekenstein-Hawking entropy \(\mathcal{S}\). Points marked by like shapes,
      respectively filled and empty, were already identified in the CRT-twisted
      black hole (\subref{fig:crtbh}). We now also identify the \(\hat{t}=0\)
      and \(\hat{t}=T\) red surfaces, such that all like shapes are identified,
      irrespective of filling. The spacetime of the constrained saddle lies in
      between these time slices and satisfies boundary conditions with
      \(T\Omega=\mathrm{Period}(\varphi)/2\). Away from \(\gamma\), this
      configuration is diffeomorphic to the one shown in
      \cref{fig:lorhelicalbh}. However, note the following differences, which we
      expect to become sharp when the singularity on \(\gamma\) is regulated ---
      see \cref{sec:regsing}. Firstly, \(\gamma\) here contains half as many
      points as in \cref{fig:lorhelicalbh}. Secondly, the contractible cycles
      shown in teal here and in \cref{fig:lorhelicalbh} differ. In particular,
      the teal cycle here is a bulk orbit of \emph{non-rotating} time
      translation \(\zeta\) and is homologous to the boundary cycle
      \(2S_{\mathrm{time}}^1-S_{\mathrm{space}}\), where \(S_{\mathrm{time}}^1\)
      is the orbit of \(\xi=\zeta+\Omega\,\varphi\) \emph{rotating} with angular
      velocity \(\Omega\). Thirdly, the hyperbolic opening angle around
      \(\gamma\) here is double that of \cref{fig:lorhelicalbh} and there is no
      helical singularity on \(\gamma\) here. More generically, one can also fix
      an \emph{off-shell} value of the angular momentum \(\mathcal{J}\)
      evaluated on \(\gamma\) here, which will lead to a helical singularity but
      still with a strength differing from the construction in
      \cref{fig:lorhelicalbh}. In fact, as before, we expect to have an
      integer-parameter family of constrained saddles with inequivalent helical
      singularities on \(\gamma\).}
    \label{fig:conicalcrtbh}
  \end{subfigure}
  \caption{A BTZ black hole, a CRT-twisted black hole \cite{Harlow:2023hjb}, and
    a constrained saddle constructed from the latter.}
  \label{fig:crtsaddle}
\end{figure}

However, let us sketch in \cref{fig:crtsaddle} an example which inspires some
hope that our construction of singular constrained saddles might also apply to
these more general geometries. For simplicity, we have assumed in the preceding
discussion that the constructions of constrained saddles in
\cref{sec:bhsaddles,sec:lorsaddles} start with a standard black hole,
specifically with a contractible \(S_{\mathrm{time}}^1\) if in Euclidean
signature. However, one can consider other possibilities. In
\cref{fig:crtsaddle}, we instead start with a Lorentzian CRT-twisted black hole
\cite{Harlow:2023hjb} which is a quotient of the standard BTZ black hole. The
resulting constrained saddle in \cref{fig:conicalcrtbh} is a Lorentzian geometry
in which the boundary cycle \(2S_{\mathrm{time}}^1-S_{\mathrm{space}}\) is
contractible in the bulk. Indeed, as refs.~\cite{Chen:2023mbc,Grabovsky:2024vnb}
point out, the Euclidean counterpart of the CRT-twisted black hole is a
Euclidean \(\mathrm{SL}(2,\mathbb{Z})\) black hole \(\mathscr{M}_{2,-1}\), which
has a contractible \(2S_{\mathrm{time}}^1-S_{\mathrm{space}}\) cycle. Running
through the same analysis as in \cref{sec:bhsaddles} and discussed above, we
might then expect to generate another integer-parameter subset of
\(\mathrm{SL}(2,\mathbb{Z})\) black holes, perhaps \(\mathscr{M}_{2,2m-1}\) or
\(\mathscr{M}_{2,4m-1}\), as saddles of the
\((\mathcal{S},\mathcal{J},\mathcal{Q})\) integrals.

\subsection{Other open problems and future directions}
\label{sec:openproblems}

Let us conclude with a discussion of some additional open problems and possible
avenues for future work.

\subsubsection{Subtleties of our new singularities and their action}

We started this paper by specifying in \cref{sec:singconfigs} what we meant by
conical, helical, and holonomic singularities on a codimension-two surface
\(\gamma\) in a Euclidean spacetime. In particular, we described the strengths
of these singularities in terms of three respective parameters
\((\kappa,v^i,\mu)\). As summarized in \cref{sec:stationarity}, we placed
various restrictions on how we allow these singularity strengths to vary around
and along \(\gamma\). While this removed some subtleties related to the cutoff
surface \(\partial\mathscr{N}_\varepsilon\) appearing in the action of singular
configurations, these ad hoc restrictions were not really based on physical
motivations. This became a poignant issue, for example, in the Lorentzian
discussion of \cref{sec:lightcones}, where we realized that a helical shift
\(v^i\) which persists for all boost times \(\tau\) around \(\gamma\) will
generically give rise to singularities on the lightcone emanating \(\gamma\).

Should we allow these lightcone singularities? Relatedly, should we allow the
singularity parameters \((\kappa,v^i,\mu)\) to vary with respect to \(\tau\)
and/or along the surface \(\gamma\)? We leave these questions for future work.
To answer these questions, one might attempt an analysis analogous to the
appendices of ref.~\cite{Dong:2019piw} and try to determine the most general
form of \((\kappa,v^i,\mu)\) and the near-\(\gamma\) expansion of the metric and
Maxwell field which would admit a solution to the Einstein-Maxwell equations
away from \(\gamma\).

In \cref{sec:action}, we wrote down an action \labelcref{eq:action} for
configurations with conical, helical, and holonomic singularities, motivated by
the study in \cref{sec:curvature} of curvatures in regulated configurations. In
these regulated configurations, the singularity on \(\gamma\) is smoothed out
over an \(\varepsilon\)-neighbourhood \(\mathscr{N}_\varepsilon\) of \(\gamma\).
In the limit where \(\varepsilon\to 0\), contact terms in the Ricci curvature
and Maxwell field strength are equated to an area term on \(\gamma\) and terms
on the cutoff surface \(\partial\mathscr{N}_\varepsilon\), as displayed in the
action \cref{eq:action}.

This derivation, however, was somewhat ad hoc because we intentionally dropped
terms which, in the \(\varepsilon\to 0\) limit, amount to ill-defined squared
\(\delta\)-functions on \(\gamma\) with coefficients quadratic in the
singularity strengths \(v^i\) and \(\mu^i\). As sketched below
\cref{eq:concontact2}, this situation is quite analogous to having conical
singularities in higher curvature theories \cite{Dong:2013qoa,Dong:2019piw} ---
indeed, the Maxwell action is, in a sense, a curvature-squared term. For states
of fixed geometric entropy in higher curvature theories,
ref.~\cite{Dong:2019piw} showed that the appropriate action for the conical
singularities \(\gamma\) is given by the geometric entropy \(\sigma\) of
\(\gamma\) times the singularity strength. Roughly speaking, this corresponds to
isolating contributions to the action, coming from a neighbourhood
\(\mathscr{N}_\varepsilon\) of \(\gamma\) in a regulated configuration, which
are \emph{linear} in the singularity strength \cite{Dong:2013qoa}. Our
derivation of the action \labelcref{eq:action} is essentially a naive
re-enactment of this procedure for helical and holonomic singularities. Of
course, ref.~\cite{Dong:2019piw} justified their proposal for the action of
fixed geometric entropy states, by showing that it leads to a good
fixed-\(\sigma\) variational principle with a careful derivation of the
near-\(\gamma\) asymptotics of (constrained) solutions. In \cref{sec:saddle}, we
similarly studied the variation of the action at fixed area, (angular)
momentum, and charge on \(\gamma\); however, as already mentioned, it still
remains to show that the near-\(\gamma\) asymptotics of Einstein-Maxwell
constrained solutions indeed take the form we have assumed.

Another subtlety of the action \labelcref{eq:action} worth further investigation
concerns the cutoff surface \(\partial\mathscr{N}_\varepsilon\) on which some of
its terms live. As explained in \cref{sec:shapeindep}, the profile of the cutoff
surface \(\partial\mathscr{N}_\varepsilon\) cannot be chosen arbitrarily without
changing the value of the action, unless some restrictions are made on how the
singularity parameters \((\kappa,v^i,\mu)\) vary around and along \(\gamma\). In
particular, the cutoff surface \(\partial\mathscr{N}_\varepsilon\) prescribed by
our derivation of the action is one which lies at constant proper separation (of
order \(\varepsilon\)) from \(\gamma\). As described in \cref{sec:lightcones},
this becomes particularly concerning when the action is continued to Lorentzian
signature, because such a cutoff surface \(\partial\mathscr{N}_\varepsilon\) can
now run affinely far away along any lightcones that \(\gamma\) might possess. Is
this a bug or a feature? In \cref{sec:lightcones}, we suggested that perhaps
this might be a feature of the action which accounts for possible lightcone
singularities. However, this claim clearly requires more careful justification.

Given the ad hoc nature of our derivation and the peculiarities of our action,
one might wonder if there is a more elegant language with which to quantify
helical singularities and their action. In earlier work where helical
singularities arose from the backreaction of spinning particle worldlines, it
was natural to describe these singularities in terms of torsion
\cite{Tod:1994iop}. Just as the localized stress energy of a particle worldline
(or brane world-volume in \(D>3\) spacetime dimensions) gives rise to a
distributional curvature recognized as a conical singularity, the localized spin
density of a spinning particle imprints a distributional torsion through
Einstein-Cartan equations of motion. Just as curvature quantifies rotational
holonomy, \eg{} conical deficit or excess, torsion similarly quantifies a
translational holonomy \cite{Petti:1986spn}, \eg{} the helical shift \(v^i\)
around a helical singularity \cite{Tod:1994iop}. Optimistically, one may
therefore expect that helical singularities and their action can be described
equally cleanly in Einstein-Cartan theory as conical singularities are in
Einstein-Hilbert theory. Even having obtained such a description, however, the
question still remains as to what it might teach us about torsion-free
gravitational theories.

\subsubsection{Higher dimensions, angular momentum, and nonconstant modes on \(\gamma\)}
\label{sec:angmom}

We expect that the majority of the formalism developed in this paper applies to
spacetimes of any dimension \(D\ge 3\). At certain points in this paper, we have
focused on \(D=3\) for simplicity, so it may be worthwhile now to review some of
the challenges that might arise in \(D>3\) --- these include divergences in the
action and difficulty defining angular momentum \(\mathcal{J}\) on a
codimension-two surface \(\gamma\). Discussion of the latter will then invite
the consideration of quantities other than
\((\mathcal{S},\mathcal{J},\mathcal{Q})\) on \(\gamma\), somewhat reminiscent of
non-constant edge modes.

Let start with the action. One new feature that appears in higher dimensions
\(D>3\) is a term \labelcref{eq:regdepterm} in the Einstein-Hilbert action
\(\int_{\mathscr{M}\setminus\mathscr{N}_\varepsilon} R\) outside an
\(\varepsilon\)-neighbourhood \(\mathscr{N}_\varepsilon\) of a helically
singular surface \(\gamma\). In \(D=3\), this term vanishes because the
quadratic scalar \labelcref{eq:divergentbulk} built from
\(\mathcal{L}_{v}h_{ij}\) is identically zero, for any metric \(h_{ij}\) on
\(\gamma\) and helical shift \(v^i\). This is no longer true generically in
\(D>3\) and the contribution \labelcref{eq:regdepterm} to the Eistein-Hilbert
action in fact diverges as \(\varepsilon\to 0\). How should we treat this
divergence? Should a counterterm be introduced to cancel it, like in the
treatment of conical singularities in higher curvature theories
\cite{Dong:2019piw}? Or, is this divergence indicating that we should restrict
to configurations where \(\mathcal{L}_v h_{ij}\) satisfies
\cref{eq:divergentbulk}? We leave these general questions for future work, but
let us note that, at least for the highly symmetric constrained saddles built
from black holes for the purpose of evaluating the thermal partition
function, we expect \(\mathcal{L}_{v}h_{ij}=0\) and this divergence to be
absent.

Another simplification of choosing \(D=3\) arose in defining a notion of angular
momentum \(\mathcal{J}\) evaluated on \(\gamma\). A distinguished role was
played by black holes in the evaluation of the thermal partition function, so,
for black holes, we would like \(\mathcal{J}\) (evaluated on the bifurcation
surface \(\gamma\)) to agree with the angular momentum of the black hole. On the
other hand, to construct constrained saddles when fixing \(\mathcal{J}\), we
needed \(\mathcal{J}\) to be defined locally on \(\gamma\), perhaps most
naturally, as an integral of the Brown-York momentum density
\((p_{\mathrm{BY}})_i\) --- see \cref{eq:pby2}. In \(D=3\), the angular momentum
of a black hole is given by one number and there is a natural constant vector
\(\chi^i\) on the one-dimensional surface \(\gamma\) with which to define
\(\mathcal{J}\propto \int_\gamma \chi^i\,(p_{\mathrm{BY}})_i\).

More generally,
the angular momenta of black holes are valued in the
\(\lfloor(D-1)/2\rfloor\)-dimensional Cartan subalgebra of
\(\mathfrak{so}(D-1)\). Additionally, in \(D\ge 4\), there might not be a
natural set of vector fields \(\chi_{(I)}^i\), where
\(I=1,\ldots,\lfloor(D-1)/2\rfloor\), which take the place of \(\chi^i\) to
define \(\mathcal{J}_{(I)}\). In highly symmetric configurations, such as
rotating black holes or constrained saddles constructed from them, the Killing
vectors on \(\gamma\) are natural candidates for \(\chi_{(I)}^i\). Indeed,
taking a Kaluza-Klein reduction, this symmetry is the reason why an analogous
problem does not arise in the definition of charges \(\mathcal{Q}_{(I)}\) in the
Yang-Mills generalization of our Maxwell analysis. The challenge, however, is to
define \(\mathcal{J}_{(I)}\) in more general configurations appearing in the
path integral, generically in the absence of such symmetry.

Rather than trying to isolate a handful of vector fields \(\chi_{(I)}^i\), one
might instead take the opposite, democratic approach of considering the infinite
set of all vector fields \(\chi^i\) on \(\gamma\) --- that is, in the path
integral, fixing then later integrating over
\(\mathcal{J}[\chi^i]\propto\int_\gamma\chi^i\,(p_{\mathrm{BY}})_i\) for all
vector fields \(\chi^i\). Equivalently, one can view this as fixing then
integrating over the local momentum density \((p_{\mathrm{BY}})_i\) at every
point on \(\gamma\). Of course, such a calculation would require careful
treatment of diffeomorphisms to ensure that gauge-equivalent momentum densities
over \(\gamma\) are not over-counted. One might also wonder whether similar
generalizations are applicable to the area, \ie{} Bekenstein-Hawking entropy
\(\mathcal{S}\), and charge \(\mathcal{Q}\), as measured by Gauss's law on
\(\gamma\). In particular, instead of \(\mathcal{S}\) and \(\mathcal{Q}\), one
might instead fix then later integrate over the volume form
\(\tensor[^{(D-2)}]{\epsilon}{}\) and the electric flux density \(*F\) at every
point on \(\gamma\).

It is interesting to note the close resemblance of such calculations to the
treatment of edge modes \cite{Donnelly:2015hxa,Donnelly:2014fua}.\footnote{For
  some more recent treatments of edge modes, see
  \cite{Donnelly:2016auv,Blommaert:2018oue,Chen:2023tvj,Ball:2024hqe,Chen:2024kuq}.}
For simplicity, let us review this for Maxwell theory in a non-dynamical
spacetime. We will consider states on a Cauchy slice \(\Sigma=R\cup L\), where
the two pieces \(R\) and \(L\) are separated by an entangling surface
\(\gamma\). Here, Gauss's law requires that the pullbacks of \(*F\) to
\(\gamma\) from \(R\) and \(L\) must agree. Consequently, the physical Hilbert
space on \(\Sigma\) is an ``entangling product'' \cite{Donnelly:2016auv} given,
roughly speaking, by the kernel of this constraint in the product
\(\mathcal{H}_R\otimes\mathcal{H}_L\) of the \(R\) and \(L\) Hilbert spaces. A
physical state, viewed as a special state in
\(\mathcal{H}_R\otimes\mathcal{H}_L\), must then reduce to a density matrix on,
say, \(R\) which commutes with the pullback of \(*F\) to \(\gamma\) from \(R\).
To isolate a given block of the density matrix for a given configuration of
\(*F\) on \(\gamma\), one then considers a path integral where \(*F\) is fixed
to said configuration. In the end, the full density matrix or partition function
is recovered by integrating over all configurations of \(*F\) on \(\gamma\)
\cite{Donnelly:2015hxa,Donnelly:2014fua}. In this paper, by fixing and then
later integrating over the total electric flux \(\int_\gamma *F\), we have
effectively focused on the constant edge mode\footnote{This constant edge mode
  is disallowed in contexts where \(\Sigma\) has no boundary for electric field
  lines to escape to. Asymptotically AdS black holes, however, have such a
  boundary.} on \(\gamma\); the other edge modes are associated to
configurations of \(*F\) which integrate to zero over \(\gamma\).

\subsubsection{Unstable variables}
\label{sec:unstablevars}

As described in \cref{sec:bhstability,sec:toyexamples}, to evaluate the thermal
partition function as an integral transform of a Lorentzian path integral,
particularly when using approximate saddle point methods, it is important to
hold certain possibly ``unstable'' variables fixed until they are integrated at
the end. In this paper, we have primarily focused on the quantities
\((\mathcal{S},\mathcal{J},\mathcal{Q})\) defined on the generically singular
codimension-two surface \(\gamma\). In principle, however, we should save the
integrals over all possibly unstable variables for last. Might there be other
such variables in the path integral beyond
\((\mathcal{S},\mathcal{J},\mathcal{Q})\), perhaps the nonconstant modes
described in \cref{sec:angmom}? Relatedly, what new types of singularities would
appear in the corresponding constrained saddles and what additional stability
conditions might arise from considering these other quantities?

For example, in certain parameter ranges, a black string (asymptotically flat
Schwarzschild times a spatial circle) has multiple Euclidean negative modes
\cite{Headrick:2009pv}: one inherited from the instability of asymptotically
flat Schwarzschild, and others from the Gregory-Laflamme instability
\cite{Gregory:1994tw}. The former is symmetric under the isometries of the
bifurcation surface and, in the formalism of ref.~\cite{Marolf:2022ybi} and this
paper, can be attributed to an instability under variations of \(\mathcal{S}\).
Indeed, asymptotically flat Schwarzschild and the black string have negative
specific heat. The latter Gregory-Laflamme instability, however, involves modes
that oscillate in the circle factor. To describe this latter type of
instability in our formalism, one might then expect to have to consider the
kinds of variables described in \cref{sec:angmom} built from nonconstant modes
on \(\gamma\). An interesting question is whether there are examples of saddles
which are stable under variations of \((\mathcal{S},\mathcal{J},\mathcal{Q})\),
but are unstable under variations of these other variables.

We will leave further study of the above questions for future work. However,
despite the conformal factor problem in Euclidean signature, let us comment
briefly that we do not expect the gravitational conformal mode to be among the
list of unstable variables that should be integrated last in our formalism.
(Indeed, if we are forced to perform a manifestly divergent integral over the
conformal mode after the integral transform \labelcref{eq:inttrans} from
Lorentzian time \(T\) to Euclidean time \(\beta\), then that would defeat the
purpose of starting in Lorentz signature to avoid the conformal factor problem.)
The reason is that instabilities due to the conformal mode seem to be eliminated
by the constraint equations associated with diffeomorphism-invariance, at least
in the linearized theories in expansions around solutions
\cite{Schleich:1987fm,Mazur:1989by,Hartle:2020glw,Dittrich:2024awu}.\footnote{Firstly,
  this is not to say that all instabilities of the linearized theories are
  eliminated by the constraints. Expected physical instabilities --- \eg{}
  associated with negative specific heat or, equivalently, variations in
  \(\mathcal{S}\) --- remain after imposing the constraints. Secondly, it may be
  the case that the Euclidean action for the \emph{nonlinear} theory is still
  unbounded from below over Euclidean configurations satisfying the constraints,
  even if the linearized theory is stable around a given solution. Simply
  imposing constraints therefore does not solve the conformal factor problem for
  path integrals over Euclidean metrics.} In the canonical form of the
gravitational action, lapse and shift appear as Lagrange multipliers for the
constraints. In a Lorentzian path integral, integrating over all\footnote{In
  some older work on ``third quantization'', \eg{} ref.~\cite{Giddings:1988wv},
  the constant lapse mode is only integrated over positive reals, leading to
  gravitational wavefunctions which are Green's functions, as opposed to
  solutions, for the Wheeler-de Witt constraint. However,
  ref.~\cite{Casali:2021ewu} instead advocates for integrating lapse over the
  full real line in the gravitational path integral. This leads to a group
  averaged inner product which projects down to the physical Hilbert space
  satisfying the Wheeler-de Witt constraint.} real\footnote{In contrast, rotated
  integration contours must be used in Euclidean signature to achieve the same
  effect. Recently, for example, ref.~\cite{Maldacena:2024spf} has highlighted
  how the rotated contour for the Euclidean lapse constant mode is needed to
  impose the Wheeler-de Witt constraint and explain an unexpected factor of
  \(i\) appearing in the gravitational path integral on a sphere (with a
  particle observer). The integral over real Lorentzian lapse and shift, on the
  other hand, seems far more natural.} values of lapse and shift results in
\(\delta\)-functionals imposing the
constraints.\footnote{Ref.~\cite{Dittrich:2024awu} makes similar comments in the
  context of a Lorentzian simplicial path integral. Instead of a canonical path
  integral, one may also consider a manifestly covariant path integral over the
  \(D\)-dimensional metric (as we implicitly have in this paper), \eg{} obtained
  by integrating out the conjugate momenta of the \((D-1)\)-metric. In this
  language, a more appropriate set of words might be that, in the linearized
  theory, the path integral over the conformal mode is cancelled by a
  Faddeev-Popov determinant or Jacobian arising from dividing out
  diffeomorphisms. (See eq.~(3.20a) in ref.~\cite{Hartle:2020glw} or the
  discussion around eqs.~(2.28)-(2.33) in ref.~\cite{Mazur:1989by}.)} We then
expect these constraints to eliminate from the Lorentzian path integral the
conformal mode variables that we might have otherwise called unstable, \ie{}
have wrong signs for the Hessian of the action around saddle points.


\acknowledgments

I am very grateful to Donald Marolf for his guidance on this project and the
time he spent reviewing the draft of this lengthy paper. I would like to thank
Jos\'e Padua-Arg\"uelles for relevant discussions about torsion. I am further
grateful for interesting conversations with Adam Ball, David Grabovsky, Jesse
Held, Gary T.~Horowitz, Maciej Kolanowski, and Gabriel Wong. I am supported by a
Fundamental Physics Fellowship through the University of California, Santa
Barbara.



\bibliographystyle{JHEP.bst}
\bibliography{references.bib}

\end{document}